\documentclass[12pt]{article}
\usepackage{jheppub}
\usepackage{amsmath}
\allowdisplaybreaks
\usepackage{subcaption}

\newcommand{\cn}{{\cal N}}

\newcommand{\co}{{\cal O}}

\newcommand{\cs}{{\cal S}}

\newcommand{\cw}{{\cal W}}

\newcommand{\nn}{\nonumber}

\def\bal#1\eal{\begin{align}#1\end{align}}
\def\alp[#1]{\begin{align}#1\end{align}}

\def\secnum[#1]{\texorpdfstring{$#1$}{TEXT}}

\def\secnuml#1\secnumr{\texorpdfstring{$#1$}{TEXT}}

\def\eqa{\begin{eqnarray}}
\def\eqae{\end{eqnarray}}
\def\eq{\begin{equation}}
\def\eqe{\end{equation}}
\def\be{\begin{equation}}
\def\ee{\end{equation}}
\def\bea{\begin{eqnarray}}
\def\eea{\end{eqnarray}}
\def\ba{\begin{array}}
\def\ea{\end{array}}
\def\bd{\begin{displaymath}}
\def\ed{\end{displaymath}}

\def\>{\rangle}
\def\<{\langle}

\def\del{\delta}
\def\e{\epsilon}
\def\f{\phi}

\def\g{\gamma}

\def\l{\lambda}
\def\m{\mu}

\def\p{\pi}
\def\q{\theta}

\def\s{\sigma}
\def\t{\tau}

\def\F{\Phi}

\def\L{\Lambda}

\def\S{\Sigma}

\def\pa{\partial}

\title{$\mathcal{N}=(0,2)$ SYK, Chaos and Higher-Spins}

\author{Cheng Peng} 
\affiliation{Department of Physics, Brown University, Providence RI 02912, USA}
\emailAdd{cheng$\underline{~}$peng@brown.edu}

\abstract{
We study a 
2-dimensional SYK-like model with $\cn=(0,2)$ supersymmetry. 
The model 
describes $N$ chiral supermultiplets and $M$ Fermi supermultiplets with a $(q+1)$-field interaction. We solve the model analytically and numerically in the $N\gg 1$, $M\gg 1$ limit with $\mu\equiv \frac{M}{N}$ being a free parameter. 
Two distinct higher-spin symmetries emerge when the $\m$ parameter approaches the two ends of its range.  
This is verified by  the appearance of conserved higher-spin operators and the vanishing of chaotic behaviors in the 
two limits. 
Therefore this model provides a manifest realization of the widely believed connection between SYK-like models and higher-spin theories.
In addition, as the parameter $\m$ varies we find the largest Lyapunov exponent of this model to be  slightly larger than that in models with non-chiral supersymmetry. 
A tensor model without random couplings that shares the same infrared physics is also introduced. 
}

\begin{document}

\maketitle

\section{Introduction}

The Sachdev-Ye-Kitaev (SYK) model ~\cite{Sachdev:1992fk,Parcollet:1997ysb,PG,KitaevTalk1,KitaevTalk2,Maldacena:2016hyu, Kitaev:2017awl} provides a simple example of strongly coupled, yet perturbatvely  solvable, models\cite{KitaevTalk1,Polchinski:2016xgd,Maldacena:2016hyu,Jevicki2016,Gross:2017hcz, Gross:2017aos,	Kitaev:2017awl}. A reparameterization symmetry emerges in the infrared of this model~\cite{KitaevTalk2,Maldacena:2016hyu, Kitaev:2017awl} and its breaking leads to soft modes that are described by a Schwarzian derivative action~\cite{KitaevTalk2,Bagrets:2016cdf,Stanford:2017thb,Mertens:2017mtv,Jevicki:2016ito, Kitaev:2017awl}. The Schwarzian derivative action also describes dilaton gravity systems on near AdS$_2$ spacetimes~\cite{Maldacena:2016hyu, Kitaev:2017awl,Strominger:1998yg,Maldacena:1998uz,Almheiri:2014cka, Maldacena:2016upp,Engelsoy:2016xyb, Cvetic:2016eiv,Grumiller:2017qao,Maldacena:2018lmt}. 
In addition, the SYK model is chaotic~\cite{KitaevTalk1,KitaevTalk2,Maldacena:2016hyu, Bagrets:2017pwq}, which 
is also a characteristic feature of gravitational theories ~\cite{Shenker:2013pqa,Shenker:2014cwa,Maldacena:2015waa,Jensen:2016pah}. 
All these properties suggest a holographic duality between the SYK model and dilaton gravity theories~\cite{Sachdev:2010um,Sachdev:2015efa,KitaevTalk1,KitaevTalk2,Maldacena:2016hyu,Maldacena:2016upp,Kitaev:2017awl}.
Properties of the Hilbert space of the SYK model are studied in~\cite{Polchinski:2016xgd,Fu:2016yrv,Jevicki:2016bwu,Maldacena:2016hyu,Garcia-Alvarez:2016wem,Jevicki:2016ito, Garcia-Garcia:2016mno, Cotler:2016fpe, Garcia-Garcia:2017pzl,Kourkoulou:2017zaj,Sonner:2017hxc,a:2018kvh}.
The operator spectrum of the model consists of a tower of operators with finite anomalous dimensions~\cite{KitaevTalk2,Polchinski:2016xgd,Maldacena:2016hyu,Kitaev:2017awl}. The finite anomalous dimensions suggest~\cite{Maldacena:2016hyu} that the SYK model could be thought of as a deformation of the vector models that have a tower of higher spin operators with small anomalous dimensions.  Such deformation from a Gross-Neveu vector model to an SYK-like model is discussed explicitly in~\cite{Peng:2017kro}. It is shown in~\cite{Peng:2017kro} that there is a transition from the vector model to the SYK-like model, which is similar to other phase transitions observed in the SYK-like models~\cite{Banerjee:2016ncu,Bi:2017yvx,Jian:2017jfl,Song:2017pfw,Luo:2017bno,Nosaka:2018iat,Mondal:2018xwy}.    
Different bulk duals of the tower of  operators are proposed in~\cite{Gross:2017hcz,Gross:2017vhb, Taylor:2017dly,Das:2017hrt,Das:2017wae},  other discussions about the relations between the two sides can be found in~\cite{ Maldacena:2017axo,Murata:2017rbp,deBoer:2017xdk,Cai:2017nwk,Kitaev:2017hnr,Qi:2018rqm,Gonzalez:2018enk,Tarnopolsky:2018env}.  
Most of the analytic results of the SYK model are derived in a new type of large-$N$ limit that is shared in particular by models without random couplings~\cite{Witten:2016iux,Gurau:2011xq, Bonzom:2012hw,Carrozza:2015adg,Gurau:2016lzk,Klebanov:2016xxf,Nishinaka:2016nxg, Krishnan:2016bvg,Ferrari:2017ryl,Gurau:2017xhf,Bonzom:2017pqs,Itoyama:2017emp,Krishnan:2017ztz,Itoyama:2017xid, Narayan:2017qtw,Chaudhuri:2017vrv,Gurau:2017qna,Dartois:2017xoe,Klebanov:2017nlk,Mironov:2017aqv,Gurau:2017qya, Krishnan:2017txw,deMelloKoch:2017bvv,Giombi:2017dtl,Azeyanagi:2017drg,Bulycheva:2017ilt,Choudhury:2017tax,Krishnan:2017lra,Azeyanagi:2017mre,Itoyama:2017wjb,Benedetti:2017fmp,Halmagyi:2017leq,BenGeloun:2017jbi,Benedetti:2017qxl,Benedetti:2018goh,Krishnan:2018jsp,Delporte:2018iyf,Maldacena:2018vsr,Klebanov:2018nfp}.

To understand the relations between the SYK model and other better known models, different generalizations of the SYK model are constructed.  
One generalization is to include supersymmetry~\cite{Gross:2016kjj,Fu:2016vas,Peng:2016mxj,Murugan:2017eto,Peng:2017spg}. 
Aspects of supersymmetric SYK models have also been studied in~\cite{Gross:2016kjj, Sannomiya:2016mnj, Li:2017hdt,Forste:2017kwy, Kanazawa:2017dpd,Hunter-Jones:2017raw, Hunter-Jones:2017crg,Narayan:2017hvh, Forste:2017apw,Garcia-Garcia:2018ruf,Bulycheva2018}.
Another generalization is to higher dimensions~\cite{Gu:2016oyy,Berkooz:2016cvq,Davison:2016ngz,Turiaci:2017zwd,Berkooz:2017efq,Gu:2017ohj,Jian:2017unn,Chen:2017dav,Chen:2017dbb,Murugan:2017eto,Zhang:2017jvh,Jian:2017tzg,Simmons-Duffin:2017nub, Cai:2017vyk,Ge:2017fix}, whose simplest  example is in 2 dimension.
Continuum theories in 2-dimensional Euclidean spacetime   are usually studied in terms of the left- and right-moving sectors due to the factorization of the isometry.  
The examples of 2d SYK-like models studied previously are all symmetric between the left- and the right-moving sectors. 

One could also consider models whose left- and right-moving sectors are not symmetric.
In this paper we study some 2d SYK-like models of this kind.  
The models have an $\cn=(0,2)$ supersymmetry in the UV. 
In the infrared, these theories are dominated by the set of melonic diagrams in the large-$N$ limit and can be solved as all other SYK-like models. 

The $\cn=(0,2)$ supersymmetry plays an important role of this model.
The $\cn=2$ supersymmetry in the right-moving sector makes the IR solution reliable. On the other hand the  absent of supersymmetry in the left-moving sector gives some room for interesting properties that are not observed in previous models.
In particular, due to the smaller number of supersymmetry it is possible to study a one parameter family of such models. As a result, one could move on the moduli space of such models and understand their peculiar features, as well as their possible connections with other well studied models. In this paper we study two examples of such interesting consequences.

Firstly, the Lyapunov exponent of the supersymmetric model considered in \cite{Murugan:2017eto}, see also \cite{Bulycheva:2017uqj}, is $\l_L=0.5824$, which does not saturate the chaotic bound \cite{Maldacena2016a}. It is then an interesting question to ask if there are other 2d SYK-like models that have larger or maximal Lyapunov exponent. In this paper, we show that in our $\cn=(0,2)$ setting, as we dial the free parameter, there is a  continuous family of theories that have slightly larger Lyapunov exponent  comparing to the supersymmetric models considered in \cite{Murugan:2017eto}. This is discussed in detail in section~\ref{sec:chaos}.

Another interesting consequence is the existence of certain higher-spin limits. By continuously tune the  parameter to some limiting values, we observe the emergence of higher-spin conserved currents explicitly. Besides, we observe the correlation of the emerging of the higher-spin symmetry and the fading of the chaotic behavior. This provides a manifestation of a connection  between higher-spin like models and SYK-like models. The details of such higher-spin limits are analyzed in section~\ref{hslimits}.

\section{An $\cn=(0,2)$ supersymmetry SYK model}\label{syk02s}

\subsection{Review of 2d $\cn=(0,2)$ supersymmetry}\label{review} 

In  this section we review some properties of 2-dimensional theories with $\cn=(0,2)$ supersymmetry. 
We work in Euclidean signature, where the two coordinates are $x^0,x^1$. 
We define
\bal
&z \equiv x^0+ix^1\,,\qquad \bar{z}\equiv x^0-ix^1\,,
\eal
and the derivatives become
\bal
&\pa_z=\frac{1}{2}(\pa_0-i \pa_1)\,,\qquad \pa_{\bar{z}}=\frac{1}{2}(\pa_0+i \pa_1) \ .
\eal
The $\cn=(0,2)$ supersymmetry is generated by 2 supercharges. In the superspace formalism they read 
\bal
& Q_+=\frac{\pa}{\pa \q^+}- 2\bar{\q}^+\pa_z\,,\qquad
\bar{Q}_+=-\frac{\pa}{\pa \bar{\q}^+}+ 2{\q}^+\pa_z\ .
\eal
The super-derivatives are
\bal
& D_+=\frac{\pa}{\pa \q^+}+2\bar{\q}^+\pa_z\,,\qquad
\bar{D}_+=-\frac{\pa}{\pa \bar{\q}^+}- 2{\q}^+\pa_z\ .
\eal
It is easy to check that the supercharges anticommute with the super-derivatives.

We consider models of two kinds of superfields. 
The chiral/anti-chiral superfields
\bal
\F&=\f + \sqrt{2} \q^+ \psi + 2\q^+\bar{\q}^+ \pa_z\f\,,\qquad
\bar\F=\bar\f - \sqrt{2} \bar\q^+\bar\psi  -  2\q^+\bar{\q}^+\pa_z\bar\f\,,
\eal
satisfy
\bal
\bar{D}_+ \F =0\,, \qquad {D}_+ \bar{\F} =0\ .
\eal
We also consider Fermi multiplets
\bal
\L
&=\l-\sqrt{2}\q^+G+2\q^+\bar{\q}^+\pa_z\l-\sqrt{2}\bar{\q}^+E\\
\bar\L
&=\bar\l -\sqrt{2}\bar\q^+\bar G -2 \q^+\bar{\q}^+ \pa_z \bar\l
-\sqrt{2}{\q}^+\bar E\,,
\eal
where
\bal
E(\F)&=E(\f_a)+\sqrt{2}\q^+\frac{\pa E}{\pa \f_a}\psi_{a}+2\q^+\bar\q^+\pa_{ \bar z}E(\f_a)\\
\bar E(\F)&=\bar E(\bar\f_a)+\sqrt{2}\bar\q^+\frac{\pa \bar E}{\pa \bar\f_a}\bar\psi_{a}-2\q^+\bar\q^+\pa_{ \bar z}\bar E(\bar\f_a)\,,
\eal
are (anti-)chiral superfields where the subscript $a$ labels different chiral superfields. The Fermi supermultiplets  satisfy
\bal
&\bar{D}_+ \L=\sqrt{2} E\,,\qquad \bar{D}_+ E=0\,,\label{dbL}\\
&{D}_+ \bar\L=\sqrt{2} \bar E\,,\qquad {D}_+ \bar E=0\ .
\eal
The supersymmetry transformation of the fields in the chiral supermultiplet are 
\bal
Q_+ \f= \sqrt{2}\psi\,, \qquad &Q_+ \psi= 0\,,\qquad 
\bar{Q}_+ \f=0\,, \qquad \bar{Q}_+ \psi= -2 \sqrt{2}\pa_z\f\\
\bar{Q}_+ \bar\f= \sqrt{2}\bar\psi\,, \qquad &\bar{Q}_+ \bar\psi= 0
\,,\qquad {Q}_+ \bar\f=0\,, \qquad  {Q}_+ \bar\psi= -2 \sqrt{2}\pa_z\bar\f\ .
\eal
The supersymmetry  transformation of the fields  in the Fermi supermultiplet are
\bal
&Q_+ \l= -\sqrt{2}G\,, \quad Q_+ G= 0\,, \quad
\bar{Q}_+ \l=\sqrt{2} E\,, \quad~~ \bar{Q}_+ G= 2 \sqrt{2}\pa_z\l+\frac{\pa E}{\pa \f_a}\psi_{a}\,,  \\
&\bar{Q}_+ \bar\l=\sqrt{2}\bar{G}\,, ~~\quad \bar{Q}_+ \bar{G}= 0\,,\quad Q_+ \bar\l= -\sqrt{2}\bar{E}\,, \quad Q_+ \bar{G}= - 2\sqrt{2}  \pa_z \bar\l+\frac{\pa \bar{E}}{\pa \bar\f_a}\bar\psi_{a}
\ .
\eal
In the rest of the paper we consider special models with $E=0$.
 
Given these transformations,  propagators of the  different components of a chiral supermultiplet are related  by 
\bal
G^\psi(z_1,z_2)=-2\pa_{z_1} G^\f(z_1,z_2)= 2\pa_{z_2} G^\f(z_1,z_2)\label{proppf}
\eal
The similar relation for the Fermi multiplet is
\bal
\<\bar{G}(z_1)G(z_2)\>&=\<\bar{Q}\bar{\l}(z_1)G(z_2)\>/\sqrt{2}=\<\bar{\l}(z_1)\bar{Q}G(z_2)\>/\sqrt{2}\\
&=\<\bar{\l}(z_1)(2 \sqrt{2}\pa_z\l_-+\frac{\pa E}{\pa \f_a}\psi_{a})(z_2)\>/\sqrt{2}\ .
\eal
For the case with $E=0$, we simply get
\bal
G^G(z_1,z_2)=-2\pa_1 G^\l(z_1,z_2)= 2\pa_2 G^\l(z_1,z_2)\ .\label{propgl}
\eal
The D-terms of a chiral and a Fermi superfields are respectively
\bal
S^0_\F&=-\int dx^2 d\q^+ d\bar{\q}^+ \bar{\F}\pa_{\bar{z}}\F\,,\\
S^0_\L&=\frac{1}{2}\int dx^2 d\q^+ d\bar{\q}^+ \bar{\L}\L\ .
\eal
In addition, we turn on holomorphic superpotentials that contribute F-term potentials. For the $\cn=(0,2)$ models, the holomorphic superpotential takes a general form
\bal
\int dx^2 d\q^+ G(x,\q^+,\bar\q^+)\,,
\eal
where $G(x,\q,\bar\q)$ is some fermionic superfield that satisfies $\bar{D}_+ G=0$. 
It is easy to check that the above results agree with the Euclidean continuation of the results~\cite{Witten:1993yc} in Lorentzian signature.

\subsection{The $\cn=(0,2)$ SYK model}\label{syk02} 

We consider a special model of $N$ chiral multiplets and $M$ Fermi multiplets with the F-term potential
\bal
G(x,\q,\bar{\q})&=\frac{J_{ia_1\ldots a_q}}{q!} \L_-^i \F^{a_1}\ldots\F^{a_q}\,,\label{J1}
\eal
where $i,j,k,\ldots$ label the Fermi  multiplets and $a,b,\ldots$ label the chiral multiplets. The $J_{ia_1\ldots a_q}$ coupling has dimension $(\frac{1}{2},\frac{1}{2})$ and is from a Gaussian distribution
\bal
\<J_{ia_1\ldots a_q}J_{ia_1\ldots a_q}\>=\frac{(q-1)!}{N^q}J^2 \ .
\eal 
In component form, the above action reads
\begin{multline}
S_{\rm int}=\int dx^2 \,  \Bigg(\frac{\sqrt{2}J_{ia_1\ldots a_q}}{(q-1)!}\,\l^i \psi^{a_1}\f^{a_2}\ldots\f^{a_q}+ \frac{\sqrt{2}J_{ia_1\ldots a_q}}{q!}\,G^i \f^{a_1}\ldots \f^{a_q}~+h.c.\Bigg) \ . \label{Sintsusy}
\end{multline}
When $M=N$ the supersymmetry is enhanced to $\cn=(2,2)$ and the theory reduces to the models discussed in \cite{Murugan2017, Bulycheva2018} whose action is recast in \eqref{n22s}.  

\begin{figure}
	\hspace{-2.1cm}\includegraphics[width=1.26\linewidth]{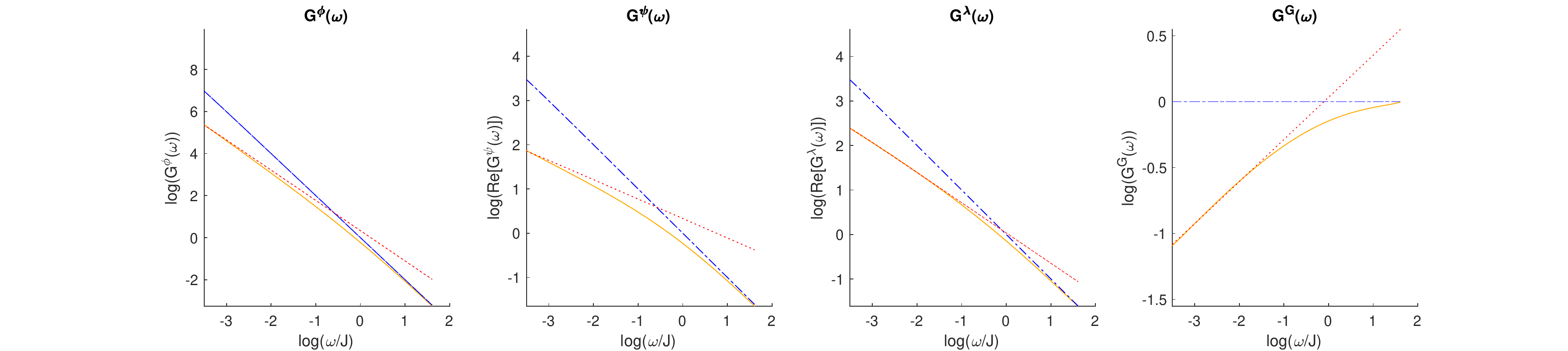}
	\caption{Numerical solutions to the Schwinger-Dyson equations. The blue dash-dot lines are the free propagators in the UV, the red dotted lines are the IR solutions. The yellow curves are result from solving the Schwinger-Dyson equations numerically. One observes that the numerical solutions interpolate between the  UV and IR behaviors. The calculation is done for $q=3$, $\m=1.5$.  }
	\label{fig:n02cl1}
\end{figure}

The self-energies of the fields are
\bal
\S^\psi(z_1,z_2)&= 2J^2 \frac{M}{N}  G^{ {\l}}(z_1,z_2) (G^{ {\f}}(z_1,z_2))^{q-1}\,,\label{sigmapsi02}\\
\S^\f(z_1,z_2)&=(q-1)2 \frac{M}{N} J^2( \nn G^{{\f}}(z_1,z_2))^{q-2}G^{{\l}}(z_1,z_2)G^{{\psi}}(z_1,z_2)\\
&\hspace{5cm} +2 { J^2} \frac{M}{N} (G^{{\f}}(z_1,z_2))^{q-1} G^{{G}}(z_1,z_2)\,,\\
\S^G(z_1,z_2)&=  \frac{2   J^2}{  q} (G^{{\f}}(z_1,z_2))^q\,, \label{sigmag02}\\
\S^\l(z_1,z_2)&=   {2  J^2}{ } (G^{ \f}(z_1,z_2))^{q-1} G^{ \psi}(z_1,z_2)\ .\label{sigmaL02}
\eal
It is easy to get the set of $\cn=(0,2)$ supersymmetric solutions of the form
\bal
G^I_c(z_1,z_2)&=\frac{n_I}{(z_1-z_2)^{2h_I} (\bar{z}_1-\bar{z}_2)^{2\tilde{h}_I} }\,, \label{02sol}
\eal
where $ n_{\lambda }n_{\phi }^{q}= -\frac{  (q-1) q }{2 \pi ^2 J^2 \left(\mu  q^2-1\right)}$ and 
\bal
& h_{\phi }= \frac{\mu  q-1}{2 \mu  q^2-2}\,,~~ h_{\psi }= \frac{\mu  q^2+\mu  q-2}{2 \mu  q^2-2}\,,~~ h_{\lambda }= \frac{q-1}{2 \mu  q^2-2}\,,~~ h_G= \frac{\mu  q^2+q-2}{2 \mu  q^2-2}\label{02sol1}\\
& \tilde{h}_{\phi }= \frac{\mu  q-1}{2 \mu  q^2-2}\,,~~\tilde{h}_{\psi }= \frac{\mu  q-1}{2 \mu  q^2-2}\,,~~\tilde{h}_{\lambda }=\frac{\mu  q^2+q-2}{2\mu  q^2-2} \,,~~\tilde{h}_G= \frac{\mu  q^2+q-2}{2\mu  q^2-2}\ .\label{02sol2}
\eal
One can solve the Schwinger-Dyson equation numerically to confirm that the model indeed flows to this IR solution. The numerical solution is shown in figure~\ref{fig:n02cl1}.

We should comment on one subtlety in this computation. Since we look for supersymmetric solutions, we only need to solve the Schwinger-Dyson equation of one component of each multiplet; the equation of the other component is then automaticaly satisfied due to supersymmetry. When we solve the equations of the chiral multiplet, we notice that in the UV regime the Fourier transform involves an integral of the form
\bal
\int rdr d\q \,r^{ 2\frac{\mu  q-1}{\mu  q^2-1}-3}e^{i\q} e^{i r \cos\q}\ .
\eal 
If we directly count the power of $r$, it seems that there is a divergence in this Fourier transformation. However, when we check the behavior near $r=0$, we can expand the $e^{ircos\q}$ factor to get
\bal
\int dr d\q r^{ 2\frac{\mu  q-1}{\mu  q^2-1}-2} (1+i r \cos\q+\ldots ) e^{i\q}\ .
\eal
We observe that the first term vanishes due to the $e^{i\q}$ in the $\q$ integral.\footnote{We thank Douglas Stanford for pointing this out.} So the leading term at $r\sim 0$ is
\bal
\int dr d\q \, i r^{ 2\frac{\mu  q-1}{\mu  q^2-1}-1} \cos\q e^{i\q}\ .\label{ftf}
\eal
Therefore as long as we focus on the models with $\m>\frac{1}{q}$, this integral converges and the model does flow to the IR solution we found above.

\section{Four-point functions}

In this section, we consider 4-point functions of this model. Because there are two different types of multiplets in the model, there will be a few different 4-point correlation functions. 
As in the 1-dimensional cases \cite{Fu:2016vas,Peng2017,Peng2017a}, the correlation function can be computed either in terms of superfields or component fields. In the rest of the paper we work in the component formalism.

\subsection{Operator spectrum}

We are interested in the 4-point function $\<\bar{\f}^i\f^i\bar{\f}^j\f^j\>$, which mixes with $\<\bar{\phi}^i\phi^i\bar{\psi}^j\psi^j\>$,  $\<\bar{\f}^i\f^i\bar{\l}^j\l^j\>$, $\<\bar{\psi}^i\psi^i\bar{\l}^j\l^j\>$ and $\<\bar{\f}^i \f^i \bar{G}^j G^j\>$. There are in total 9 kernels that contribute to these 4-point functions. 

\begin{figure}[t]
	\centering
	\includegraphics[width=0.5\linewidth]{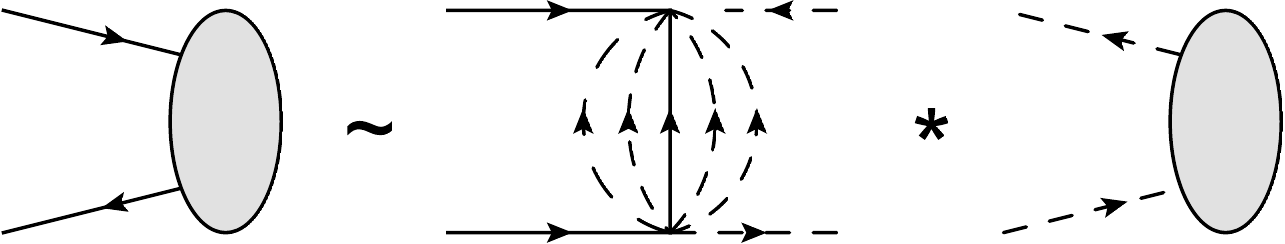}
	\caption{Action of a single kernel.}
	\label{fig:ker1}
\end{figure}

\begin{figure}[t]
	\centering
	\includegraphics[width=0.7\linewidth]{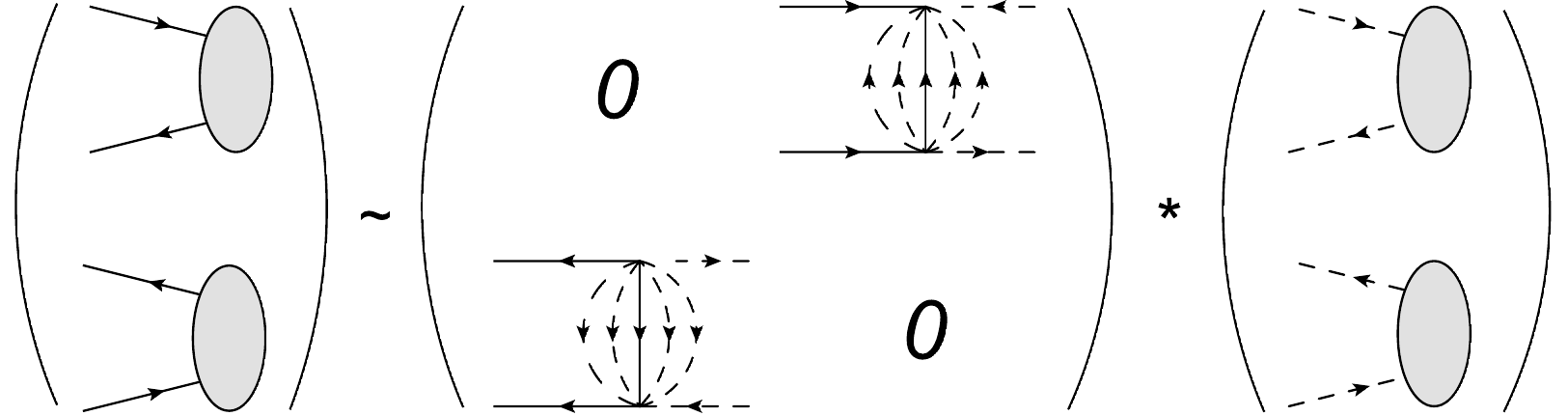}
	\caption{The eigenfunctions come in pairs. The two kernels in the  $2\times 2$ matrix give identical contributions to the eigenvalue matrix 	\eqref{evmx}, which leads to the extra $\s_1$
	factor.}
	\label{fig:ker2}
\end{figure}

The kernels take the following expressions
\bal
&K^{\f\f}(z_1,z_2,z_3,z_4)=2(q-1) J^2 \frac{M}{N} G^\f(z_{13}) G^\f(z_{24}) G^G(z_{34}) (G^\f(z_{34}) )^{q-2}\label{Kff}\\
&\qquad +2(q-1)(q-2) J^2 \frac{M}{N} G^\f(z_{13}) G^\f(z_{24}) G^\psi(z_{34}) G^\l(z_{34}) (G^\f(z_{34}) )^{q-3}\\
&K^{\f\psi}(z_1,z_2,z_3,z_4)=2(q-1) J^2 \frac{M}{N} G^\f(z_{13}) G^\f(z_{24}) G^\l(z_{34}) (G^\f(z_{34}) )^{q-2}\\
&K^{\f\l}(z_1,z_2,z_3,z_4)=2(q-1) J^2  G^\f(z_{13}) G^\f(z_{24}) G^\psi(z_{34}) (G^\f(z_{34}) )^{q-2}\\
&K^{\f G}(z_1,z_2,z_3,z_4)=2 J^2  G^\f(z_{13}) G^\f(z_{24})  (G^\f(z_{34}) )^{q-1}\\
&K^{\psi\f}(z_1,z_2,z_3,z_4)=-2(q-1) J^2 \frac{M}{N} G^\psi(z_{13}) G^\psi(z_{24}) G^\l(z_{34}) (G^\f(z_{34}) )^{q-2}\\
&K^{\psi\l}(z_1,z_2,z_3,z_4)=-2  J^2   G^\psi(z_{13}) G^\psi(z_{24}) (G^\f(z_{34}) )^{q-1}\\
&K^{\l\f}(z_1,z_2,z_3,z_4)=-2(q-1) J^2 \frac{M}{N}  G^\l(z_{13}) G^\l(z_{34})G^\psi(z_{34}) (G^\f(z_{34}) )^{q-2}\\
&K^{\l\psi}(z_1,z_2,z_3,z_4)=-2  J^2  \frac{M}{N} G^\l(z_{13}) G^\l(z_{24}) (G^\f(z_{34}) )^{q-1}\\
&K^{G\f}(z_1,z_2,z_3,z_4)=-2  J^2  \frac{M}{N} G^G(z_{13}) G^G(z_{24}) (G^\f(z_{34}) )^{q-1}\,,\label{KGf}
\eal
where we use the short hand notation $z_{ij}=z_i-z_j$. The following ansatz  \bal
\F^i(z_1,z_2)=(z_{12})^{h-2h_i} (\bar{z}_{12})^{\tilde{h}-2\tilde{h}_i}\,,\qquad i=\f,\psi,\l, G\,,
\eal
turns out to be the eigenfunctions of the above kernels
\bal
K^{(ij)}*\F^j=k^{ij} \F^i\ .\label{eigK}
\eal
where the $*$ denotes a convolution in position space. Making use of the following integral formula \cite{Murugan2017}
\bal
&\int d^2 y (y-t_0)^{a+n}(\bar{y}-\bar{t}_0)^a(t_1-y)^{b+m}(\bar{t}_1-\bar{y})^b\\
&=(t_0-t_1)^{a+n+b+m+1}(\bar{t}_0-\bar{t}_1)^{a+b+1}\p \frac{\Gamma (a+1) \Gamma (b+1) \Gamma (-a-b-m-n-1)}{\Gamma (a+b+2) \Gamma (-a-n) \Gamma (-b-m)}\,,
\eal
one finds the non-vanishing eigenvalues to be
\bal
k^{\f\f}&=\frac{\mu  (q-1)^2 q \left(\mu  q^2-2 \mu  q+1\right) \Gamma \left(\frac{(q-1) q \mu }{q^2 \mu -1}\right)^2 \Gamma \left(\frac{-h \mu  q^2+\mu  q+h-1}{q^2 \mu -1}\right) \Gamma \left(\tilde{h}-\frac{(q-1) q \mu }{q^2 \mu -1}\right)}{\left(\mu  q^2-1\right)^2 \Gamma \left(\frac{q \mu -1}{q^2 \mu -1}\right)^2 \Gamma \left(\frac{h \mu  q^2-2 \mu  q^2+\mu  q-h+1}{1-q^2 \mu }\right) \Gamma \left(\tilde{h}+\frac{(q-1) q \mu }{q^2 \mu -1}\right)}\label{kn1}\\
k^{\f\psi}&=-\frac{\mu  (q-1)^2 q \Gamma \left(\frac{(q-1) q \mu }{q^2 \mu -1}\right)^2 \Gamma \left(\frac{-h \mu  q^2+\mu  q+h-1}{q^2 \mu -1}\right) \Gamma \left(\tilde{h}-\frac{(q-1) q \mu }{q^2 \mu -1}\right)}{2 \left(\mu  q^2-1\right) \Gamma \left(\frac{q \mu -1}{q^2 \mu -1}\right)^2 \Gamma \left(\frac{h \mu  q^2-2 \mu  q^2+\mu  q-h+1}{1-q^2 \mu }\right) \Gamma \left(\tilde{h}+\frac{(q-1) q \mu }{q^2 \mu -1}\right)}\\
k^{\f\l}&=-\frac{4 \pi ^2 J^2 (q-1) n_{\f}^{q+1} (\mu  q-1) \Gamma \left(\frac{(q-1) q \mu }{q^2 \mu -1}\right)^2 \Gamma \left(\frac{-h \mu  q^2+\mu  q+h-1}{q^2 \mu -1}\right) \Gamma \left(\tilde{h}-\frac{(q-1) q \mu }{q^2 \mu -1}\right)}{\left(\mu  q^2-1\right) \Gamma \left(\frac{q \mu -1}{q^2 \mu -1}\right)^2 \Gamma \left(\frac{h \mu  q^2-2 \mu  q^2+\mu  q-h+1}{1-q^2 \mu }\right) \Gamma \left(\tilde{h}+\frac{(q-1) q \mu }{q^2 \mu -1}\right)}\\
k^{\f G}&=\frac{2 \pi ^2 J^2 n_{\f}^{q+1} \Gamma \left(\frac{(q-1) q \mu }{q^2 \mu -1}\right)^2 \Gamma \left(\frac{-h \mu  q^2+\mu  q+h-1}{q^2 \mu -1}\right) \Gamma \left(\tilde{h}-\frac{(q-1) q \mu }{q^2 \mu -1}\right)}{\Gamma \left(\frac{q \mu -1}{q^2 \mu -1}\right)^2 \Gamma \left(\frac{h \mu  q^2-2 \mu  q^2+\mu  q-h+1}{1-q^2 \mu }\right) \Gamma \left(\tilde{h}+\frac{(q-1) q \mu }{q^2 \mu -1}\right)}\\
k^{\psi\f}&=-\frac{2 \mu  (q-1)^2 q (\mu  q-1)^2 \Gamma \left(\frac{(q-1) q \mu }{q^2 \mu -1}\right)^2 \Gamma \left(\frac{-h \mu  q^2+\mu  q^2+\mu  q+h-2}{q^2 \mu -1}\right) \Gamma \left(\tilde{h}-\frac{(q-1) q \mu }{q^2 \mu -1}\right)}{\left(\mu  q^2-1\right)^3 \Gamma \left(\frac{\mu  q^2+\mu  q-2}{q^2 \mu -1}\right)^2 \Gamma \left(\frac{-h \mu  q^2+(q-1) \mu  q+h}{q^2 \mu -1}\right) \Gamma \left(\tilde{h}+\frac{(q-1) q \mu }{q^2 \mu -1}\right)}\\
k^{\psi\l}&=\frac{8 \pi ^2 J^2 n_{\f}^{q+1} (\mu  q-1)^2 \Gamma \left(\frac{(q-1) q \mu }{q^2 \mu -1}\right)^2 \Gamma \left(\frac{-h \mu  q^2+\mu  q^2+\mu  q+h-2}{q^2 \mu -1}\right) \Gamma \left(\tilde{h}-\frac{(q-1) q \mu }{q^2 \mu -1}\right)}{\left(\mu  q^2-1\right)^2 \Gamma \left(\frac{\mu  q^2+\mu  q-2}{q^2 \mu -1}\right)^2 \Gamma \left(\frac{-h \mu  q^2+(q-1) \mu  q+h}{q^2 \mu -1}\right) \Gamma \left(\tilde{h}+\frac{(q-1) q \mu }{q^2 \mu -1}\right)}\\
k^{\l\f}&=-\frac{\mu  (q-1)^3 q^2 n_{\f}^{-q-1} (\mu  q-1) \Gamma \left(\frac{1-q}{q^2 \mu -1}\right)^2 \Gamma \left(\frac{-h \mu  q^2+q+h-1}{q^2 \mu -1}\right) \Gamma \left(\frac{q+\tilde{h} \left(q^2 \mu -1\right)-1}{q^2 \mu -1}\right)}{4 \pi ^2 J^2 \left(\mu  q^2-1\right)^3 \Gamma \left(\frac{q-1}{q^2 \mu -1}\right)^2 \Gamma \left(\frac{-h \mu  q^2+2 \mu  q^2-q+h-1}{q^2 \mu -1}\right) \Gamma \left(\frac{-q+\tilde{h} \left(q^2 \mu -1\right)+1}{q^2 \mu -1}\right)}\\
k^{\l\psi}&=\frac{\mu  (q-1)^2 q^2 n_{\f}^{-q-1} \Gamma \left(\frac{1-q}{q^2 \mu -1}\right)^2 \Gamma \left(\frac{-h \mu  q^2+q+h-1}{q^2 \mu -1}\right) \Gamma \left(\frac{q+\tilde{h} \left(q^2 \mu -1\right)-1}{q^2 \mu -1}\right)}{8 \pi ^2 J^2 \left(\mu  q^2-1\right)^2 \Gamma \left(\frac{q-1}{q^2 \mu -1}\right)^2 \Gamma \left(\frac{-h \mu  q^2+2 \mu  q^2-q+h-1}{q^2 \mu -1}\right) \Gamma \left(\frac{-q+\tilde{h} \left(q^2 \mu -1\right)+1}{q^2 \mu -1}\right)}\\
k^{G\f}&=\frac{\mu  (q-1)^4 q^2 n_{\f}^{-q-1} \Gamma \left(\frac{1-q}{q^2 \mu -1}\right)^2 \Gamma \left(\frac{-h \mu  q^2+\mu  q^2+q+h-2}{q^2 \mu -1}\right) \Gamma \left(\frac{q+\tilde{h} \left(q^2 \mu -1\right)-1}{q^2 \mu -1}\right)}{2 \pi ^2 J^2 \left(\mu  q^2-1\right)^4 \Gamma \left(\frac{\mu  q^2+q-2}{q^2 \mu -1}\right)^2 \Gamma \left(\frac{-h \mu  q^2+\mu  q^2-q+h}{q^2 \mu -1}\right) \Gamma \left(\frac{-q+\tilde{h} \left(q^2 \mu -1\right)+1}{q^2 \mu -1}\right)}\ .\label{kn9}
\eal
It is convenient to organize the eigenvalues into an $8\times 8$ matrix 
\bal
\begin{pmatrix}
	0 &1 \\
	1 & 0
\end{pmatrix}
\otimes 
\begin{pmatrix}
	k^{\f\f} & k^{\f\psi} & k^{\f\l} & k^{\f G} \\
	k^{\psi\f} & 0 & k^{\psi\l} & 0\\
	k^{\l\f} & k^{\l\psi} & 0 & 0\\
	k^{G\f} & 0 & 0 & 0
\end{pmatrix} 
\label{evmx}
\eal
and the final eigenvalues come from diagonalizing this matrix. Here the presence of the extra $\s_1=\begin{pmatrix}
0 &1 \\
1 & 0
\end{pmatrix}$ matrix is due to the form of the interaction in \eqref{Sintsusy}. To illustrate it, we consider, for example, the action of the kernel $K^{\f\psi}$ on the eigenfunction $\F^{\psi}$. As shown in figure~\ref{fig:ker1}, after the action of the kernel, the upper leg becomes a conjugate field, namely the direction of the arrow is flipped. Therefore the actual eigenfunctions come in conjugate pairs. This means the actual action of the kernel should be like in figure~\eqref{fig:ker2}. Further notice that due to the form of the IR propagators, the two entries in the kernel matrix share the same expression. This leads to the extra tensor product with the $\s_1$ matrix in \eqref{evmx}. 

Diagonalizing this matrix, there are 4 eigenvalues being the 4 roots of the following equation
\begin{multline}
E_c(x,h,\tilde{h},\m,q)=x^4- k^{\phi \phi } x^3-\left( k^{  \phi G} k^{  G \phi }+ k^{\phi \psi } k^{\psi \phi }+ k^{\phi \lambda } k^{\lambda \phi }+ k^{\psi \lambda } k^{\lambda \psi }\right) x^2\\
+\left(k^{\phi \phi } k^{\psi \lambda } k^{\lambda \psi }- k^{\phi \psi } k^{\psi \lambda } k^{\lambda \phi }- k^{\phi \lambda } k^{\psi \phi } k^{\lambda \psi }\right) x+k^{  \phi G} k^{\psi \lambda } k^{\lambda \psi } k^{  G\phi }=0\,,\label{eigens}
\end{multline}
which we will call the symmetric eigenvalues.
There are another 4 eigenvalues being the solution of the equation
\begin{multline}
E'_c(x,h,\tilde{h},\m,q)=x^4+ k^{\phi \phi } x^3-\left( k^{  \phi G} k^{  G \phi }+ k^{\phi \psi } k^{\psi \phi }+ k^{\phi \lambda } k^{\lambda \phi }+ k^{\psi \lambda } k^{\lambda \psi }\right) x^2\\
-\left(k^{\phi \phi } k^{\psi \lambda } k^{\lambda \psi }- k^{\phi \psi } k^{\psi \lambda } k^{\lambda \phi }- k^{\phi \lambda } k^{\psi \phi } k^{\lambda \psi }\right) x+k^{  \phi G} k^{\psi \lambda } k^{\lambda \psi } k^{  G\phi }=0\,,\label{eigena}
\end{multline}
which we will call the antisymmetric eigenvalues. Their presence is a result of the complex fundamental fields in our model \eqref{Sintsusy}, similar to the 1-dimensional cases \cite{Fu:2016vas,Peng2017a,Bulycheva:2017uqj}.
We have not succeeded in getting  simple  expressions of the eigenvalues. But the equation \eqref{eigens} and \eqref{eigena} pass a few consistency checks.  

First, as we discussed above we expect the result to  reduce to that of the $\cn=(2,2)$ model in the $\m\to 1$ limit. Indeed, we can solve \eqref{eigens} and \eqref{eigena} at $\m=1$ to find the following 8 eigenvalues 
\bal
 \pm k^{FB}(h -\frac{1}{2},\tilde{h}-1)\,,\quad  \pm k^{FB}(h+\frac{1}{2} ,\tilde{h}-1)\,,\quad \pm k^{FB}(h-\frac{1}{2},\tilde{h})\,,\quad \pm k^{FB}(h+\frac{1}{2},\tilde{h})\,, \label{N22eigen}
\eal 
which are consistent with the $\cn=(2,2)$ result. At generic $\m$, the eigenfunctions can be considered as deformations of the eigenvalues \eqref{N22eigen}. 

Another consistency check is the presence of the stress-energy tensor at any generic $\m$. To see this, recall that in this model the 4-point functions are sums of ladder diagrams. Therefore there is always a factor
\bal
\frac{1}{1-k_i}\,, \label{ladder}
\eal
where $i=1,\ldots,8$ are the 8 eigenvalues obtained from solving \eqref{eigens} and \eqref{eigena}. Then an operators with dimension $(h_*,\tilde{h}_*)$ running in each channel can be represented by a pole in the factor \eqref{ladder} at $(h,\tilde{h})=(h_*,\tilde{h}_*)$. For the special case of the stress-energy tensor, we simply expand the coefficients of \eqref{eigens} and \eqref{eigena} to the first order of $h-2$ and then find the eigenvalues to the first order of $h-2$. It turns out that there is always a solution
behaves like
\bal
k^1=1+\frac{  \left(\mu  q^2-1\right)^2}{q \left(\mu ^2 q^2-1\right)}(h-2)+\co\left((h-2)^2\right)\,,
\eal 
which corresponds to the deformation of $k^{FB}(h-\frac{1}{2},\tilde{h})$. 
Therefore $(h,\tilde{h})=(2,0)$ is always a solution that sets some eigenvalue to 1. 
This can also be confirmed by explicitly verifying that $E(1,2,0,\m,q)=0$.

To get the complete spectrum of the operators in the IR limit of the model, we need to find their corresponding $(h,\tilde{h})$ that makes some eigenvalues to 1. Therefore the dimension of the operators should solve
\bal
E_c(1,h_s,\tilde{h}_s,\m,q)=0\,, \qquad E'_c(1,h_a,\tilde{h}_a,\m,q)=E_c(-1,h_a,\tilde{h}_a,\m,q)=0\,,\label{keyker}
\eal
where we use the subscript $a,s$ to present operators in the symmetry and antisymmetric channels.

\begin{figure}[t!]
	\centering
	\begin{subfigure}[t]{0.5\textwidth}
		\centering
		\includegraphics[width=0.9\linewidth]{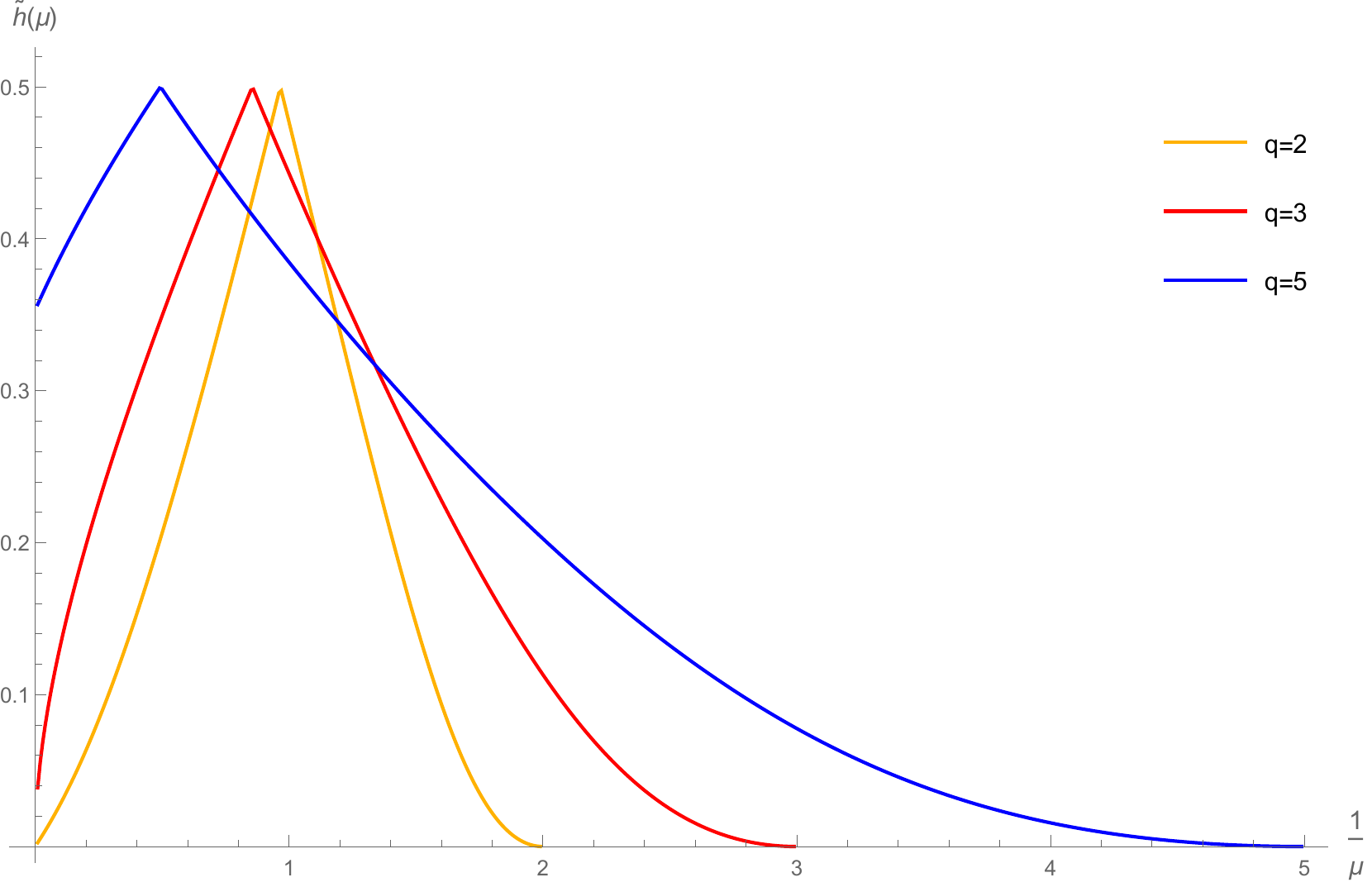}
		\caption{The symmetric channel.}\label{fig:scalarDel1mu}
		
	\end{subfigure}%
	~ 
	\begin{subfigure}[t]{0.5\textwidth}
		\centering
		\includegraphics[width=0.9\linewidth]{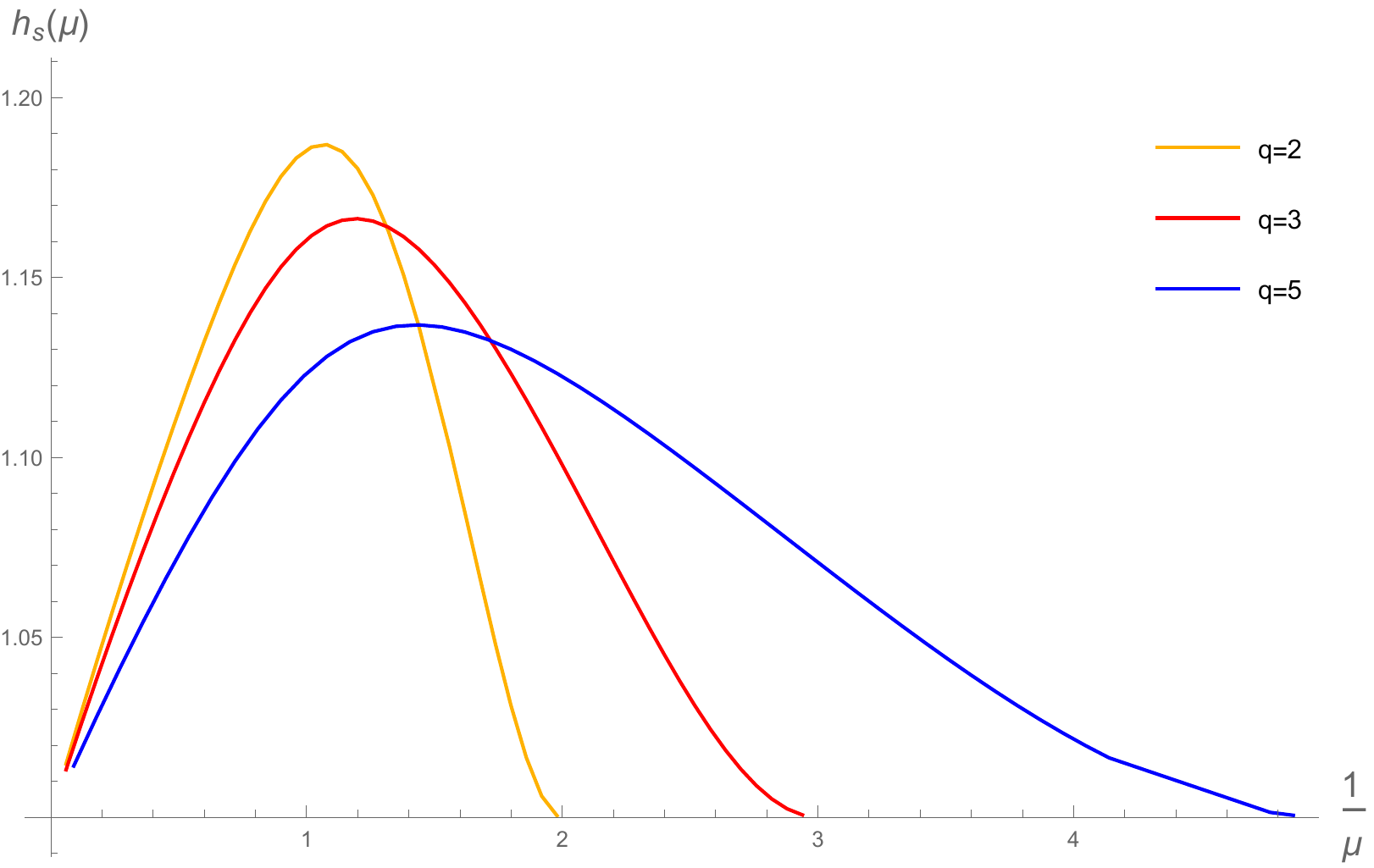}
		\caption{The antisymetric channel.}\label{fig:scalarADel1mu}
	\end{subfigure}
		\caption{The lowest dimensions of the scalar operators in the four point function. The plots illustrate how does the dimension change as a function of $\m$. The kinks on each curve in figure~\ref{fig:scalarDel1mu} is the crossover point of the operator and its ``shadow" operator due to the symmetry $h\leftrightarrow 1-h$ and $\tilde{h} \leftrightarrow 1-\tilde{h}$.}
	\label{fig:scalardelmu}
\end{figure}

For instance, we get the lowest dimensions of scalar operators running in the 4-point functions of the fundamental fields by solving
\bal
E_c(1,h_s,h_s,\m,q)=0\,, \qquad  E_c(-1,h_a,h_a,\m,q)=0\,,
\eal
for generic $\m $ and $q$. The equations are easily solved numerically. The solutions to the equation $E_c(\pm 1,h_s,h_s,\m,q)=0$ is shown in figure~\ref{fig:scalardelmu}.
As we can see the dimension of the scalar operators  in the symmetric channel approach zero as the $\m$ approaches~$\frac{1}{q}$. 
In addition, the dimension of the operators in the antisymmetric channel is larger than those in the symmetric channel. This is as expected since the operators in the antisymmetric channel involves one more derivative.

We also computed other 4-point functions where fermionic operators runs in the ladder. The details are elaborated in appendix~\label{sec:fermi}. We only outline here that one can check explicitly that at the $\m=1$ point, we do observe both $(h,\tilde{h})=(\frac{3}{2},0)$ and $(h,\tilde{h})=(0,\frac{3}{2})$ operators that correspond to the supercharges in the holomorphic and antiholomorphic sectors. This again confirms that at $\m=1$ our model has an enhanced $\cn=(2,2)$ supersymmetry. We further check that as long as $\m\neq 1$ the left-moving supercharges are lifted and stop generating supersymetry in the left-moving sector. This is as we expected since the model only has $\cn=(0,2)$ supersymmetry at generic $\m$.

\subsection{Chaotic behavior}\label{sec:chaos}

One can further go to the chaos region and study the out-of-time-ordered correlators. For this we simply consider retarded kernels.  We start from the Euclidean propagators on a periodic $\t$ direction and a noncompact spatial direction $x$
\bal
G^I_{th}(\t_1,x_1,\t_2,x_2)&=\frac{n_I}{\left(2\sinh(\frac{x_{12}+i \t_{12}}{2})\right)^{2h_I} \left(2\sinh(\frac{x_{12}-i \t_{12}}{2})\right)^{2\tilde{h}_I} }\,,\label{ther}
\eal
The retarded propagators can be computed from analytic continuation of \eqref{ther} 
\bal
G^b_R(t_1,x_1,t_2,x_2)&=-\frac{2 i \sin \left( \pi  (h_{b }+\tilde{h}_{b })\right) \q(t_{12}-|x_{12}|) n_b}{ \left( 2 \sinh \left(\frac{t_{12}-x_{12}}{2}\right)\right)^{2 h_{b}}  \left(2 \sinh  \left(\frac{t_{12}+x_{12}}{2}\right)\right)^{2 \tilde{h}_{b }}}\,,\quad b=\f,G\label{ret1}\\
G^f_R(t_1,x_1,t_2,x_2)&=\frac{2 \cos \left( \pi ( h_{f }+ \tilde{h}_{f }) \right) \q(t_{12}-|x_{12}|) n_f}{ \left( 2 \sinh \left(\frac{t_{12}-x_{12}}{2}\right)\right)^{2 h_{f }}  \left(2 \sinh  \left(\frac{t_{12}+x_{12}}{2}\right)\right)^{2 \tilde{h}_{f }}}\,,~\quad f=\psi,\l\ .\label{ret2}
\eal
We also need the set of ladder rung propagators between the rails. They can be obtained from a simple analytic continuation
\begin{multline}
G_{lr}^I(t_1,x_1;t_2,x_2)=G_{th}(i t_1,x_1;it_2+\p,x_2)\\
=\frac{n_I}{\left(2\cosh(\frac{x_{12}-  t_{12}}{2})\right)^{2h_I} \left(2\cosh(\frac{x_{12}+ t_{12}}{2})\right)^{2\tilde{h}_I} }\label{lrpp}
\end{multline}
The set of retarded kernels can be obtained from \eqref{kn1}-\eqref{kn9} by replacing the propagators on the rails by the corresponding retarded ones \eqref{ret1} or \eqref{ret2}  and replacing the ladder rung propagators by the ones in \eqref{lrpp}.

 Following  \cite{Murugan:2017eto}, we introduce the new variable
 \bal
 u=e^{x-t}\,,\qquad v=e^{-x-t}\,,
 \eal
for the retarded propagators on the upper rail and 
 \bal
 u=-e^{x-t}\,,\qquad v=-e^{-x-t}\,,
 \eal
for the retarded propagators on the lower rail. 
We consider the following ansatz 
\bal
\Psi_R^I(3,4)=(-u_3u_4)^{\frac{h_3+h_4}{2}}(-v_3v_4)^{\frac{\tilde{h}_3+\tilde{h}_4}{2}} u_{34}^{h-h_3-h_4}v_{34}^{\tilde{h}-\tilde{h}_3-\tilde{h}_4}\,,\label{aztu}
\eal
where 
$h_i$, $\tilde{h}_i$ labels the dimensions of the operators at $t_i,x_i$. In terms of the $t$ and $x$ coordinate, \eqref{aztu} becomes
\bal
\Psi_R^I(3,4)&=\frac{e^{-\frac{1}{2}(h+\tilde{h})(t_1+t_2)-\frac{1}{2}(h-\tilde{h})(x_1+x_2)}}{(2\cosh\frac{x_{12}-t_{12}}{2})^{ {h}_1+ {h}_2- {h}}(2\cosh\frac{x_{12}+t_{12}}{2})^{\tilde{h}_1+\tilde{h}_2-\tilde{h}} }\ .\label{aztz}
\eal
Our goal is to find eigenfunctions that grows exponentially with time but remain normalizable in the spatial direction. This requires $h-\tilde{h}$ to be imaginary and we can follow \cite{Murugan:2017eto} to reparametrize $h$ and $\tilde{h}$ as
\bal
h=-\frac{\l_L}{2}+i\frac{p}{2}\,\qquad \tilde{h}=-\frac{\l_L}{2}-i\frac{p}{2}\ .
\eal
Finding the largest chaotic behavior then means to find the largest $\l_L$ that renders at least one eigenvalues to 1. 

As in the case discussed in \cite{Murugan:2017eto},
 the convolution integral in the eigenequation
\bal
K_R^{(ij)}*\Psi_R^j=k_R^{ij} \Psi_R^i\,,\label{reteig}
\eal
factorizes into two 1-dimensional integrals in the $u$, $v$ variables with the eigenfunction \eqref{aztu}, each of which can be carried out straightforwardly. The resulting eigenvalues are
\bal
k^{\f\f}_R&=\frac{2 \mu  (q-1)^2 q \left(\mu  q^2-2 \mu  q+1\right)\Gamma \left(\frac{(q-1) q \mu }{q^2 \mu -1}\right)^4 \sin ^2\left(\frac{\pi(1 -  \mu  q)}{\mu  q^2-1}\right)}{\pi ^4 \left(\mu  q^2-1\right)^2}\\
&\times \sin \left(\frac{\pi  h}{2}+\frac{\pi  \left(\mu  q-1\right)}{2 \mu  q^2-2}\right) \Gamma \left(h-\frac{(q-1) \mu q  }{\mu q^2 -1}\right) \Gamma \left(\frac{\mu q  -1}{\mu q^2  -1}-h\right) \\
&\times \sin \left(\frac{\pi   \tilde{h} }{2}+\frac{\pi  \left(\mu  q-1\right)}{2 \mu  q^2-2}\right) \Gamma \left( \tilde{h} -\frac{(q-1) \mu q  }{\mu q^2 -1}\right) \Gamma \left(\frac{\mu q -1}{\mu q^2  -1}- \tilde{h} \right)\\
&\times \left(\cos \left(\frac{1}{2} \pi  \left(h+ \tilde{h} +\frac{2 (\mu  q-1)}{\mu  q^2-1}\right)\right)+\cos \left(\frac{1}{2} \pi  (h- \tilde{h} )\right)\right)\\
k^{\f\psi}_R&=\frac{\mu  q^2-1}{2 \mu  q^2-4 \mu  q+2}k^{\f\f}_R\\
k^{\f\l}_R&=\frac{4 \pi ^2 J^2 n_{\f}^{q+1} (\mu  q-1) \left(\mu  q^2-1\right)}{\mu  (q-1) q \left(\mu  q^2-2 \mu  q+1\right)}k^{\f\f}_R\\
k^{\f G}_R&= \frac{2 \pi ^2 J^2 n_{\f}^{q+1} \left(\mu  q^2-1\right)^2}{\mu  (q-1)^2 q \left(\mu  q^2-2 \mu  q+1\right)} k^{\f\f}_R\\
k^{\psi\f}_R&=2 \left(h -\frac{ \mu  q -1 }{ \mu  q^2-1 }\right)\frac{ \left(h \left(\mu  q^2-1\right)-\mu  (q-1) q\right) \tan \left(\frac{\pi  }{2}(h+\frac{  \mu  q-1}{  \mu  q^2-1})\right) }{  \left(\mu  q^2-2 \mu  q+1\right)\tan \left(\frac{\pi  }{2}(h+\frac{   \mu  q-1}{  \mu  q^2-1})\right)}k^{\f\f}_R\\
k^{\psi\l}_R&=\frac{8 \pi ^2 J^2 n_{\f}^{q+1}(\mu  q^2-1)^2 \left(h -\frac{ \mu  q-1}{\mu  q^2-1}\right) \left(h -\frac{\mu  (q-1) q}{\mu  q^2-1}\right) \tan \left(\frac{1}{2} \pi  \left(h+\frac{\mu  q-1}{\mu  q^2-1}\right)\right) }{\mu q  (q-1)^2  \left(\mu  q^2-2 \mu  q+1\right)\tan \left(\frac{\pi  h}{2}+\frac{\pi  \left(\mu  q-1\right)}{2 \mu  q^2-2}\right)}k^{\f\f}_R\\
k^{\l\f}_R&=-\frac{\mu  (q-1)^5 q^2  (\mu  q-1) \sin ^2\left(\frac{\pi  \left(\mu  q^2+q-2\right)}{\mu  q^2-1}\right) }{\pi ^6n_{\f}^{q+1} J^2 \left(\mu  q^2-1\right)^5}\Gamma \left(\frac{1-q}{q^2 \mu -1}\right)^4\\
&\times  \sin \left(\frac{\pi  h}{2}+\frac{\pi  (q-1)}{2 \mu  q^2-2}\right) \cos \left(\frac{\pi  h}{2}-\frac{\pi  (q-1)}{2-2 \mu  q^2}\right) \Gamma \left(\frac{q-1}{q^2 \mu -1}-h\right) \Gamma \left(h+\frac{q-\mu q^2   }{\mu q^2  -1}\right)\\
&\times \sin \left(\frac{\pi \tilde{h} }{2}   -\frac{\pi (q- 1)}{2- 2 \mu  q^2} \right) \cos \left(\frac{\pi  \tilde{h}}{2} +\frac{\pi (q-1) }{  2\mu  q^2-2}\right) \Gamma \left( \tilde{h} +\frac{q-1}{\mu q^2  -1}\right) \Gamma \left(\frac{\mu  q^2+q-2}{\mu q^2  -1}- \tilde{h} \right)\\
k^{\l\psi}_R&=\frac{\mu  q^2-1}{2 (q-1) (\mu  q-1)} k^{\l\f}_R\\
k^{g\f}_R&=-\frac{2 \left(h -\frac{q-1}{\mu  q^2-1}\right) \left(h -\frac{\mu  q^2-q}{\mu  q^2-1}\right) \tan \left(\frac{\pi  h}{2}-\frac{\pi  (q-1)}{2-2 \mu  q^2}\right) \tan \left(\frac{\pi  h}{2}-\frac{\pi  q (\mu  q-1)}{2 \mu  q^2-2}\right)}{(q-1) (\mu  q-1) \left(\mu  q^2-1\right)^{-1}}k^{\l\f}_R\ .
\eal

In principle, we diagonalize the matrix of retarded kernels, which is the retarded version of \eqref{evmx}, to get the equation of the eigenvalues
\begin{multline}
E_R(x,h,\tilde{h},\m,q)=x^4- k_R^{\phi \phi } x^3-\left( k_R^{  \phi G} k_R^{  G \phi }+ k_R^{\phi \psi } k_R^{\psi \phi }+ k_R^{\phi \lambda } k_R^{\lambda \phi }+ k_R^{\psi \lambda } k_R^{\lambda \psi }\right) x^2\\
+\left(k_R^{\phi \phi } k_R^{\psi \lambda } k_R^{\lambda \psi }- k_R^{\phi \psi } k_R^{\psi \lambda } k_R^{\lambda \phi }- k_R^{\phi \lambda } k_R^{\psi \phi } k_R^{\lambda \psi }\right) x+k_R^{  \phi G} k_R^{\psi \lambda } k_R^{\lambda \psi } k_R^{  G\phi }=0\ .\label{eigenR}
\end{multline} 
Then we solve for the $h$ and $\tilde{h}$ that set the eigenvalue to 1. In our case we do not get a simple expression of the eigenfunctions. But we  can still find the maximal values of $\l_L$ by a direct analysis of the eigenfunction equation: 
since we are interested in eigenvalue 1, we can simply set $x=1$ of the equation \eqref{eigenR} that determines $\l_L$ as  an implicit function of $\m$ and $p$. We can then find the largest value of $\l_L$ by tuning $\m$ and $p$. 

To proceed we start with a sanity check by focusing on the $\m=1$ case and check if the result  agrees with the $\cn=(2,2)$ result. There are two ways to do such a check. The first approach is noticing that at $\m=1$ we can solve the retarded eigenequation \eqref{eigenR} directly to find that there are 4 eigenvalues  
\bal
k^{\m=1}_1&=-\frac{\Gamma \left(\frac{q}{q+1}\right)^2 \Gamma \left(\frac{1}{q+1}-h\right) \Gamma \left(\frac{1}{q+1}-\tilde{h}\right)}{\Gamma \left(\frac{1}{q+1}-1\right) \Gamma \left(1+\frac{1}{q+1}\right) \Gamma \left(\frac{q}{q+1}-h\right) \Gamma \left(\frac{q}{q+1}-\tilde{h}\right)}\\
k^{\m=1}_2&=-\frac{h q+h-1}{h q+h-q}k^{\m=1}_1\\
k^{\m=1}_3&=-\frac{\tilde{h} q+\tilde{h}-1}{\tilde{h} q+\tilde{h}-q}k^{\m=1}_1\\
k^{\m=1}_4&=\frac{(h q+h-1) (\tilde{h} q+\tilde{h}-1)}{(h q+h-q) (\tilde{h} q+\tilde{h}-q)} k^{\m=1}_1\,,
\eal
and indeed $k^{\m=1}_1$ is identical to the $k^{BB}_R$ function in \cite{Murugan:2017eto}. Notice that the other $k^{\m=1}_i$ are due to the super-descendents of the eigenfunction corresponding to $k^{\m=1}_1$. Hence they are not expected to be related to the  $k^{BF}_R$, $k^{FB}_R$, $k^{FF}_R$ functions in \cite{Murugan:2017eto} that are due to different primaries of the third operators in the eigenfunctions. 

The second approach is to follow the method we discribed above, namely plug $x=1$ into the equation \eqref{eigenR}, then setting $\m=1$ and looking for maximal $\l_L$ by tuning $p$. We indeed find a miximal $\l_L=0.5824$ at $p=0$. This agrees with the result in \cite{Murugan:2017eto} and confirms the validity of our procedure.

\begin{figure}[t]
	\centering
	\begin{subfigure}[t]{0.5\textwidth}
		\centering
		\includegraphics[width=0.9\linewidth]{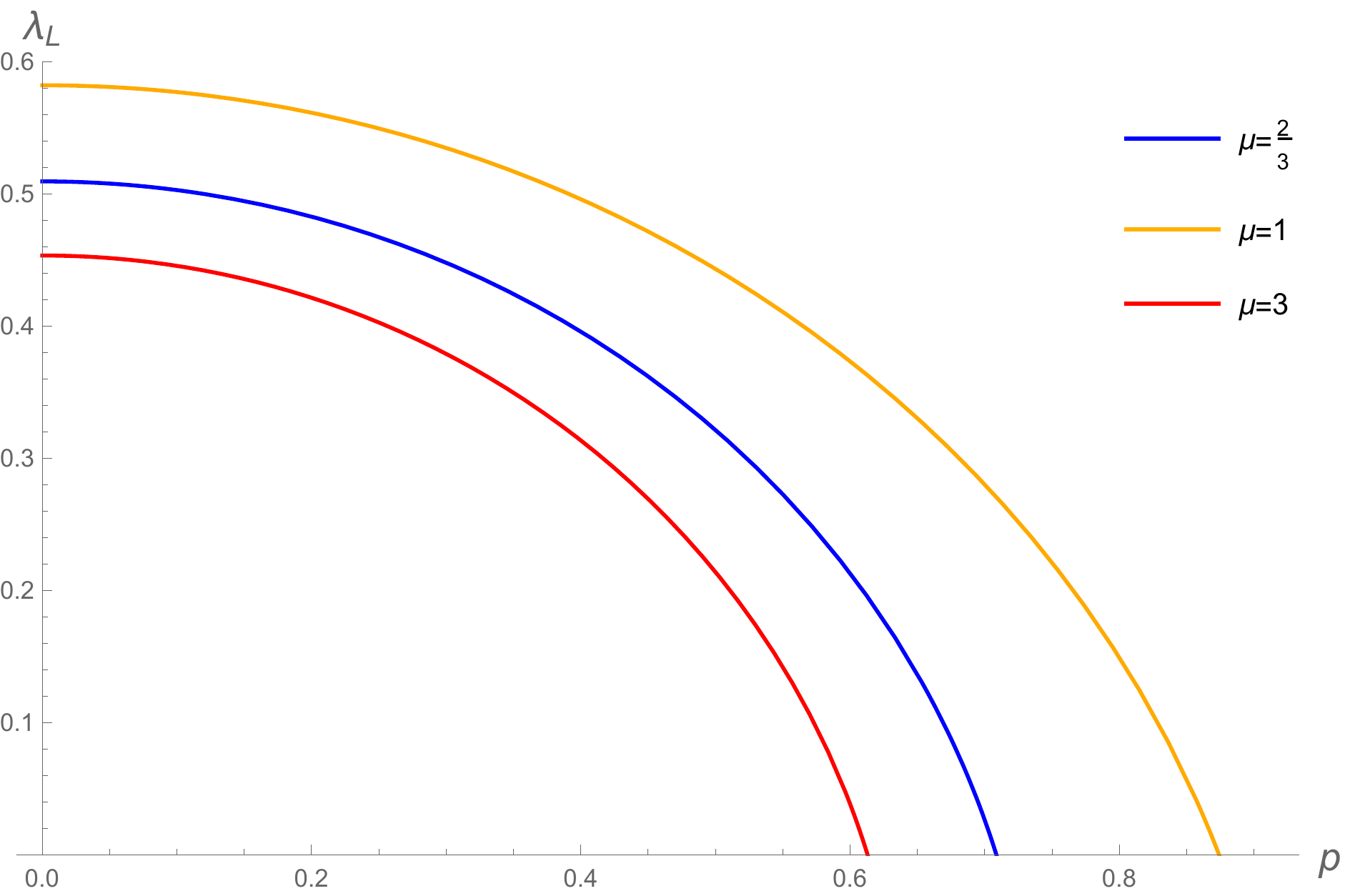}
		\caption{$p$ dependence.}\label{fig:lp}
	\end{subfigure}%
	~ 
	\begin{subfigure}[t]{0.5\textwidth}
		\centering
		\includegraphics[width=0.9\linewidth]{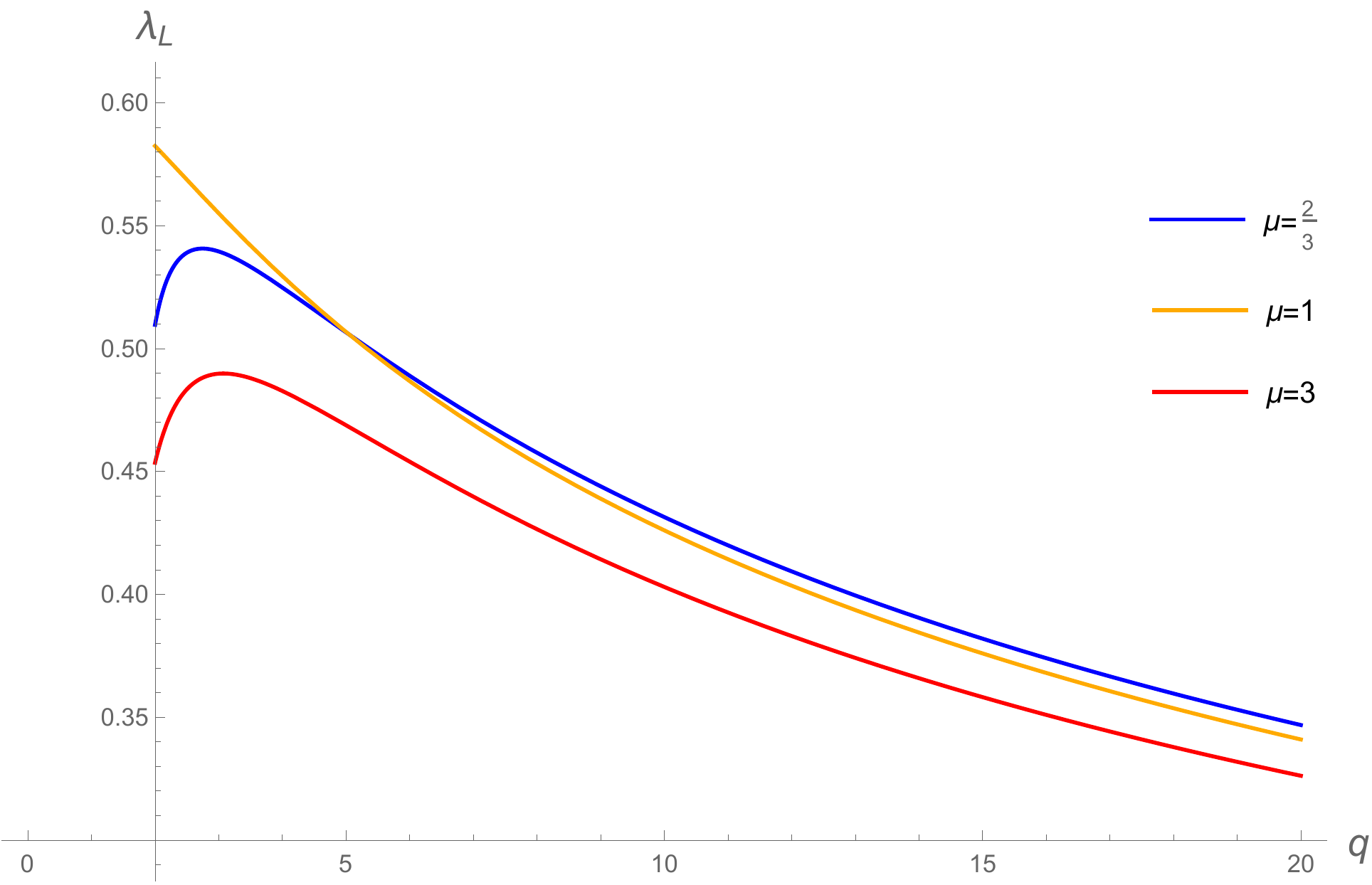}
		\caption{$q$ dependence.}\label{fig:lq}
	\end{subfigure}
	\caption{Functional dependence of the Lyapunov exponent~$\l_L$.}\label{fig:lplq}
\end{figure}

We now move to general $\m$. Because the integral \eqref{reteig} again factorize into two 1-dimensional ones, by a similar monotonic argument as in \cite{Murugan:2017eto} we expect the largest $\l_L$ is reached at $p=0$. This is indeed true as one can check explicitly. We present the $p$ dependence for some special values of $\m$ in figure~\ref{fig:lp}.
It is also straightforward to check the $q$ dependence of $\l_L$, a few examples of which are presented in figure~\ref{fig:lq}. We find the largest $\l_L$ appears close to the smallest $q$ that leads to nontrivial interactions, namely $q=2$. 

\begin{figure}[t]
	\centering
	\begin{subfigure}[t]{0.5\textwidth}
		\centering
		\includegraphics[width=0.9\linewidth]{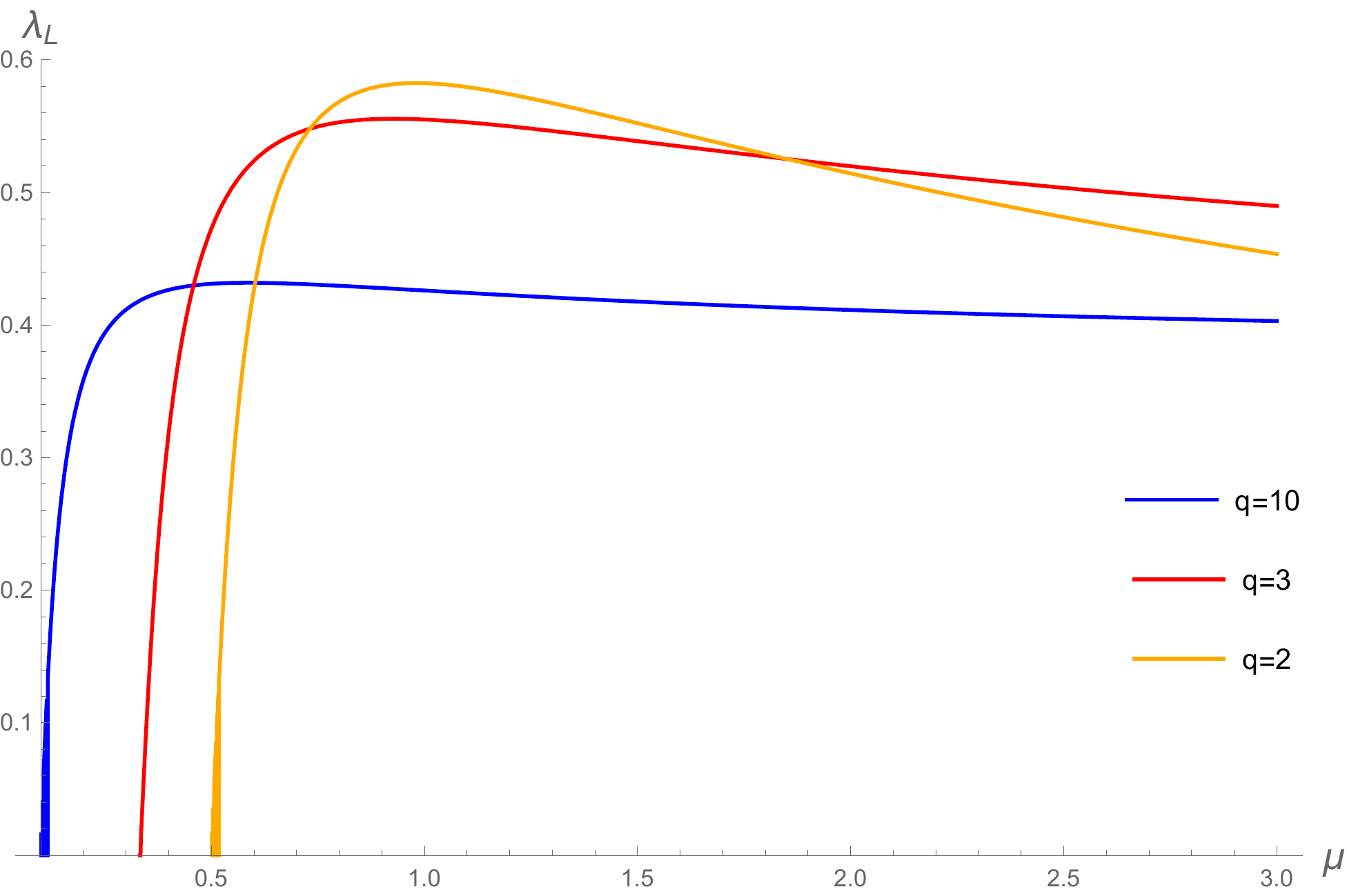}
		\caption{Small $\m$ region}\label{fig:lmu1}
		
	\end{subfigure}%
	~ 
	\begin{subfigure}[t]{0.5\textwidth}
		\centering
		\includegraphics[width=0.9\linewidth]{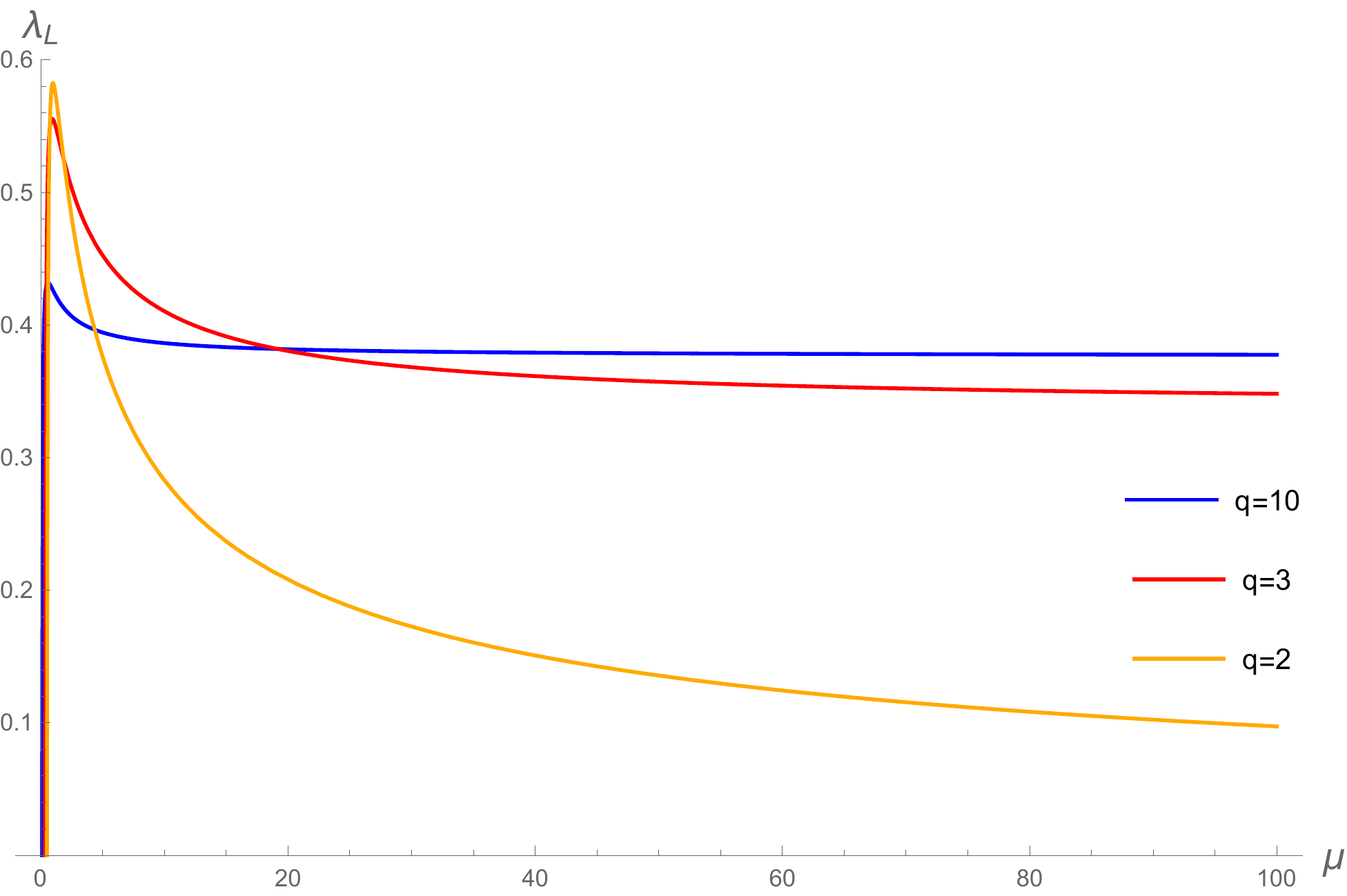}
		\caption{Large $\m$ region}\label{fig:lmu2}
	\end{subfigure}
	\caption{$\m$ dependence of the Lyapunov exponent $\l_L$.  The yellow, red and blue curves are evaluated at  $p=0$ and $q=2$, 3, 10 respectively.}\label{fig:lmu}
\end{figure}

Then we look for maximal $\l_L$ as a function of $\m$. The general dependence is shown in figure~\ref{fig:lmu}.
One finds a maximum of $\l_L$ as we change $\m$. Interestingly, this maximal value does not appear at the special point $\m=1$;  rather it appears at $\m\simeq 0.9802$ where the maximal value is $\l_L(p=0,\m=0.9802)=0.5825$. This maximal value is only slightly larger than the value for the $\cn=(1,1)$ and $\cn=(2,2)$ model where $\l_L(p=0,\m=1)\simeq 0.5824$. Notice that in determining this Lyapunov exponent we do all the computation analytically, except for the very last step where we find the solution to a given equation numerically. Since the error of this last step is very well controlled, our result is  genuinely different from the exponent in the $\cn=(1,1)$ or $\cn=(2,2)$ model.

It is not clear what is the physical reason of why the maximal value of $\l_L$ in the class of models we consider here is only slightly higher than that found in the special case $\m=1$, and why the correspnding $\m$ is only slightly smaller than 1.  The slightly larger Lyapunov exponent probably indicates that there should be a wider class of similar models that are continuously related. Our model, and the $\cn=(1,1)$, $\cn=(2,2)$ models are only examples that sit on a generic point on the moduli space. It is conceivable that there are special theories on (some corners) the moduli space that have larger, or even maximal, Lyapunov exponent.  

\section{Two higher-spin limits}\label{hslimits}

It is widely believed that the SYK-like models have close relations with higher-spin theories; higher-spin theories should be thought as a subsector of some tensionless limit of string theory, while the SYK model should be holographically dual to some string theory with finite tension \cite{Maldacena:2016hyu}.  
Therefore it is tempting to find a direct relation between SYK-like models and models with higher-spin symmetry. In 1-dimension, one example of such relation is discussed in \cite{Peng2017}. In this section we give another explicit example of such connection in 2 dimension. The basic idea is to tune the free parameter to some critical/singular value where the model develops some properties that is characteristic for models with higher-spin symmetry. In particular, we find two singular limits at the two ends of the range of $\m$, where the model \eqref{Sintsusy} develops emergent higher-spin symmetries. In the following we discuss the two limits separately.

\subsection{The $\m q \to 1^+ $ (``classical chiral") limit}

The Fourier transform \eqref{ftf} at the special value $\mu=\frac{1}{q}$ has a logarithmic divergence.  This means a proper renormalization analysis of the model at $ \mu=\frac{1}{q}$ is needed. We will not do it here and will postpone this in future work. Nevertheless, we consider the limit 
\bal
\m\to  \left(\frac{1}{q}\right)^+ \ .\label{freechiral}
\eal 
Taking the limit in this manner, all our previous computation are valid since there is no divergence in the limiting process. 
The IR dimensions of the various fields in this limit are
\bal
& \lim_{\m\to  \left({1}/{q}\right)^+}h_{\phi }= 0\,,~~  \lim_{\m\to  \left({1}/{q}\right)^+} h_{\psi }= \frac{1}{2 }\,,~~  \lim_{\m\to  \left({1}/{q}\right)^+} h_{\lambda }= \frac{ 1}{ 2}\,,~~  \lim_{\m\to  \left({1}/{q}\right)^+}h_G= 1\\
&  \lim_{\m\to  \left({1}/{q}\right)^+}\tilde{h}_{\phi }= 0\,,~~ \lim_{\m\to  \left({1}/{q}\right)^+}\tilde{h}_{\psi }= 0\,,~~ \lim_{\m\to  \left({1}/{q}\right)^+}\tilde{h}_{\lambda }= 1\,,~~ \lim_{\m\to  \left({1}/{q}\right)^+}\tilde{h}_G= 1\,,
\eal 
where the dimension of the $\f$ and $\psi$ fields take the values in a free chiral multiplet. This is a first hind that we should expect a larger higher spin type symmetry to emerge in  this limit. This is consistent with the result from the chaos analysis in the previous section; as shown in figure~\ref{fig:lmu} the Lyapunov exponents all vanish as $\m\to  \left(\frac{1}{q}\right)^+$ for any $q>1$. In the following we confirm the existence of a higher-spin symmetry from a few   different aspects.\\

\begin{figure}[t!]
 	\centering
 	\begin{subfigure}[t]{0.5\textwidth}
 		\centering
 		\includegraphics[width=0.9\linewidth]{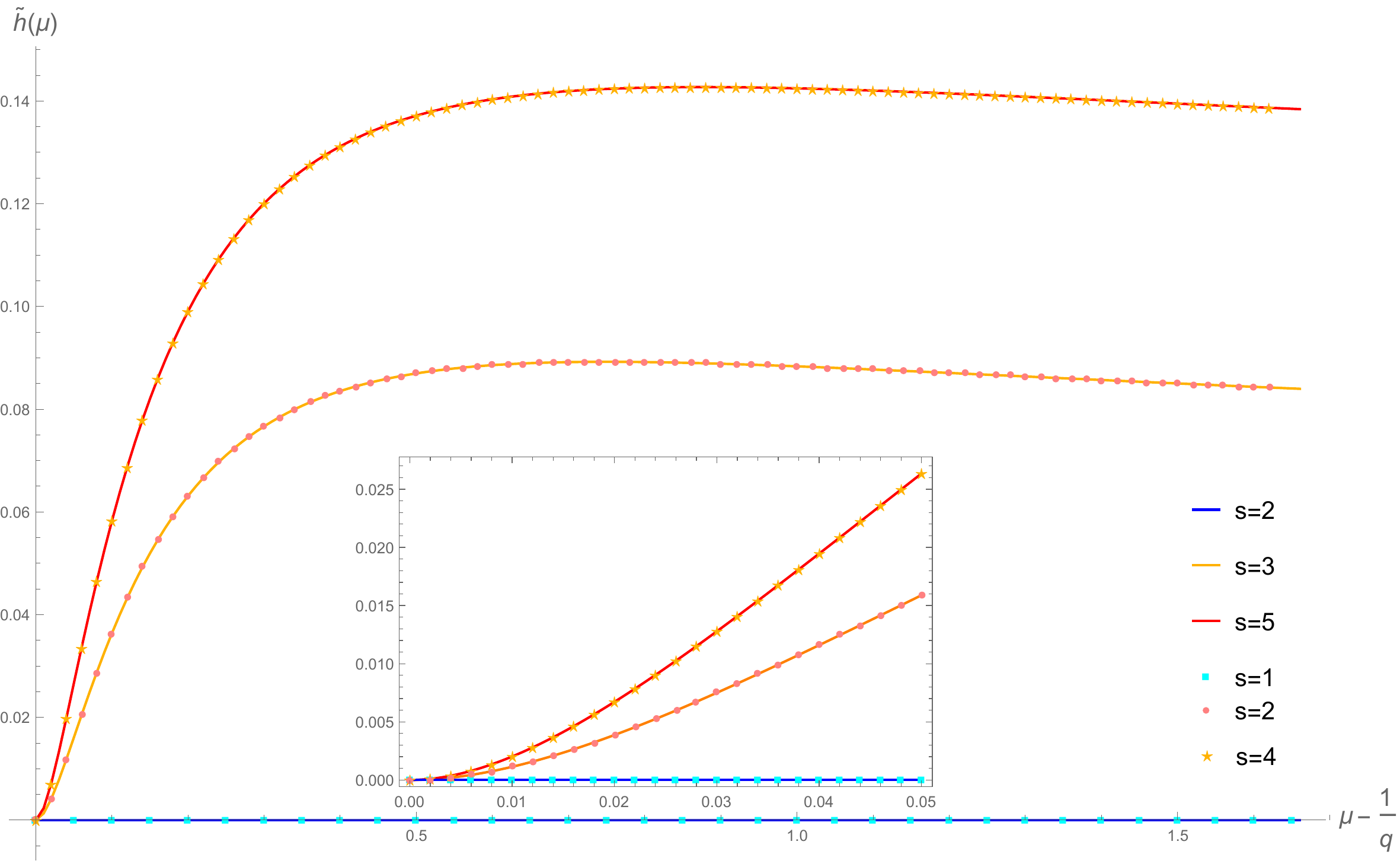}
 		\caption{Anomalous dimensions of the holomorphic higher-spin operators. Here we plot the left dimension $\tilde{h}$ of the operators and the right dimension is $h=\tilde{h}+s$. 
 		}\label{fig:hshbmu}
 	\end{subfigure}%
 	~ 
 	\begin{subfigure}[t]{0.5\textwidth}
 		\centering
 		\includegraphics[width=0.9\linewidth]{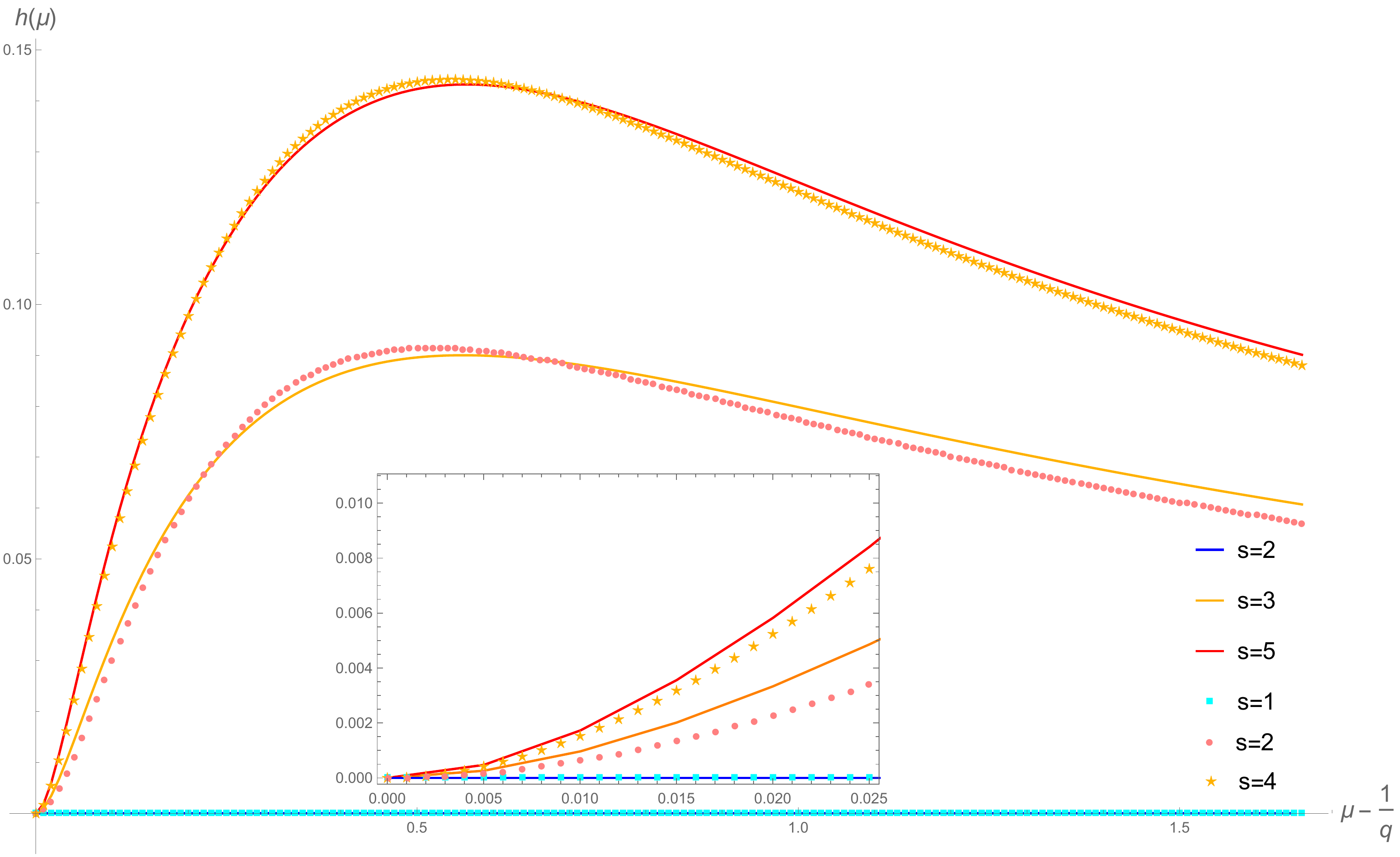}
 		\caption{Anomalous dimensions of the anti-holomorphic higher-spin operators. Here we plot the right dimension $ {h}$ of the operators and the left dimension is $\tilde{h}={h}+s$. }\label{fig:hshbmua}
 	\end{subfigure}
 	\caption{The dimensions of operators with integer spins as a function of $\m$. The smaller figure in the frame zooms in to the bottom left corner of each large figure and shows that the anomalous dimensions of these operators approach zero in the limit $\m\to \left(\frac{1}{q}\right)^+$. The solid curves denotes operators in the symmetric channel, the discrete data points denote the operators in the antisymmetric channel.	The plots are computed for $q=3$. The $\cn=(0,2)$ supersymmetry is manifest in the plots. }\label{fig:hshbmu1}
 \end{figure}

\noindent $\bullet~$ A tower of conserved higher-spin operators.\\
Given the above motivation, we look for a tower of higher-spin operators in the limit \eqref{freechiral}.
Recall that in 2 dimensions, conserved higher-spin operators are represented by (anti-)holomorphic primary operators with vanishing (right) left conformal dimensions. Therefore we extend the computation in the previous subsection to find such (anti-)holomorphic primary operators in the limit \eqref{freechiral}. For this we go back to \eqref{keyker} and look for solutions of
\bal
\lim_{\m\to (1/q)^+}E_c(1,\tilde{h}+s,\tilde{h},\m,q)=0\,,\quad \lim_{\m\to (1/q)^+}E_c(-1,\tilde{h}+s,\tilde{h},\m,q)=0\,, \quad s>0\,,~\tilde{h}\to 0\,,\label{tosol1}
\eal
that correspond to the holomorphic higher-spin operators in the symmetric and antisymmetric channel. Because our model has a manifest $\cn=(0,2)$ supersymmetry, the operator spectrum of the holomorphic and anti-holomorphic operators could be different. Therefore we need to find the spectrum of the anti-holomorphic  operators seperately, which corresponds to solving  
\bal
\lim_{\m\to (1/q)^+}E_c(1, {h}, {h}+s,\m,q)=0\,,\quad \lim_{\m\to (1/q)^+}E_c(-1, {h} , {h}+s,\m,q)=0\,, \quad s>0\,,~ {h}\to 0\ .\label{tosol2}
\eal
Furthermore,  the operators running in the channel tha is detected by  our ladder diagrams \eqref{Kff}-\eqref{KGf} are all bosonic, so we only look for solutions with integer $s$, at  any $q$. The equations are again easily solved numerically, and we get a function $\tilde{h}(\m)$ or ${h}(\m)$ of any given $s$ and $q$. We summarize the operators that become (anti)-holomorphic in the limit \eqref{freechiral} in the following table
\begin{center}
\renewcommand{\arraystretch}{1.5}
\begin{tabular}{|c|c|c|}
	\hline 
	 & holomorphic operators & anti-holomorphic operators \\ 
	\hline 
	symmetric channel & $(h,\tilde{h})= (s,0)$, $s\geq 1$  & $(h,\tilde{h})= (0,s)$, $s\geq 1$  \\ 
	\hline 
	antisymmetric channel & $(h,\tilde{h})= (s,0)$, $s\geq 1$  & $(h,\tilde{h})= (0,s)$, $s\geq 1$  \\ 
	\hline 
\end{tabular} 
\renewcommand{\arraystretch}{1}
\end{center}
Some example solutions of \eqref{tosol1} and \eqref{tosol2} near the limit \eqref{freechiral} are illustrated in figure~\ref{fig:hshbmu1}. 
From this result we indeed observe that in the $\m \to \left(\frac{1}{q}\right)^+$ limit the dimensions of some operators in 4-point functions approaches the dimensions of the (anti-)holomorphic higher-spin operators shown in the above table. This result  indicates that a tower of conserved higher spin currents emerges in the limit \eqref{freechiral}.

The above holomorphic operators close among themselves under Operator Product Expansion (OPE) and hence generate a higher-spin type $\cw$-algebra. A similar result holds for the anti-holomorphic operators. Moreover, for each spin we find an operator in the symmetric channel and one in the antisymmetric channel, which is identical\footnote{The is one subtlety: we have one more spin-1 field in our model comparing to the generators of the $\cn=2$ $\mathcal{S}\mathcal{W}_{1+\infty}$ algebra. This spin-1 extended $\cn=2$ $\mathcal{S}\mathcal{W}_{1+\infty}$ algebra is very similar with the one withour the extra spin-1 field and is recently discussed in the context of Affine Yangian \cite{Gaberdiel:2017hcn, Gaberdiel:2017toap}.  } to the structure of the  generators of the bosonic subalgebra of the $\cn=2$ supersymmetric $\cs\mathcal{W}_{1+\infty}$ algebra that governs the symmetry of many supersymmetric higher-spin theories in 2 dimension \cite{
	Henneaux:2012ny,Hanaki:2012yf,Peng:2012ae,Gaberdiel:2014yla,Gaberdiel:2017dbk,Gaberdiel:2017hcn}.
This spectrum is as we expected since the model has a right-moving $\cn=2$ supersymmetry.

It is slightly more surprising that the anti-holomorphic operators have a same spectrum as an $\cn=2$ supersymmetric $\cs\cw_{1+\infty}$ algebra. We have checked that in the limit \eqref{freechiral} the left-moving sector does not have supersymmetry as well. Therefore the left-moving section only has bosonic higher-spin symmetry emerging, which is consistent with the $\cn=(0,2)$ supersymmetry of our model.

\noindent $\bullet~$ Anomalous dimensions\\
As we go away from the limit \eqref{freechiral}, namely as $\m$ becomes larger, this emergent higher-spin symmetry is broken, which is characterized by the anomalous dimensions acquired to those (anti-)holomorphic operators. This is another way to interpret  figure~\ref{fig:hshbmu1}, in particular the small figures inside figure~\ref{fig:hshbmu1}. 
Before analyzing the results, we first remind the reader that in the following we call the operators that become the (anti-)holomorphic in the higher-spin limit $\m \to \left(\frac{1}{q}\right)^+$ ``almost (anti-)holomorphic higher-spin operators",  even after we move away from the higher-spin limit \eqref{freechiral} at generic $\m$. We keep in mind that they are only strictly related to conserved currents in the limit \eqref{freechiral}.

Coming back to the results, in figure~\eqref{fig:hshbmu} we plot the anomalous dimension of the holomorphic higher-spin operators in the symmetric channel (solid lines) and the antisymmetric channel (discrete points). We observe that for the holomorphic higher-spin operators, the anomalous dimension of a spin-$s$ operator in the antisymmetric channel is identical to the anomalous dimension of a spin-$(s+1)$ operator in the symmetric channel. This is consistent with the $\cn=2$ supersymmetry in the right-moving section: the two operators are respectively the top and the bottom component of a single multiplet that consists of operators with spin $(s,s+\frac{1}{2},s+1)$.\footnote{Notice that we have implicitly used the shadow representation of the 4-point functions. So each solution to the equation \eqref{keyker} correspond to an $SL(2)$ primary field. On the other hand, since we are not using the superspace formalism and directly work with the component fields in each supermultiplet, the eigenfunctions/operators we  found could be supersymmetric descendant fields.} This confirms that the right-moving $\cn=2$ supersymmetry is always perserved at generic value of $\m$. 

On the other hand, from figure~\eqref{fig:hshbmua} we observe that there is no such relation among the almost anti-holomorphic operators in the symmetric and antisymmetric channels away from the limit \eqref{freechiral}. This is again compatible with the fact that there is  no supersymmetry in the left-moving sector.

\begin{figure}
	\centering
	\includegraphics[width=0.7\linewidth]{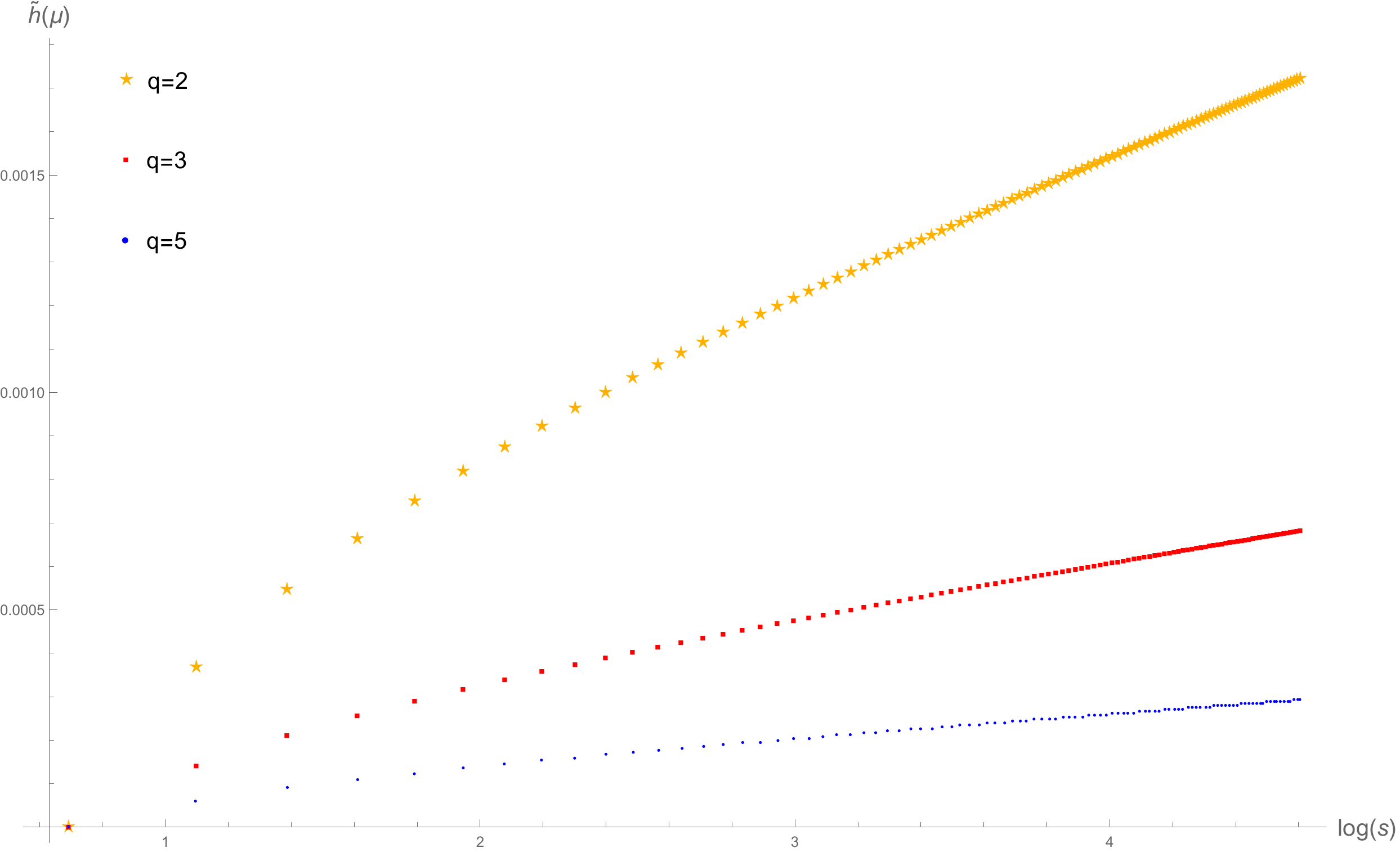}
	\caption{The anomalous dimension of the higher spin operators as a function of the spin of the operators. The yellow stars, red dots and the blue crosses are for $q=2,3,5$ respectively and $\m q-1=\e=0.01$. The horizontal axis is $\log(s)$ where $s$ is the spin of the operator. The vertical axis is the anomalous dimension. It is observed that when $s$ is relatively large, the anomalous dimension is propotional to the log of the spin. This is the same behavior as the result from a perturbative higher-spin CFT computation \cite{Gaberdiel:2015uca}, as well as the dispersion relation of a classical rotating string in AdS spacetime in the large-spin limit \cite{Gubser:2002tv}.}
	\label{fig:hblogs}
\end{figure}

It is also illustrative to study the dispersion relation, namely to understand how do the anomalous dimensions, which is the same as $\tilde{h}$ for the anti-holomorphic higher-spin operators or ${h}$ for the holomorphic higher-spin operators, depend on the spin once we move away from the higher spin limit. One can find this by solving  
\bal
E_c(\pm 1,\tilde{h}+s,\tilde{h},\frac{1+\e}{q},q)=0\,,\qquad \e\ll 1\,, s>0\,,
\eal
to determine the implicit function $\tilde{h}^\e(s)$. And similarly the implicit function ${h}^\e(s)$ for the anti-holomorphic operators. This is again easily achieved numerically. The result for the anomalous dimension of the almost holomorphic higher-spin operators in the symmetric channel  is plotted in figure~\ref{fig:hblogs}.
 The results for the operators in the antisymmetric channel and the anti-holomorphic operators have  similar structure.
In figure~\ref{fig:hblogs} we observe that for each given $q$, the anomalous dimension $\g_\e(s)=\tilde{h}_\e(s)$ behaves like 
\bal
\tilde{h}_\e(s) \sim \log(s)\,,
\eal 
for relative large spin $s$ near the higher-spin limit \eqref{freechiral}.  This result, which comes from diagonalizing the kernels of the SYK model and taking a higher-spin limit  \eqref{freechiral}, agrees well with the result from a direct higher-spin CFT perturbation computation \cite{Gaberdiel:2015uca}, as well as the dispersion relation of a classical rotating string in AdS spacetime in the large-spin regime \cite{Gubser:2002tv}. 
This agreement means we have a consistent picture describing the properties of a model with an approximate higher-spin symmetry: we get the same results by either going away from a theory with exact higher-spin symmetry or going towards a higher-spin limit from a model that does not have a higher-spin symmetry. 
Since our computation is from a different approach, it serves as an independent evidence to support the general picture that the SYK-like models can be related to some finite tension string theory, while the higher-spin theories can be regarded as some (truncation of) tensionless string theory. In our model tuning the $\m$ parameter away from the limit \eqref{freechiral} drives the model from a higher-spin-like regime to SYK-like regime, which mimics the process of turning on the string tension.\\

\noindent $\bullet~$ The Lyapunov exponent\\
The higher-spin limit $\m\to \left(\frac{1}{q}\right)^+$ is also detected from the chaotic behavior. Indeed the Lyapunov exponent vanishes in the limit $\m\to \left(\frac{1}{q}\right)^+$ as can be seen from figure~\ref{fig:lmu1}. This vanishing Lyapunov exponent agrees with previous expectations that theories with an infinite dimensional higher-spin symmetry is not chaotic~\cite{Shenker2015,Perlmutter:2016pkf,Maldacena:2016hyu}.\\

\noindent $\bullet~$ The $q$-dependence\\
A further comment is about the relative magnitude of the anomalous dimensions for different $q$. We notice that the anomalous dimensions become smaller as $q$ becomes larger. 
This is what we expected for many previously studied SYK-like models; the larger the value of $q$ the less relevant the interaction. Consequently, we expect the anomalous dimensions to be smaller for a larger $q$. \\

\noindent $\bullet~$ This is a singular limit\\
A last comment is that this $\m\to (\frac{1}{q})^+$ limit is singular: one cannot naively take all the formula and plug in $\m=\frac{1}{q}$ since the Fourier transform \eqref{ftf} diverges. Instead one has to consider setting $\m=\frac{1}{q}+\del$ with $1\gg\del>0$ and extract the result by taking the limit $\e\to 0$, which is how all the above results are computed.  It is not very surprising that the higher-spin limit is singular: 
we have learned from many other cases that the higher-spin limit of different models are often singular \cite{Seiberg:1999xz, Maldacena:2000hw, Aharony:2006th} , see \cite{Gaberdiel:2017oqg,Ferreira:2017pgt,Giribet:2018ada,Gaberdiel:2018rqv} for recent development. A different way to phase the singular nature of this limit is that the system undergoes a phase transition when going into/away from this limit. This is a first order phase transition because the free energy of the system, which can be expressed in terms of the IR propagators similar to \cite{KitaevTalk2,Maldacena:2016hyu,Kitaev:2017awl},  is discontinuous in this limit.
We call this a ``classical chiral" limit since the chiral mulitplet has classical dimension. But this  is not quite a free field limit since the coupling remains large in this limit and the dimension of the Fermi multiplet does not take its classical/free value.

\subsection{The $\m \to +\infty $ (``classical Fermi") limit}

We can consider a different limit 
\bal
\m \to +\infty\,,  \label{freefermi}
\eal
where the IR dimension of the various fields are
\bal
& \lim_{\m\to +\infty }h_{\phi }= \frac{1}{2 q}\,,~~  \lim_{\m\to +\infty } h_{\psi }= \frac{1+q}{2 q}\,,~~  \lim_{\m\to +\infty } h_{\lambda }= 0\,,~~  \lim_{\m\to +\infty }h_G= \frac{ 1}{ 2}\\
&  \lim_{\m\to +\infty }\tilde{h}_{\phi }= \frac{1}{2 q}\,,~~ \lim_{\m\to +\infty }\tilde{h}_{\psi }= \frac{1}{2 q}\,,~~ \lim_{\m\to +\infty }\tilde{h}_{\lambda }= \frac{1}{2 }\,,~~ \lim_{\m\to +\infty }\tilde{h}_G= \frac{1}{2}\,,
\eal 
In this limit another tower of higher-spin operators emerges. To illustrate this we carry out a set of computation that is in parallel with what we have done in the previous subsection. \\

\noindent $\bullet~$ A tower of higher-spin operators\\
As in the previous limit \eqref{freechiral}, we look for holomorphic operators with dimension $(h,\tilde{h})=(s,0)$ and anti-holomorphic operators with dimension $(h,\tilde{h})=(0,s)$ in both the symmetric and antisymmetric channels of the limit \eqref{freefermi}. This amounts to solve
\bal
\lim_{\m\to +\infty}E_c(\pm 1,\tilde{h}+s,\tilde{h},\m,q)=0\,,\quad \lim_{\m\to +\infty}E_c(\pm 1,{h},{h}+s,\m,q)=0\,, \quad s\in \mathbb{Z}_+\ . 
\eal
Numerically we  get a function $\tilde{h}(\m)$ or  ${h}(\m)$ for any given $s$ and $q$.  
The emergent conserved higher-spin operators are summarized in the following table
\begin{center}
	\renewcommand{\arraystretch}{1.5}
	\begin{tabular}{|c|c|c|}
		\hline 
		& holomorphic operators & anti-holomorphic operators \\ 
		\hline 
		symmetric channel & $(h,\tilde{h})= (s,0)$, $s\geq 2$, $q=2$  & $(h,\tilde{h})= (0,s)$ , $s\geq 1$ \\ 
		\hline 
		antisymmetric channel & $(h,\tilde{h})= (s,0)$, $s\geq 1$, $q=2$  & $(h,\tilde{h})= (0,s)$ , $s\geq 1$ \\ 
		\hline 
	\end{tabular} 
	\renewcommand{\arraystretch}{1}
\end{center}
Therefore, we find a set of holomorphic higher-spin operators in the right-moving sector only at $q=2$. This spectrum matches with the spectrum of generators of the bosonic subalgebra of the $\cn=2$ $\cs\cw_{1+\infty}$ algebra. This is again consistent with the $\cn=2$ supersymmetry in the right-moving sector. 
In addition, we find a tower of higher-spin operators in the left-moving sector for any $q>1$. Therefore we expect an $\cn=(0,2)$ supersymmetric higher-spin symmetric in the limit \eqref{freefermi} at $q=2$. Therefore at $q=2$ we find both the holomorphic and anti-holomorphic conserved higher spin operators at each integer spin $s$. But for other $q>2$, we only find anti-holomorphic higher-spin operators that generate an anti-chiral bosonic $ \cw_{1+\infty}$ algebra in the left-moving sector. Notice that this tower of chiral higher-spin operators is slightly unfamiliar since a general conserved higher-spin currents should have both the holomorphic and anti-holomorphic components. Therefore we believe the tower of higher-spin operators at $q=2$ are related to the usual higher-spin currents that generate the higher-spin symmetry. 

\begin{figure}[t!]
	\centering
	\begin{subfigure}[t]{0.5\textwidth}
		\centering
	\includegraphics[width=0.9\linewidth]{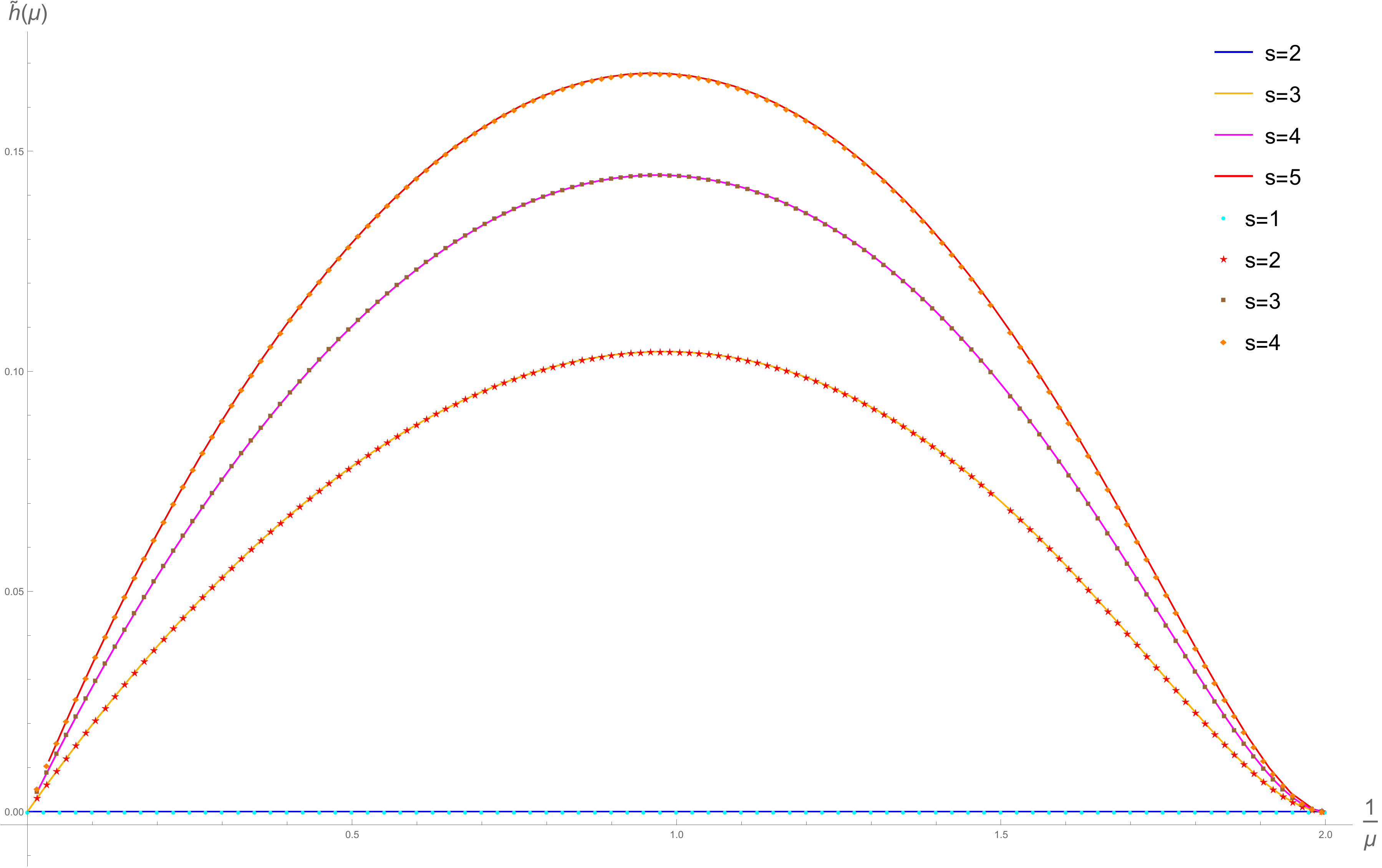}
	\caption{The $q=2$ case. The anomalous dimension of the higher-spin operators vanish in the limit $\m\to \infty$. They become conserved higher-spin currents in this limit.  The spin-2 operator, namely the stress-energy tensor, is always conserved for the whole range of $\m$ where our SYK solution is reliable. }
	\label{fig:hshbmu2}	
	\end{subfigure}%
	~ 
	\begin{subfigure}[t]{0.5\textwidth}
		\centering
		\includegraphics[width=0.9\linewidth]{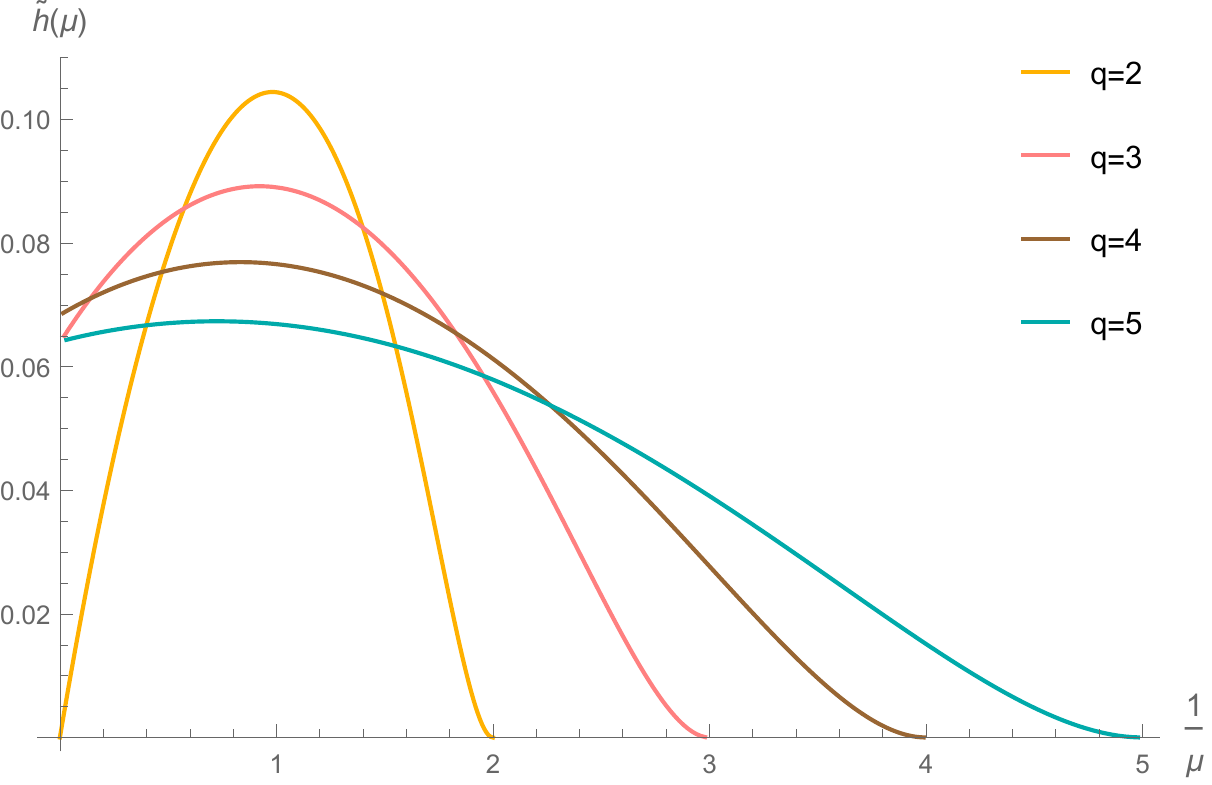}
		\caption{The $q\neq 2$ cases. The curves are the anomalous dimension of the spin $s=3$ operator as a function of $\m$. The operator has nonvanishing anomalous dimension and are not conserved as $\mu\to +\infty$. The behavior of the operators in the antisymmetric channel are similar.}\label{fig:hshbmuq}
	\end{subfigure}
	\caption{The anomalous dimension of the  almost holomorphic higher-spin operators as a function of $\frac{1}{\m}$ for the entire range of $\m$.}\label{fig:hshbmuent}
\end{figure}

\begin{figure}[t!]
	\centering
	\begin{subfigure}[t]{0.5\textwidth}
		\centering
		\includegraphics[width=0.9\linewidth]{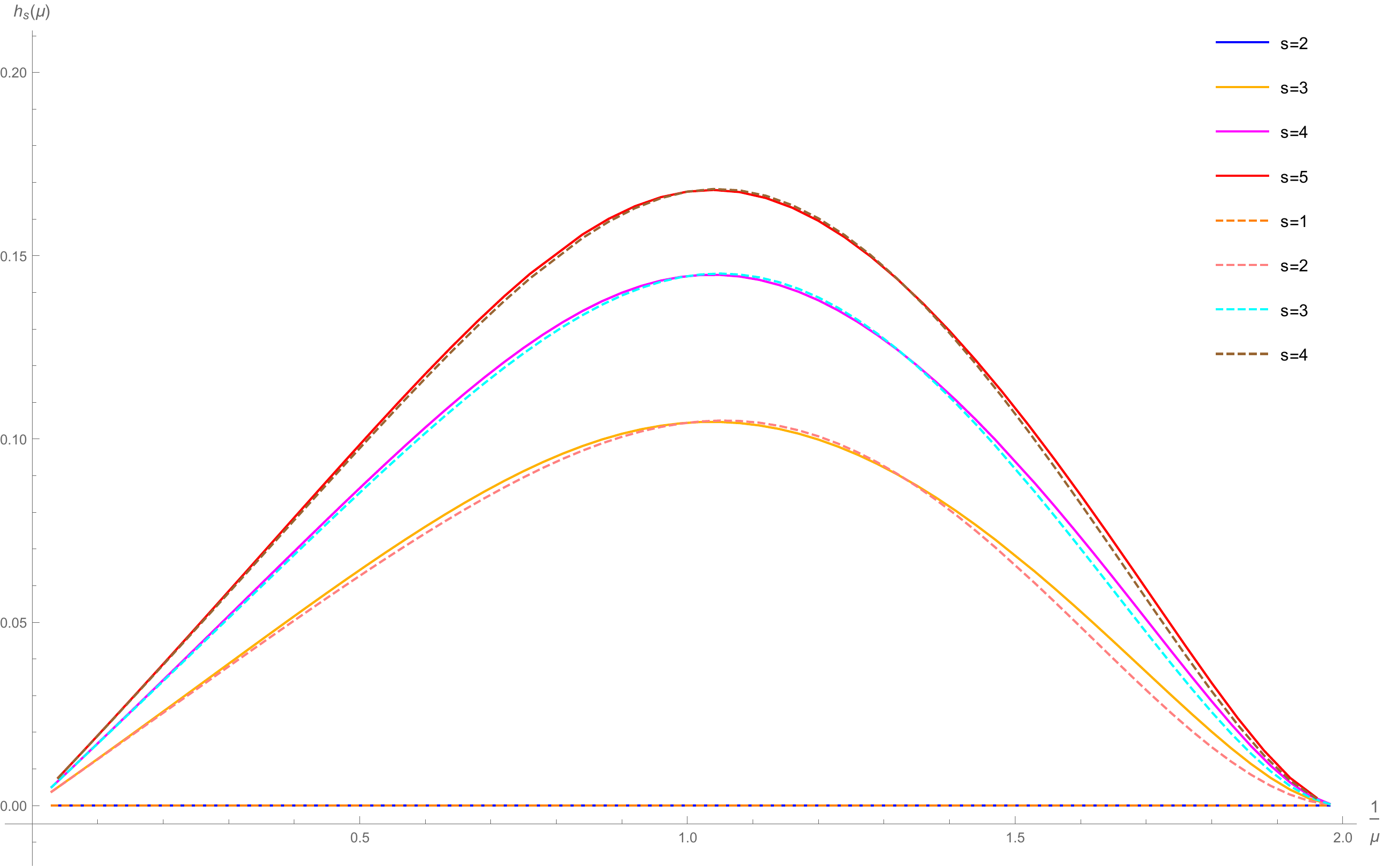}
		\caption{The $q=2$ case.}\label{fig:hshbmuqa1}
	\end{subfigure}%
	~ 
	\begin{subfigure}[t]{0.5\textwidth}
		\centering
		\includegraphics[width=0.9\linewidth]{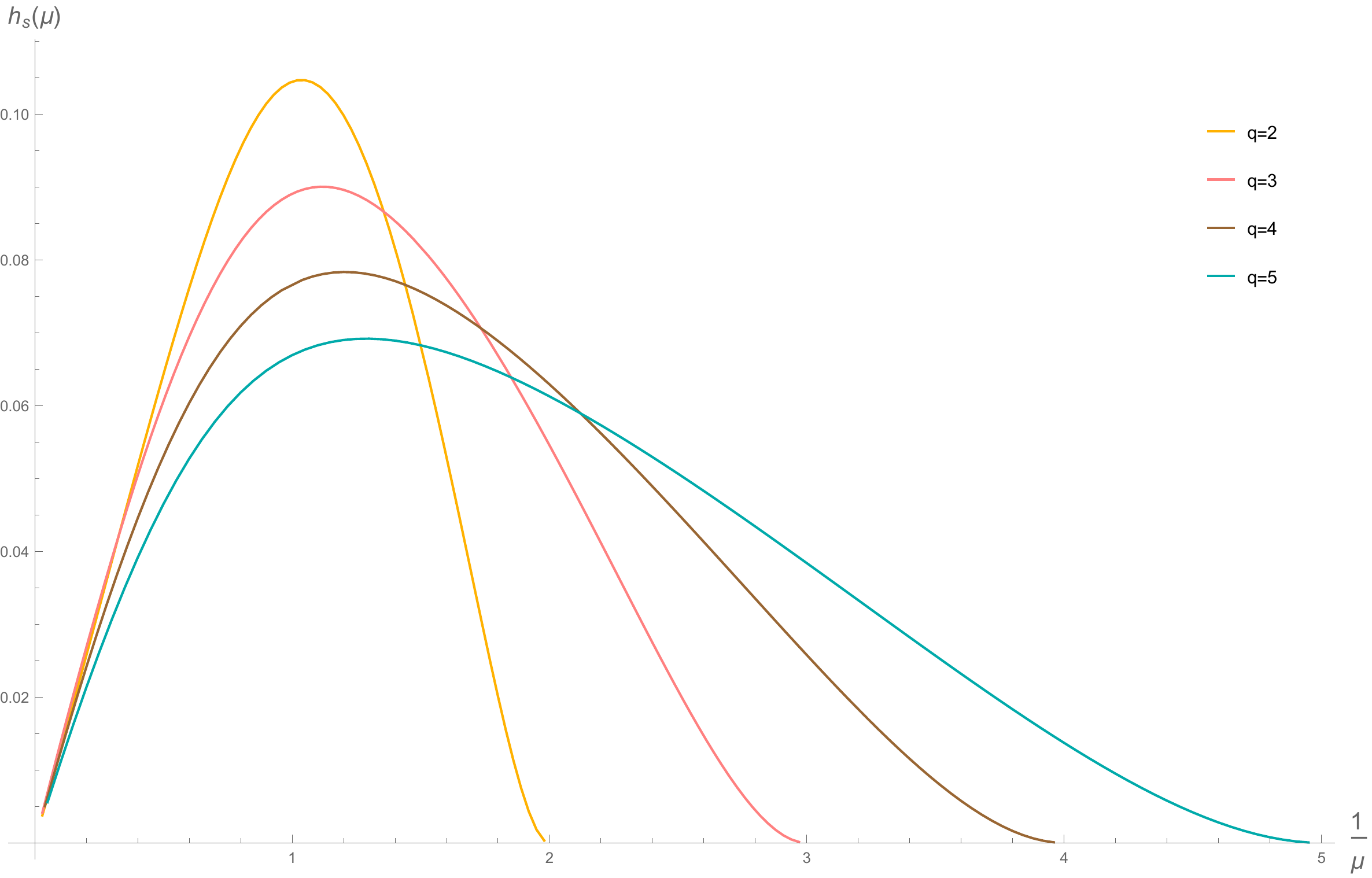}
		\caption{General $q$.}\label{fig:hshbmuqa2}
	\end{subfigure}
	\caption{Anomalous dimension of the anti-holomrphic higher-spin operators. The solid curves are for operators in the symmetric channel and the dashed lines are for operators in the antisymmetric channel.  The plot is for $q=3$. No overlapping of the solid and dashed curves are obvserved, which reflects the absence of supersymmetry in the left-moving sector.}
	\label{fig:hshbmuqa}
\end{figure}

Some examaples of the dimensions of the holomorphic higher-spin operators are shown in figure~\ref{fig:hshbmu2}, where we plot the left dimension $\tilde{h}$ of the spin-$s$ operators both in the symmetric (solid lines) and the antisymmetric (discrete points) channels as a function of $\frac{1}{\m}$ for the whole range of $\frac{1}{\m}$.
 From this result we  indeed see that in the $\m \to +\infty$ limit the left dimension $\tilde{h}_s(\m)$ all approach zero, which indicates that  there is a tower of holomorphic   operators with dimension $(h,\tilde{h})=(s,0)$. 
We also plot the anomalous dimension of the anti-holomorphic operators in figure~\ref{fig:hshbmuqa}. We do not see any coincidence of the dimensions between the symmetric and antisymmetric channels. This indicates that in the infrared the model only has $\cn=(0,2)$ supersymmetry, namely there is no supersymmetry in the left-moving section, away from the higher-spin limits \eqref{freechiral} and \eqref{freefermi}.  \\

\noindent $\bullet~$ The $q$-dependence\\
We notice that the tower of holomorphic higher-spin operators only emerges at $q=2$. We have solved the equation
\bal
\lim_{\m\to +\infty}E_c(1,\tilde{h}+s,\tilde{h},\m,q)=0\,,
\eal 
for a few different $q$'s and we find that the operators with higher spin have nonvanishing anomalous dimension for all $q>2$. This is shown in figure~\ref{fig:hshbmuq}. 
 We do not have a good physical explanation of why $q=2$ is special in this sense.\footnote{We remind the reads that our model \eqref{Sintsusy} is not free at $q=2$. Instead the interaction becomes a random mass term at $q=1$.} For completeness we list the dimensions of the chiral and Fermi multiplet in the limit \eqref{freefermi} at $q=2$ 
\bal
& \lim_{\m\to +\infty,q=2 }h_{\phi }= \frac{1}{4 }\,,~~  \lim_{\m\to +\infty,q=2} h_{\psi }= \frac{3}{4 }\,,~~  \lim_{\m\to +\infty,q=2} h_{\lambda }= 0\,,~~  \lim_{\m\to +\infty,q=2}h_G= \frac{1}{2  }\\
&  \lim_{\m\to +\infty,q=2}\tilde{h}_{\phi }= \frac{1}{4}\,,~~ \lim_{\m\to +\infty,q=2}\tilde{h}_{\psi }= \frac{1}{4}\,,~~ \lim_{\m\to +\infty,q=2}\tilde{h}_{\lambda }=\frac{1}{2 }\,,~~ \lim_{\m\to +\infty,q=2}\tilde{h}_G= \frac{1}{2 }\ .
\eal 
We notice that in the limit \eqref{freefermi} the chiral multplet gets its largest possible  dimension at $q=2$.\\

\begin{figure}
	\centering
	\includegraphics[width=0.7\linewidth]{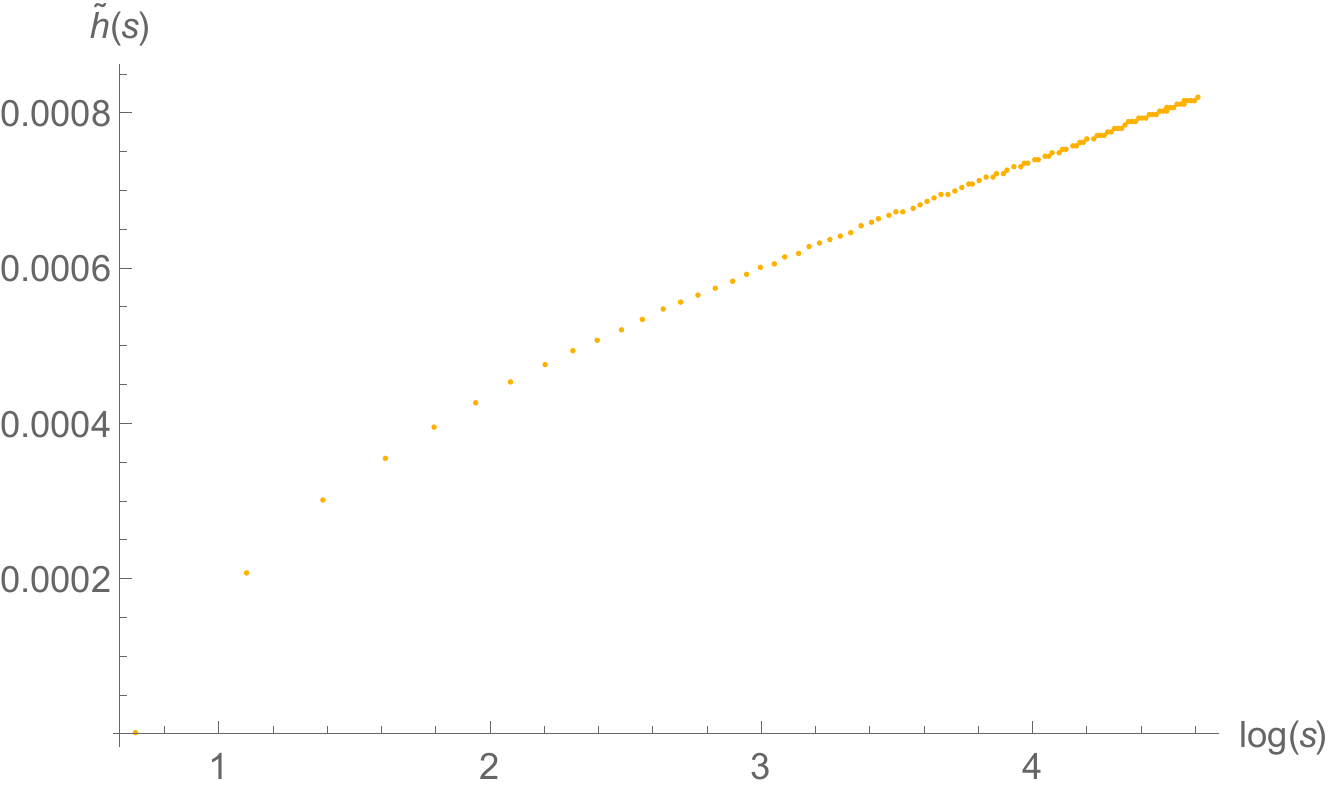}
	\caption{The anomalous dimension of the higher-spin operators as a function of their spin. The plot is around the higher-spin limit \eqref{freefermi}. We observe that for large enough spin the anomalous dimension is proportional to the log of the spin.}
	\label{fig:hblogs2}
\end{figure}

\noindent $\bullet$ Anomalous dimensions\\
The anomalous dimensions again have a logarithmic dependence on the spin, as can be seen from figure~\ref{fig:hblogs2}. This is similar to the behavior in the other limit \eqref{freechiral} and agrees with the result from pure higher-spin theory computation \cite{Gaberdiel:2015uca}. We only plot the holomorphic operators in the symmetric channel; the holomorphic operators in the anti-symmetric channel are in the same supermultplets and have identical anomalous dimensions.
\\

\noindent $\bullet$ The Lyapunov exponent\\
We can compute the Lyaponov exponent of our model for the whole range of $\m$. We find vanishing Lyaponov exponent at $\m\to \infty$ only at $q=2$. This is compatible with the fact that a normal higher-spin symmetry emerges in this limit only at $q=2$, as we have discussed above.  \\
\begin{figure}
	\centering
	\includegraphics[width=0.6\linewidth]{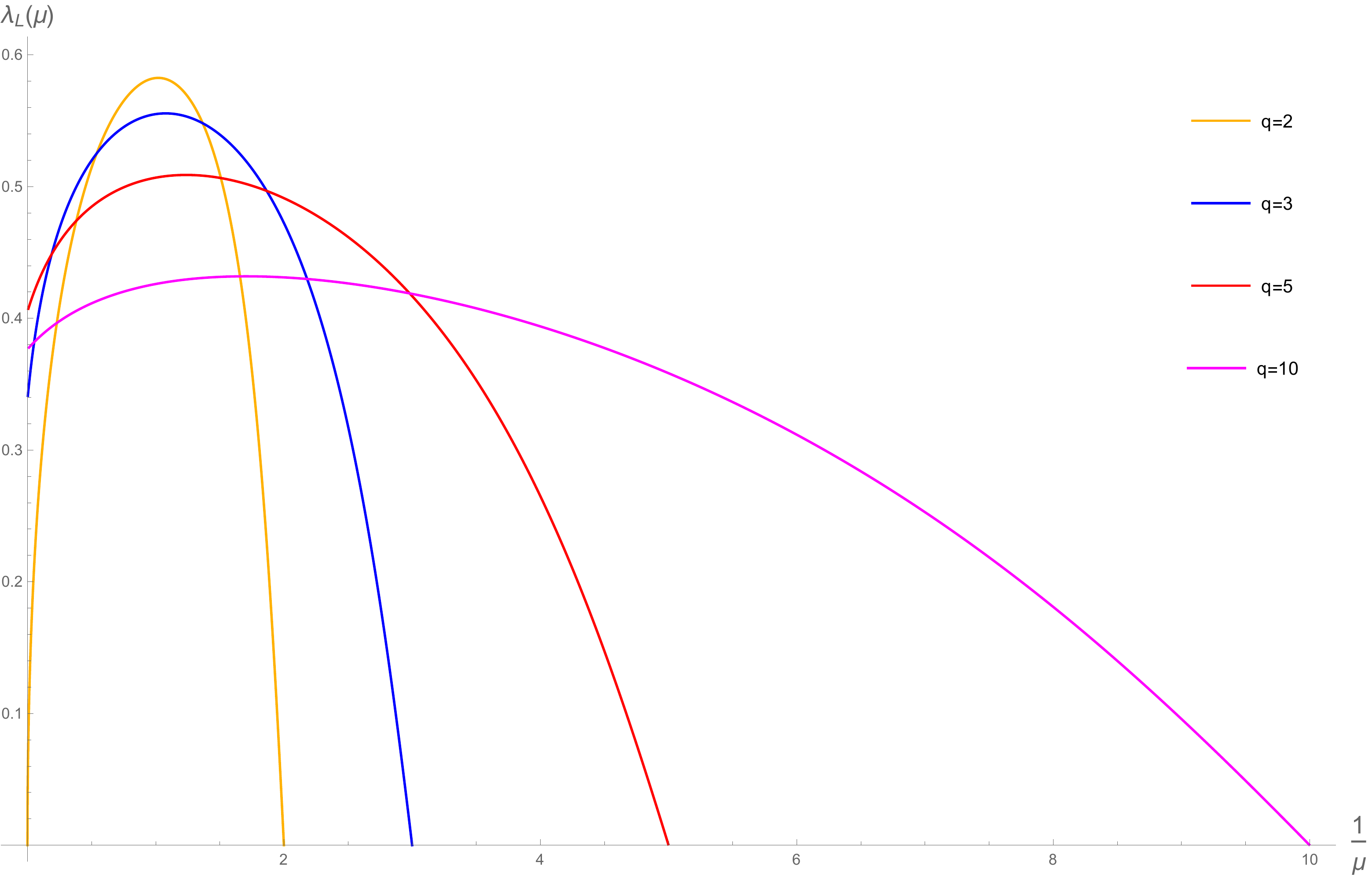}
	\caption{The Lyapunov exponent as a function of $\m$ for the whole range of $\m$. We observe that near the $\m\to+\infty$ limit only the model with $q=2$ has a vanishing Lyapunov exponent. This is consistent with the fact that only at $q=2$ the set of higher-spin operators have vanishing anomalous dimension and generate a higher-spin symmetry.}
	\label{fig:l1mu2}
\end{figure}

\noindent $\bullet~$ This is a singular limit\\
The limit \eqref{freefermi} is singular for a similar reason as the limit \eqref{freechiral}: the result depends on how the limit is taken. In the above analysis, we solve for the anomalous dimension at a given $\m$ around $\m \to +\infty$ and track how does this anomalous dimension behave as $\m$ approaches $+\infty$. If we instead take an unphysical approach by simply plugging $\frac{1}{\m}=0$ and $h=0 $ (or $\tilde{h}=0$) into the equation \eqref{keyker}, we are not guaranteed to get the same result. In addition, in this computation we have first assumed the $\m$ to be finite, solve the set of Schwinger-Dyson equations, and then take the $\m\to \infty$ limit of the solution. We do not expect to recover the same result if we first take a naive $\m\to+\infty$ limit of the  Schwinger-Dyson equation and then solve it. As in the previous limit \eqref{freechiral}, we believe  this subtlety reflects the common feature that a tensionless limit of string theory is usually singular and the model undergoes a first order phase transition in this limit.

\subsection{Relations with higher-spin theories}

\begin{figure}
	\centering
	\includegraphics[width=0.7\linewidth]{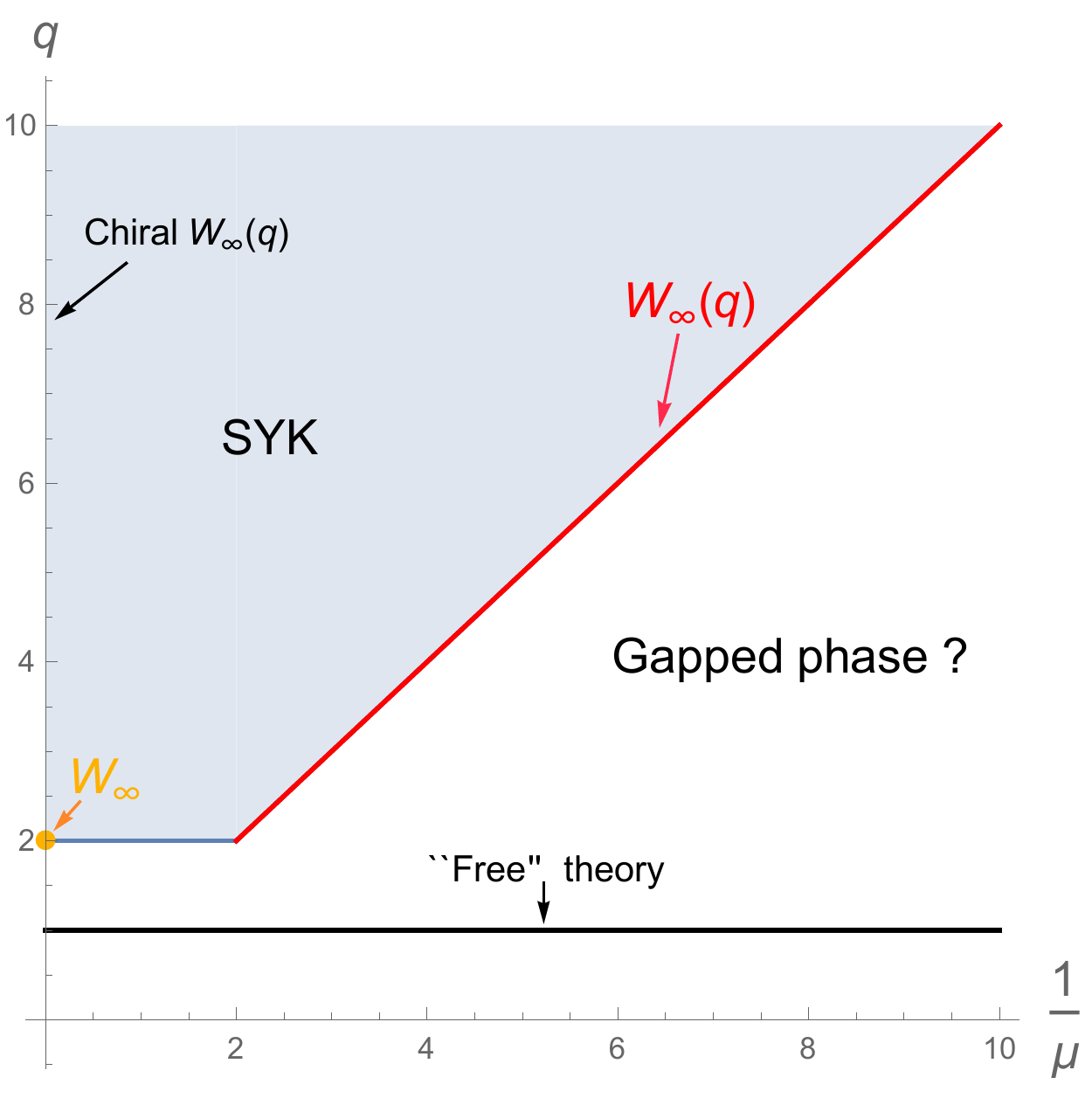}
	\caption{An illustration of the moduli space of our model. In the shaded region the model has SYK-like behavior. In the two limits, which are on the boundary of the shaded region, one observe emergent higher-spin symmetries. At $q=1$, the interaction is quadratic, which can be regarded as a free ``mass" term and the model is essentially free. Outside the shaded region the Fourier transform in \eqref{ftf} diverges. As suggested in \cite{Murugan:2017eto} one probably has to turn on a negative mass counterterm to reach another fixed point. This is an indication that in this region the model might be gapped.}
	\label{moduli}
\end{figure}

Our model is defined in 2-dimensions where  conformal field theories with higher-spin symmetries and their holographic dual have been extensively studied, see e.g. \cite{Gaberdiel:2012uj} for a review. In a supersymmetric context, the higher-spin conserved currents generate a supersymmetric $\cs\cw_{\infty}$-algebras. 
It is well known that the $\cs\cw_{\infty}$ algebra 
allows a continuous deformation that preserves the higher-spin symmetry; there is a family of higher-spin $\cs\cw_{\infty}[\l]$ algebra. 
In our model there are two parameters. As we have shown in previous sections, the $\m$ parameter  controls  the breaking of the higher-spin symmetry. While in the higher-spin limit \eqref{freechiral}, there is no requrement of the $q$ paramter: the higher-spin symmetry emerges at any given $q>2$. It is thus natural to conjecture that the different higher-spin algebras emerging in the models with different $q$ should be identified with the $\cs\cw_{\infty}[\l(q)]$ algebras via an explict map $\l=\l(q)$. At the moment we have not yet identified this map, but a natural conjecture is 
\bal
\lambda=\frac{1}{\m q}\ .
\eal
We plan to study this point in detail in future works. In figure~\ref{moduli} we draw a cartoon summarizing the properties of the model \eqref{Sintsusy} on the moduli space spanned by the $q$ and $\mu$ paramerters. On the graph we mark out the two limits where higher-spin symmetries emerge. Outside the shaded region the Fourier transform in \eqref{ftf} diverges. As suggested in \cite{Murugan:2017eto} one probably has to turn on a negative mass counterterm to reach another fixed point. This is an indication that in this region the model might be gapped. It is interested to clarify this point further, and we will defer this into future works.

The existence of the $\m$ parameter and the appearance of a higher-spin limit as we vary $\m$ has some resemblance to the ABJ triality \cite{Chang:2012kt} in one dimension higher. In our model there are indeed two global symmetry $U(N)\times U(M)$ corresponding to the rotations of the $N$ chiral multiplets and the $M$ Fermi multiplets respectively. Notice that this symmetry is only manifest once we average over the random coupling. Furthermore since we consider only singlets under these two global symmetry groups, we do not expect the results discussed in this paper to alter significantly once we gauge the $U(N)\times U(M)$ group \cite{Maldacena:2018vsr}. As a result, tuning our parameter $\m$ to reach limits with higher-spin symmetries has a qualitative  similarity with, in the context of the ABJ triality, dialing the ratio $\frac{M}{N}$ to reach a conjectured dual of $\cn=6$ matrix extended Vasiliev higher-spin theory \cite{Chang:2012kt}. But there are also qualitative differences between our model and the case of the ABJ triality. Firstly, the ABJ theory consists of  (anti-)bifundamental fields, while our model \eqref{Sintsusy} is built from (anti-)fundamental fields of the two global symmetry groups. Therefore we expect a different mechanism of the emergence of the higher-spin fields. Secondly, as the ratio $\frac{M}{N}$ increases there are two phase transitions separating three different phases with different thermodynamic properties \cite{Chang:2012kt}.  But in our model, at least for $q=2$, as we increase $\m$ the model moves from a higher-spin behavior to a chaotic SYK-like behavior and then back to a higher-spin behavior (although a different one comparing to where the model started).

There could be other parameters that we can turn on while maintaining the properties of the models we have discussed in this paper. 
Notice that this possibility goes along with our previous analysis based on the chaotic behavior of this model, where the fact that the Lyapunov exponent does not saturate the chaos bound \cite{Maldacena2016a} could be a hint that there should be other directions on the moduli space of 2d SYK-like models along which the model ets more and more chaotic.

\section{Tensor models}

As in 1 dimension, the SYK-like model we have discussed in this paper has tensor analogues that do not involve any random coupling. The construction of the tensor models is similar to the 1d case~\cite{Peng:2016mxj} where we promote each multiplet to carry tensor indices.
In this section we construct a tensor model 
that reproduces the physics of the model \eqref{Sintsusy}, 
which is a direct generalization of~\cite{Witten:2016iux,Peng:2016mxj}.

The model has a global symmetry
\bal
G=\prod_{a,b=0,\ldots,3}H_{ab}(n_{ab})
\,,\label{tensorG}
\eal
where 
\bal
&H_{ab}(n_{ab})=U(N)\,, \quad \text{ if } ab\in \{01,12,13,23\}\\
&H_{ab}(n_{ab})=U(M)\,, \quad \text{ if } ab\in \{02,03\}\ .
\eal
It consists of the following chiral and Fermi supermultiplets
\bal
\L_{01}^{i_{02}i_{03}i_{12}i_{13}}\,,\qquad \F_{02}^{i_{01}i_{03}i_{12}i_{23}}\,, \qquad \F_{03}^{i_{01}i_{02}i_{13}i_{23}} \,,
\eal
where the $ab$ subscripts are color indices labeling the different supermutliplets.
The  $i_{ab}$ superscripts are tensor indices labeling different components of the fundamental representation of $H_{ab}(n_{ab})$.  Similarly, there are conjugate fields
\bal
(\bar{\L}_{01})_{i_{02}i_{03}i_{12}i_{13}}\,,\qquad (\bar{\F}_{02})_{i_{01}i_{03}i_{12}i_{23}}
\,, \qquad (\bar{\F}_{03})_{i_{01}i_{02}i_{13}i_{23}} \,,
\eal 
The action of the tensor model is
\begin{multline}
S=\int dx^2 d\q^+ d\bar{\q}^+ \left(\frac{1}{2} \bar{\L}_{01}\L_{01}-\bar{\F}_{02}\pa_{\bar{z}}\F_{02}-\bar{\F}_{03}\pa_{\bar{z}}\F_{03}\right)\\
-\int dx^2d\q^+ \left(\frac{J}{ N} \L_{01} \F_{02}\F_{03}+h.c.\right) \,,\label{Sintsusyt}
\end{multline}
where we have omitted all the tensor indices that are contracted in the unique way so that each term in the action is a singlet under the global symmetry \eqref{tensorG}. This model is the $q=2$ analogue of \eqref{Sintsusy}, and it is easy to generalize to models with higher $q$. In the large-$N,M$ limit, the model is dominated by the same set of the melonic diagrams as in the model \eqref{Sintsusy}. This can be proved in  essentially the same as that in~\cite{Peng:2016mxj}. 
The 2-point functions in the infrared can be determined  again by solving the  set of the SD equations 
with the self energies the same as \eqref{sigmapsi02}-\eqref{sigmaL02}.
Therefore the infrared physics is the same as the model \eqref{Sintsusy} to the leading order of $\frac{1}{N}$, $\frac{1}{M}$.

\acknowledgments

We thank Ksenia Bulycheva, Anthony Charles,  Wenbo Fu, Ping Gao, Harsha Hampapura,  Thomas Hartman, Antal Jevicki, Shota Komatsu, Eric Marcus, Daniel Mayerson, Leopoldo Pando Zayas, Eric Perlmutter, Vimal Rathee, Marcus Spradlin, Stefan Stanojevic, Bogdan Stoica, Grigory Tarnopolsky, Anastasia Volovich,  Cenke Xu, Zhenbin Yang and Alexander Zhiboedov for discussions on related topics. We are especially grateful to Juan Maldacena and Douglas Stanford for illuminating discussions. In particular we thank Douglas Stanford for sharing a code based on which the numerical analysis in section~\ref{syk02} is developed.  
We thanks the Leinweber Center for Theoretical Physics of  University of Michigan, Ann Arbor and the Institute for Advanced Study for hospitality at different stages of this project. This work was supported by the US Department of Energy under contract DE-SC0010010 Task A.

\appendix

\section{2d SYK model with $\cn=(2,2)$ supersymmetry}\label{syk22} 

In 2 dimension, the $\cn=(2,2)$ superfields are constructed with the help of two complex fermionic coordinates $\q^+$ and ${\q}^-$ with
\bal
\bar{\q}^\pm=(\q^\pm)^*\ .
\eal
A bosonic chiral superfield is defined as
\bal
\Phi(x^i,\q^\pm,\bar{\q}^\pm)&=\f(y^\pm)+\sqrt{2}\q^+ \psi(y^\pm)+\sqrt{2}\q^- \l(y^\pm)+2\q^+\q^-F(y^\pm)\,,
\eal
where $y^\pm=x^\pm - i \q^\pm\bar{\q}^\pm$.
A simple 2d SYK-like model with $\cn=(2,2)$ supersymmetry is~\cite{Murugan:2017eto,Bulycheva2018}
\bal
S&=\int dx^2d^2\q^+ d^2\q^- \F\bar{\F}+\int dx^2 d\q^+d\q^- \left(\frac{J_{i_0\ldots i_q}}{(q+1)!} \F^{i_0}\ldots \F^{i_q} + h.c.\right)\label{n22s}
\eal
In component form the interaction is
\bal
S^{(0,0)}_{\rm int}
&=\int dx^2 \left(\frac{J_{i_0\ldots i_q}}{q!} F^{i_0}\f^{i_1}\ldots \f^{i_q}+\frac{J_{i_0\ldots i_q}}{q!} \psi^{i_0}\l^{i_1}\f^{i_3}\ldots \f^{i_q} + h.c\right)\ . \label{n22comp}
\eal
It is straightforward to solve the Schwinger-Dyson equation of the two point functions in the IR limit 
\bal
\int d^2z_2G_c^I(z_1,z_2)\S_c^I(z_2,z_3)=-\delta^2(z_1-z_3)\,, \qquad I=\f\,,\psi\,,\l\,,G\,,\label{sd22}
\eal
where $G_c^I(z_1,z_2)$ are the propagators in the infrared
and $\S_c^I(z_1,z_2)$ are the corresponding self-energies of the various fields
\bal
\S^\psi(z_1,z_2)&= 2J^2    G^{ {\l}}(z_1,z_2)  (G^{ {\f}}(z_1,z_2))^{q-1}\label{sigmapsi}\\
\S^\f(z_1,z_2)&=(q-1)2  J^2( \nn G^{{\f}}(z_1,z_2))^{q-2}G^{{\l}}(z_1,z_2)G^{{\psi}}(z_1,z_2)\\
&\hspace{5cm} +2 { J^2}  (G^{{\f}}(z_1,z_2))^{q-1} G^{{G}}(z_1,z_2)\,,\\
\S^G(z_1,z_2)&=  \frac{2   J^2}{  q} (G_c^{{\f}}(z_1,z_2))^q \,, \label{sigmaG}\\
\S^\l(z_1,z_2)&=   {2  J^2}{ } (G_c^{ \f}(z_1,z_2))^{q-1} G_c^{ \psi}(z_1,z_2) \,,\label{sigmaL}
\eal
A supersymmetric solution of these equations reads
\bal
G^\f_c(z_1,z_2)&=\left(\frac{q}{8 \p^2J^2}\right)^{\frac{1}{q+1}}  {(z_1-z_2)^{-\frac{1}{q+1}} (\bar{z}_1-\bar{z}_2)^{-\frac{1}{q+1}}}\,,
\eal
and the other propagators are related by supersymmetry
\bal
&G^\psi(z_1,z_2)=-2\pa_{z_1} G^\f(z_1,z_2)\,,\quad G^\l(z_1,z_2)=-2\pa_{\bar{z}_1} G^\f(z_1,z_2)\\
&G^G(z_1,z_2)=-2\pa_{z_1}  G^\l(z_1,z_2)= -2\pa_{\bar{z}_1} G^\psi(z_1,z_2)\ .
\eal
The existence of such solutions is nontrivial. Indeed one can write down a similar set of Schwinger-Dyson equations for various models without supersymmetry \cite{Murugan:2017eto}, but  it is not clear if the theory actually flows to the assumed SYK-like IR fixed points.
The presence of supersymmetry improves the UV behavior and the potential divergences are avoided. 
In practice, the supersymmetry allows a natural regularization and renormalization of the possible divergences in a naive direct solution.

For later purpose it is useful to recast 
the $\cn=(2,2)$ model \eqref{n22s}  as a model with a smaller number of supersymmetry.
In terms of the $\cn=(0,2)$  superfields 
\bal
\text{chiral:  }~\F'^{i}(x,\q^+)&=\F^i(y,\q^\pm,\bar\q^\pm)\big|_{\q^-,\, \bar{\q}^-\to 0}=\f(x)+\sqrt{2}\q^+\psi(x)+2\q^+\bar{\q}^+\pa_z \f(x)\\
\text{Fermi:  }~\L(x,\q^+)^i&=D_-\F^i(y,\q^\pm)\big|_{\q^-\to 0}=\l(x)-\sqrt{2}\q^+F(x)+2\q^+\bar{\q}^+ \,\pa_{z}\psi_-(x) \,,\label{dec2}
\eal
where $ D_-=\frac{\pa}{\pa \q^-}+ 2\bar{\q}^-\pa_z$ and 
the action can be written as
\bal
S^{(0,2)}&= \int dx^2 d\q^+ d\bar{\q}^+ \left(\frac{1}{2}\bar{\L}\L-\bar{\F}\pa_{\bar{z}}\F\right)
+\int d\q^+ \frac{J_{i_0\ldots i_q}}{q!} \L^{i_0}\F^{i_1}\ldots \F^{i_q} + h.c.\label{m02} \ .
\eal

\section{Perturbing the $\cn=(2,2)$ model to the $\cn=(0,2)$ models}\label{pert}

In this section, we treat the $\cn=(0,2)$ model as a perturbation of the $\cn=(2,2)$ model. We then solve the $\cn=(0,2)$ model perturbatively around the $\cn=(2,2)$ fixed point. We will show 
that the result from this perturbative analysis does agree with the $\m\to 1$ expansion of the result in section~\ref{syk02}. Since the computation in this section is independent from directly solving the Schwinger-Dyson equations in section~\ref{syk02}, the results in this section provide another consistency check of the results in the main text.

The perturbed theory has the same action as in section~\ref{syk02}, the only difference is that now we take the ratio $\m=M/N$ to be $\m=1+\e$\,, where $\e\ll 1$ is treated as a small parameter, and we expand the  propagators of the  deformed model as
\bal
G^I(x)=G^I_c(x)(1+\e g^I(x)+\ldots)\,,\qquad I=\f,\psi,\l,G\ .\label{Gexp}
\eal
In frequency domain, the perturbation can be denoted as
\bal
G^I(p)=G_c^I(p)+\e \left(\widetilde{G_c^I\times g^I}\right)+\ldots\label{Gdef}\,,
\eal
where $\widetilde{~}$ indicates a Fourier transform. 
The self energies of the deformed model are, to the leading order in $\e$
\bal
\S^\psi(z_1,z_2)&= 2J^2 (1+\e)  G^{ {\l}}(z_1,z_2)(1+\e g^\l(z_1,z_2)) (G^{ {\f}}(z_1,z_2))^{q-1}(1+\e g^\f(z_1,z_2))^{q-1}\\
&=\S^\psi(z_1,z_2)_c\left(1+\e(1+g^\l+(q-1)g^\f))\right)\,,\label{sigmapsip}\\
\S^\f(z_1,z_2)&=(q-1)2 (1+\e) J^2( \nn G^{{\f}}(z_1,z_2))^{q-2}G^{{\l}}(z_1,z_2)G^{{\psi}}(z_1,z_2)\\
&\hspace{5cm} +2 { J^2} (1+\e) (G^{{\f}}(z_1,z_2))^{q-1} G^{{G}}(z_1,z_2)\,,\\
&=\frac{q-1}{q}\S^{\f}_c\left(1+\e(1+g^\l+g^\psi+(q-2)g^\f )\right)+\frac{1}{q}\S_c^{\f}\left(1+\e(1+g^G+(q-1)g^\f)\right)\\
\S^G(z_1,z_2)&=  \frac{2   J^2}{  q} (G_c^{{\f}}(z_1,z_2))^q\left(1+\e q g^\f)\right)=\S_c^G(z_1,z_2)\left(1+\e q g^\f)\right)\,, \label{sigmagp}\\
\S^\l(z_1,z_2)&=   {2  J^2}{ } (G_c^{ \f}(z_1,z_2))^{q-1} G_c^{ \psi}(z_1,z_2)\left(1+\e( g^\psi+(q-1)g^\f))\right)\\
&=\S^\l(z_1,z_2)\left(1+\e( g^\psi+(q-1)g^\f))\right)\,,\label{sigmapL}
\eal
Because we are perturbing around the SYK fix point in the IR, we solve the Schwinger-Dyson equations of the perturbed model
\bal
G^I(p)\S^I(p)=-1\,,
\eal 
with the known results of the unperturbed ones
\bal
G_c^I(p)\S_c^I(p)=-1\ .
\eal
This leads to
\bal
G_c^I(p)\delta\S^I(p)+\delta G^I(p) \S^I_c(p)=0\,,
\eal
where $\delta G^I(p)=G^I(p)-G^I_c(p)$ and $\delta\S^I(p)=\S^I(p)-\S^I_c(p)$. In components, the equations are
\bal
-1&=G_c^\psi(p)\left(\widetilde{\S_c^\psi\times g^\l}+(q-1)\widetilde{\S_c^\psi\times g^\f}\right)+\S_c^\psi(p) \widetilde{G_c^\psi\times g^\psi}\label{modpsi}\\
-1&=G_c^\f(p)\left(\frac{q-1}{q}\widetilde{\S_c^\f\times g^\l}+\frac{q-1}{q}\widetilde{\S_c^\f\times g^\psi}+\frac{1}{q}\widetilde{\S_c^\f\times g^G}\right.\\
&\hspace{5cm }\left. +\frac{q^2-2q+1}{q}\widetilde{\S_c^\f\times g^\f}\right)+\S_c^\f(p) \widetilde{G_c^\f\times g^\f}\label{modpsi1}\\
0&=G_c^\l(p)\left(\widetilde{\S_c^\l\times g^\psi}+(q-1)\widetilde{\S_c^\l\times g^\f}\right)+\S_c^\l(p) \widetilde{G_c^\l\times g^\l}\label{modl1}\\
0&=G_c^G(p)\left(q\widetilde{\S_c^G\times g^\f}\right)+\S_c^G(p) \widetilde{G_c^G\times g^G}\ .\label{modG1}
\eal
Now we look for a solution of the above equations with $\cn=(0,2)$ supersymmetry, which means
\bal
\widetilde{G_c^G\times g^G}&=i\bar{p}\widetilde{G_c^\l\times g^\l}\,,\qquad 
\widetilde{G_c^\psi\times g^\psi}=i\bar{p}\widetilde{G_c^\f\times g^\f}\ .
\eal
These relations simplify \eqref{modl1} to
\bal
\left(\widetilde{\S_c^\l\times g^\psi}+(q-1)\widetilde{\S_c^\l\times g^\f}\right)&=i\bar{p} \left(q\widetilde{\S_c^G\times g^\f}\right)\ .
\eal
Now we can Fourier transform back to get
\bal
\S_c^\l\times g^\psi +(q-1) \S_c^\l\times g^\f &=-2q\pa_z \left(\S_c^G\times g^\f \right)= -2q\pa_z\S_c^G\times g^\f -2q\S^G_c \pa_z  g^\f\ .
\eal
Further using
\bal
\S_c^\l=-2 \pa_z\S_c^G\,,\qquad \S_c^\f=-2 \pa_{\bar{z}}\S_c^{\psi}\,,
\eal
we get
\bal
\S_c^\l\times g^\psi - \S_c^\l\times g^\f &=  -2q\S^G_c \pa_z  g^\f\ .
\eal
Multiplying by $(\S_c^\l)^{-1}G^\psi_c$ and using the known results for the $\cn=(2,2)$ conformal propagators and the self energies, we get
\bal
G^\psi_c\times g^\psi - G^\psi_c\times g^\f &=  -2 b^\f (z\bar{z})^{-\frac{1}{q+1}} \pa_z  g^\f\ .
\eal
Then using
\bal
G^\psi_c\times g^\psi=-2\pa_z (G^\f_c\times g^\f)\,,\eal
the above equation becomes
\bal
-2\pa_z (G^\f_c\times g^\f) - G^\psi_c\times g^\f &=  -2 b^\f (z\bar{z})^{-\frac{1}{q+1}} \pa_z  g^\f\ .
\eal
Plugging in the known results for the $\cn=(2,2)$ conformal  propagators, we get a trivial identity. 

Similarly, the  equations \eqref{modpsi1} turns out to be a trivial identity as well. 
This means that the $\cn=(0,2)$ SUSY is compatible with the 4 equations and it further reduces those to two equations. To solve the rest 2 equations we consider the following ansatz
\bal
g^\f= {n^\f+\del^\f \log(z\bar{z})} \,,\qquad g^\l= {n^\l+\del^\l \log(z\bar{z})} \ .\label{pllazt}
\eal

In the later computation, the following Fourier transform
\begin{multline}
\int_{-\infty}^{\infty}d^2z \frac{\log(z\bar{z})}{z^a \bar{z}^{a }}e^{i p\cdot z}=\int_{0}^{\infty} dr \frac{2\log(r)}{r^{2a-1}}\int_{0}^{2\p}d\q\,e^{i pr\cos(\q)}=4\p \int_{0}^{\infty} dr \frac{\log(r)}{r^{2a-1}}J_0(pr)\\
=\frac{\pi  4^{1-a} p^{2 a-2} \Gamma (1-a) (\psi ^{(0)}(1-a)+\psi ^{(0)}(a)-2 \log (p/2))}{\Gamma (a)}\ .\label{int1}
\end{multline}
Strictly speaking, the integral on the LHS is only convergent when $\frac{1}{4}<a<1$. However,  we do not worry to much about the $\frac{1}{4}$ end since it is an IR divergence and can be regulated by not going very deep in the IR. We only makes sure that $a<1$ is satisfied in the following computation.

Using \eqref{int1} we immediately compute
\bal
\widetilde{G_c^\f\times g^\f}&=a^\f G_c^\f(p)+d^\f  \frac{\pi  4^{ \frac{q}{q+1}}  \Gamma (\frac{q}{q+1}) (\psi ^{(0)}( \frac{q}{q+1})+\psi ^{(0)}(\frac{1}{q+1})-2 \log (p/2))}{k^{\frac{2}{q+1}} p^{ \frac{2q}{q+1}}\Gamma (\frac{1}{q+1})}\\
\widetilde{G_c^\f\times g^\l}&=a^\l G_c^\f(p)+d^\l  \frac{\pi  4^{ \frac{q}{q+1}}  \Gamma (\frac{q}{q+1}) (\psi ^{(0)}( \frac{q}{q+1})+\psi ^{(0)}(\frac{1}{q+1})-2 \log (p/2))}{k^{\frac{2}{q+1}} p^{ \frac{2q}{q+1}}\Gamma (\frac{1}{q+1})}\ .
\eal
Similarly, we have
\bal
\widetilde{\S_c^G\times g^\f}&=a^\f \S_c^G(p)+d^\f    \frac{k^{\frac{2}{q+1}} \Gamma (\frac{1}{q+1}) (\psi ^{(0)}(\frac{1}{q+1})+\psi ^{(0)}(\frac{q}{q+1})-2 \log (p/2))}{\pi  4^{ \frac{q}{q+1}}  p^{\frac{2}{q+1}}\Gamma (\frac{q}{q+1})}\\
\widetilde{\S_c^G\times g^\l}&=a^\l\S_c^G(p)+d^\l    \frac{k^{\frac{2}{q+1}} \Gamma (\frac{1}{q+1}) (\psi ^{(0)}(\frac{1}{q+1})+\psi ^{(0)}(\frac{q}{q+1})-2 \log (p/2))}{\pi  4^{ \frac{q}{q+1}}  p^{\frac{2}{q+1}}\Gamma (\frac{q}{q+1})}
\eal
In addition, we get
\bal
&\widetilde{\S_c^\psi\times g^\f}=\frac{1+q}{ q}d^\f \S_c^\psi(p)  + i\, p \int_{-\infty}^{\infty}d^2z   ( \S_c^G g^\f   e^{i p\cdot z} )=d^\f \S_c^\psi(p) \frac{1+q}{ q} + i\, p \widetilde{\S_c^G\times g^\f}\\
&\widetilde{\S_c^\psi\times g^\l}=\frac{1+q}{ q}d^\f \S_c^\psi(p)  + i\, p \int_{-\infty}^{\infty}d^2z   ( \S_c^G g^\f   e^{i p\cdot z} )=d^\l \S_c^\psi(p) \frac{1+q}{ q} + i\, p \widetilde{\S_c^G\times g^\l}\\
&\widetilde{G_c^\l\times g^\l}
=(1+q) d^\l G_c^{\l}(p)+i\, p \int_{-\infty}^{\infty}d^2z    (  G_c^{\f}   e^{i p\cdot z} g^\l )=(1+q) d^\l G_c^{\l}(p)+i\, p \,\widetilde{G_c^{\f} \times   g^\l}
\eal
where we have used
\bal
G_c^\l=-2 \pa_{\bar{z}}G_c^{\f}\,,\qquad \S_c^\psi=-2 \pa_{\bar{z}}\S_c^G\ .
\eal
With these Fourier transformed quantities it is straightforward to solve \eqref{modpsi},  \eqref{modG1} and the solution is
\bal
\delta^\f=-\frac{q}{(q-1)(q+1)^2}\,,\qquad 
n^\l =   -\frac{q^2}{q^2-1}-q n^\f     \,, \qquad \delta^\l= \frac{q^2}{(q-1)(q+1)^2} \ .  
\eal

Notice that the form of the correction is compatible with the infinitesimal form of the correction to the conformal 2-point function
\bal
G_\e(z_1,z_2)&=\frac{b(1+\e n)}{(z_1-z_2)^{2h-\e \del} (\bar{z}_1-\bar{z}_2)^{2\tilde{h}-\e \bar{\del}} }\\
&=\frac{b}{(z_1-z_2)^{2h } (\bar{z}_1-\bar{z}_2)^{2\tilde{h}} }(1+\e (n+   \del \log(|z_1-z_2|^2))+\mathcal{O}(\e^2))\ .
\eal
where we have taken $\bar{\del}^I=\del^I$ to retain locality. Therefore, we can rewrite the above perturbative result into an equivalent form
\bal
G^\f_\e(z_1,z_2)&=\frac{b^\f(1+\e n^\f)}{(z_1-z_2)^{2h^\f-\e \del^\f} (\bar{z}_1-\bar{z}_2)^{2\tilde{h}^\f-\e  {\del^\f} } }\,,\quad G^\psi_\e(z_1,z_2)=-2\pa_{z_1} G^\f_\e(z_1,z_2)\\
G^\l_\e(z_1,z_2)&=\frac{b^\l(1+\e n^\l)}{(z_1-z_2)^{2h^\l-\e \del^\l} (\bar{z}_1-\bar{z}_2)^{2\tilde{h}^\l-\e  {\del^\l} } }\,,\quad G^G_\e(z_1,z_2)=-2\pa_{z_1} G^\l_\e(z_1,z_2)\ .
\eal
This is compatible with the solution \eqref{02sol1}, \eqref{02sol2} when expanded around $\m=1$.

\section{Fermionic operators in the model}

In this appendix we compute some other 4-point functions in which fermionic operators propagate. From this computation we get the fermionic spectrum of the model \eqref{Sintsusy} and observe the emergence of fermionic higher-spin operators in the two limits \eqref{freechiral} and \eqref{freefermi}. 

\subsection{The $\bar\f\psi$ and $\f\bar\psi$ sector}

\begin{figure}[t!]
	\centering
	\begin{subfigure}[t]{0.5\textwidth}
		\centering
		\includegraphics[width=0.9\linewidth]{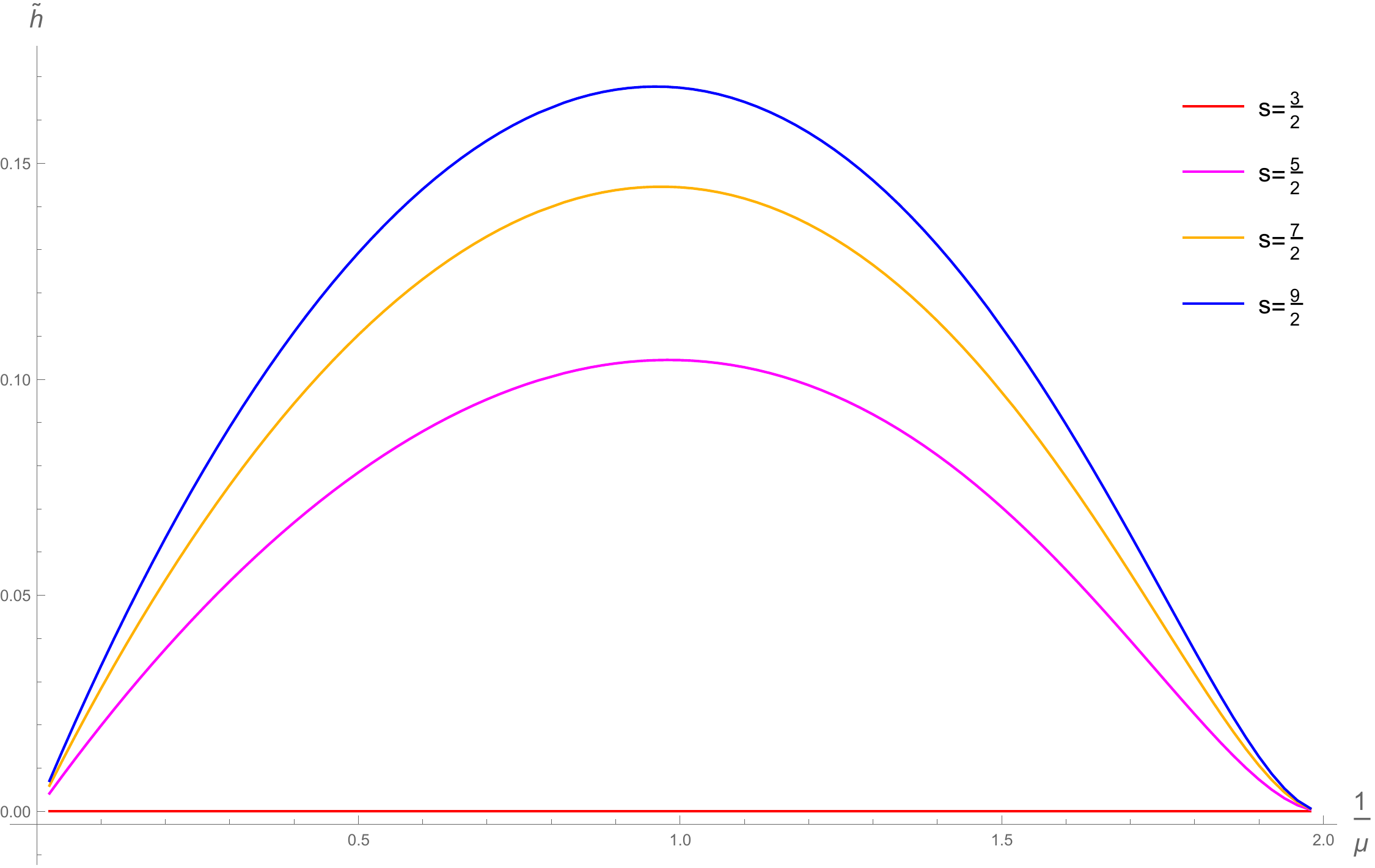}
		\caption{The $\bar\f\psi$ channel at $q=2$. We plot the anomalous dimensions of the higher-spin operators with half-integer spin.  }\label{fig:fermifpsi}
		
	\end{subfigure}%
	~ 
	\begin{subfigure}[t]{0.5\textwidth}
		\centering
		\includegraphics[width=0.9\linewidth]{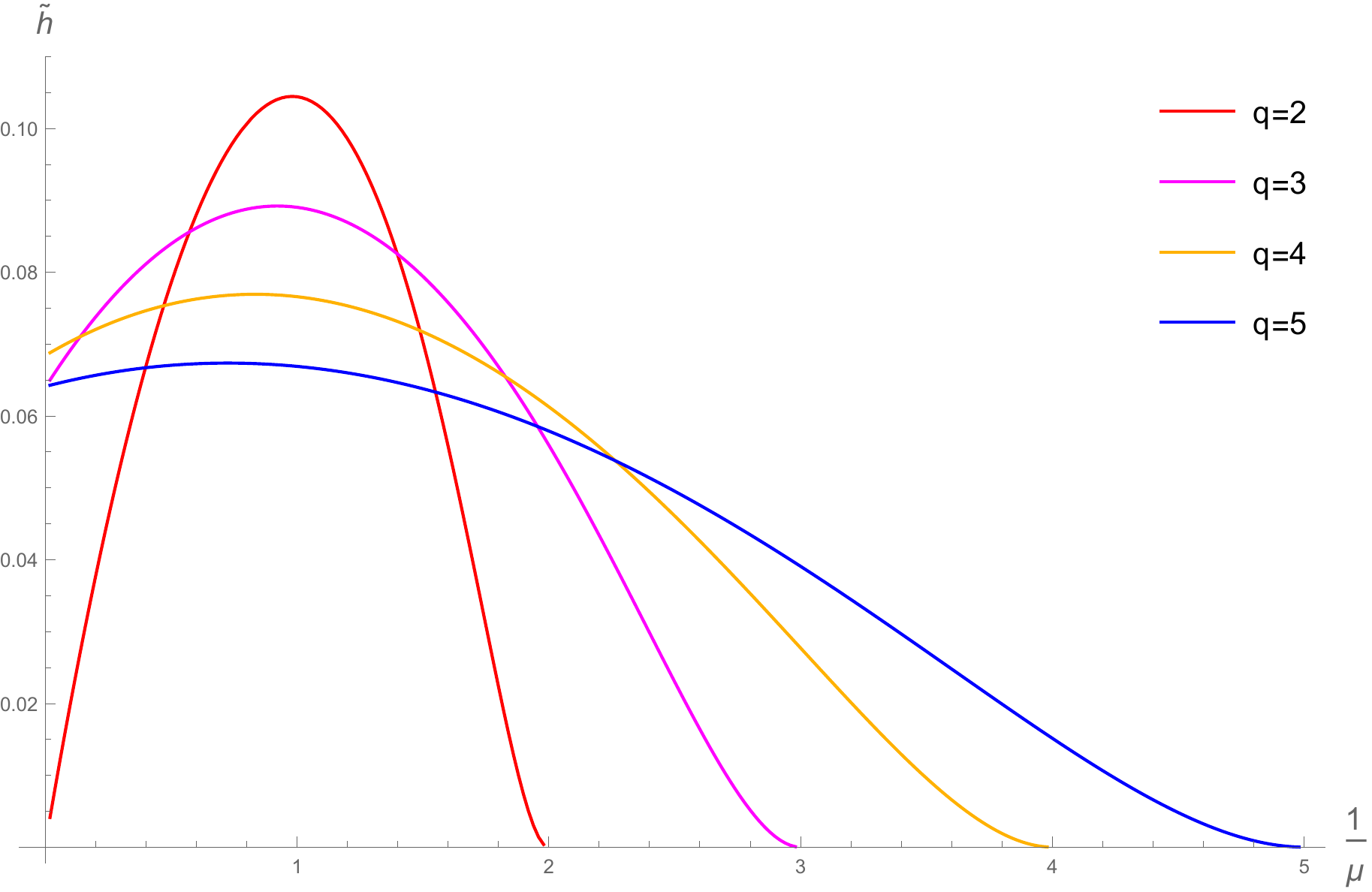}
		\caption{The $\bar\f\psi$ channel at general $q$. The curves represent the dimensions of the spin-3 operators in models with different $q$. }\label{fig:fermifpsi2}
	\end{subfigure}
	\caption{The lowest dimensions of the fermionic operators in the $\<\f^i\bar{\psi}^i{\psi}^j\bar{\f}^j\>$ and   $\<\f^i\bar{\psi}^i{G}^j\bar{\l}^j\>$ four point function. The plots illustrate how does the dimension change as a function of $\m$. }
	\label{fig:fermifpsiall}
\end{figure}

We first consider the $\<\f^i\bar{\psi}^i{\psi}^j\bar{\f}^j\>$ and   $\<\f^i\bar{\psi}^i{G}^j\bar{\l}^j\>$ correlators. There are 6 contributing kernels 
\bal
&K^{\bar\f\psi\f\bar\psi}(z_1,z_2,z_3,z_4)=-2(q-1)  J^2  \frac{M}{N} G^\f(z_{13}) G^\psi(z_{24}) (G^\f(z_{34}) )^{q-2} G^\l(z_{34})\\
&K^{\psi\bar\f\bar\psi\f}(z_1,z_2,z_3,z_4)=2(q-1)  J^2  \frac{M}{N} G^\psi(z_{13}) G^\f(z_{24})  (G^\f(z_{34}) )^{q-2} G^\l(z_{34})\\
&K^{\bar\f\psi\l\bar{G}}(z_1,z_2,z_3,z_4)= 2   J^2  \frac{M}{N} G^\f(z_{13}) G^\psi(z_{24}) (G^\f(z_{34}) )^{q-1} \\
&K^{\psi\bar\f\bar{G}\l}(z_1,z_2,z_3,z_4)=2   J^2  \frac{M}{N} G^\psi(z_{13}) G^\f(z_{24})  (G^\f(z_{34}) )^{q-1} \\
&K^{\bar\l G\f\bar{\psi}}(z_1,z_2,z_3,z_4)= 2   J^2    G^\l(z_{13}) G^G(z_{24}) (G^\f(z_{34}) )^{q-1} \\
&K^{G \bar\l\bar{\psi}\f}(z_1,z_2,z_3,z_4)=2   J^2    G^G(z_{13}) G^\l(z_{24})  (G^\f(z_{34}) )^{q-1} \ .
\eal
The functions that diagonalize the above kernels are of the form
\bal
\F^{ij}(z_1,z_2)=(z_{12})^{h-h_i-h_j}(\bar{z}_{12})^{\bar{h}-\bar{h}_i-\bar{h}_j}\,,
\eal
which satisfy
\bal
K^{ijkl}*\F^{\bar{i}\bar{j}} =k^{ijkl} \F^{ij} \ .
\eal
One can explicitly evaluate the integral and get
\bal
k^{\bar\f\psi\f\bar\psi}&=\frac{\mu  (q-1)^2 q (\mu  q-1) \Gamma \left(\frac{(q-1) q \mu }{q^2 \mu -1}\right)^2 \Gamma \left(\frac{\mu  q^2+2 \mu  q+h \left(2-2 q^2 \mu \right)-3}{2 q^2 \mu -2}\right) \Gamma \left(\tilde{h}-\frac{(q-1) q \mu }{q^2 \mu -1}\right)}{\left(\mu  q^2-1\right)^2 \Gamma \left(\frac{q \mu -1}{q^2 \mu -1}\right) \Gamma \left(\frac{\mu  q^2+\mu  q-2}{q^2 \mu -1}\right) \Gamma \left(\frac{2 h \mu  q^2-3 \mu  q^2+2 \mu  q-2 h+1}{2-2 q^2 \mu }\right) \Gamma \left(\tilde{h}+\frac{(q-1) q \mu }{q^2 \mu -1}\right)}\\
k^{\bar\f\psi\l\bar{G}}&=-\frac{4 \pi ^2 J^2 \mu  n_{\f }^{q+1} (\mu  q-1) \Gamma \left(\frac{(q-1) q \mu }{q^2 \mu -1}\right)^2 \Gamma \left(\frac{\mu  q^2+2 \mu  q+h \left(2-2 q^2 \mu \right)-3}{2 q^2 \mu -2}\right) \Gamma \left(\tilde{h}-\frac{(q-1) q \mu }{q^2 \mu -1}\right)}{\left(\mu  q^2-1\right) \Gamma \left(\frac{q \mu -1}{q^2 \mu -1}\right) \Gamma \left(\frac{\mu  q^2+\mu  q-2}{q^2 \mu -1}\right) \Gamma \left(\frac{2 h \mu  q^2-3 \mu  q^2+2 \mu  q-2 h+1}{2-2 q^2 \mu }\right) \Gamma \left(\tilde{h}+\frac{(q-1) q \mu }{q^2 \mu -1}\right)}\\
k^{\bar\l G\f\bar{\psi}}&=\frac{(q-1)^3 q^2 n_{\f }^{-q-1} \Gamma \left(\frac{1-q}{q^2 \mu -1}\right)^2 \Gamma \left(\frac{\mu  q^2+2 q+h \left(2-2 q^2 \mu \right)-3}{2 q^2 \mu -2}\right) \Gamma \left(\frac{q+\tilde{h} \left(q^2 \mu -1\right)-1}{q^2 \mu -1}\right)}{4 \pi ^2 J^2 \left(\mu  q^2-1\right)^3 \Gamma \left(\frac{q-1}{q^2 \mu -1}\right) \Gamma \left(\frac{\mu  q^2+q-2}{q^2 \mu -1}\right) \Gamma \left(\frac{2 h \mu  q^2-3 \mu  q^2+2 q-2 h+1}{2-2 q^2 \mu }\right) \Gamma \left(\frac{-q+\tilde{h} \left(q^2 \mu -1\right)+1}{q^2 \mu -1}\right)}\\
k^{\psi\bar\f\bar\psi\f}&=\frac{\mu  (q-1)^2 q (\mu  q-1) \Gamma \left(\frac{(q-1) q \mu }{q^2 \mu -1}\right)^2 \Gamma \left(\frac{\mu  q^2+2 \mu  q+h \left(2-2 q^2 \mu \right)-3}{2 q^2 \mu -2}\right) \Gamma \left(\tilde{h}-\frac{(q-1) q \mu }{q^2 \mu -1}\right)}{\left(\mu  q^2-1\right)^2 \Gamma \left(\frac{q \mu -1}{q^2 \mu -1}\right) \Gamma \left(\frac{\mu  q^2+\mu  q-2}{q^2 \mu -1}\right) \Gamma \left(\frac{2 h \mu  q^2-3 \mu  q^2+2 \mu  q-2 h+1}{2-2 q^2 \mu }\right) \Gamma \left(\tilde{h}+\frac{(q-1) q \mu }{q^2 \mu -1}\right)}\\
k^{\psi\bar\f\bar{G}\l}&=\frac{4 \pi ^2 J^2 \mu  n_{\f }^{q+1} (\mu  q-1) \Gamma \left(\frac{(q-1) q \mu }{q^2 \mu -1}\right)^2 \Gamma \left(\frac{\mu  q^2+2 \mu  q+h \left(2-2 q^2 \mu \right)-3}{2 q^2 \mu -2}\right) \Gamma \left(\tilde{h}-\frac{(q-1) q \mu }{q^2 \mu -1}\right)}{\left(\mu  q^2-1\right) \Gamma \left(\frac{q \mu -1}{q^2 \mu -1}\right) \Gamma \left(\frac{\mu  q^2+\mu  q-2}{q^2 \mu -1}\right) \Gamma \left(\frac{2 h \mu  q^2-3 \mu  q^2+2 \mu  q-2 h+1}{2-2 q^2 \mu }\right) \Gamma \left(\tilde{h}+\frac{(q-1) q \mu }{q^2 \mu -1}\right)}\\
k^{G \bar\l\bar{\psi}\f}&=-\frac{(q-1)^3 q^2 n_{\f }^{-q-1} \Gamma \left(\frac{1-q}{q^2 \mu -1}\right)^2 \Gamma \left(\frac{\mu  q^2+2 q+h \left(2-2 q^2 \mu \right)-3}{2 q^2 \mu -2}\right) \Gamma \left(\frac{q+\tilde{h} \left(q^2 \mu -1\right)-1}{q^2 \mu -1}\right)}{4 \pi ^2 J^2 \left(\mu  q^2-1\right)^3 \Gamma \left(\frac{q-1}{q^2 \mu -1}\right) \Gamma \left(\frac{\mu  q^2+q-2}{q^2 \mu -1}\right) \Gamma \left(\frac{2 h \mu  q^2-3 \mu  q^2+2 q-2 h+1}{2-2 q^2 \mu }\right) \Gamma \left(\frac{-q+\tilde{h} \left(q^2 \mu -1\right)+1}{q^2 \mu -1}\right)}
\eal
The final eigenvalues are obtained from diagonalizing the following matrix
\bal
\begin{pmatrix}
	0 & 0 & k^{\bar\f\psi\f\bar\psi} & k^{\bar\f\psi\l\bar{G}} \\
	0 & 0 & k^{\bar\l G\f\bar{\psi}} & 0\\
	k^{\psi\bar\f\bar\psi\f} & k^{\psi\bar\f\bar{G}\l} & 0 & 0\\
	k^{G \bar\l\bar{\psi}\f} & 0 & 0 & 0
\end{pmatrix}
\eal
The expressions of the eigenvalues are not very eluminating. Here we only give the dimensions of the higher-spin operators that is of most interest to us. This is obtained in a similar manner as what we did to find the behaviors of the bosonic higher-spin operators. Namely we find the lowest dimensions, as a function of $\m$, that set some eigenvalue to 1 at a given spin, then we check the behavior of the dimensions as $\m$ approaches the two limits \eqref{freechiral} and \eqref{freefermi}. Instead of give the explicit expression of the dimensions of the operators, we simply present the results in figure~\ref{fig:fermifpsiall}. From the plot, we indeed observe that there is a tower of half-interger spin operators become conserved fermionic higher-spin operators in the limit \eqref{freechiral} at any $q$. One the other hand, we only observe the emergence of the tower of conserved operators in the limit \eqref{freefermi} at $q=2$. These behavior are identical to those observed in the bosonic operators, as we expected due to supersymmetry. 
In adddition, the right-moving supercharges is present at any value of $\m$, in accord with the fact that the model has $\cn=(0,2)$ supersymmetry at any $\m$.   We call the tower of higher-spin operator running in these 4-point function in the $\bar\f\psi$ sector since they are expect to have the form
\bal
\co^{s+\frac{1}{2}}_{\bar\f\psi}&\sim\bar\f^i\pa^s\psi^i+\ldots+\bar\l\pa^{s+1} G+\ldots\,,
\eal
where ``$\ldots$" represents other terms with different distributions of the derivatives on the two fields; the correct combination is determined by requiring the operator to be a primary field.

There is a set of conjugate kernels that gives identical eigenvalues.  Hence there is a second tower of fermionic higher-spin operators appearing in the two limits \eqref{freechiral} and \eqref{freefermi} with identical behaviors as the parameters change. They are referred as in the $\f\bar\psi$ sector. They complete the right-moving $\cn=2$ multiplets at each spin.

\subsection{The $\bar\f\l$ and $\f\bar\l$ sector}
\begin{figure}[t!]
	\centering
	\begin{subfigure}[t]{0.5\textwidth}
		\centering
		\includegraphics[width=0.9\linewidth]{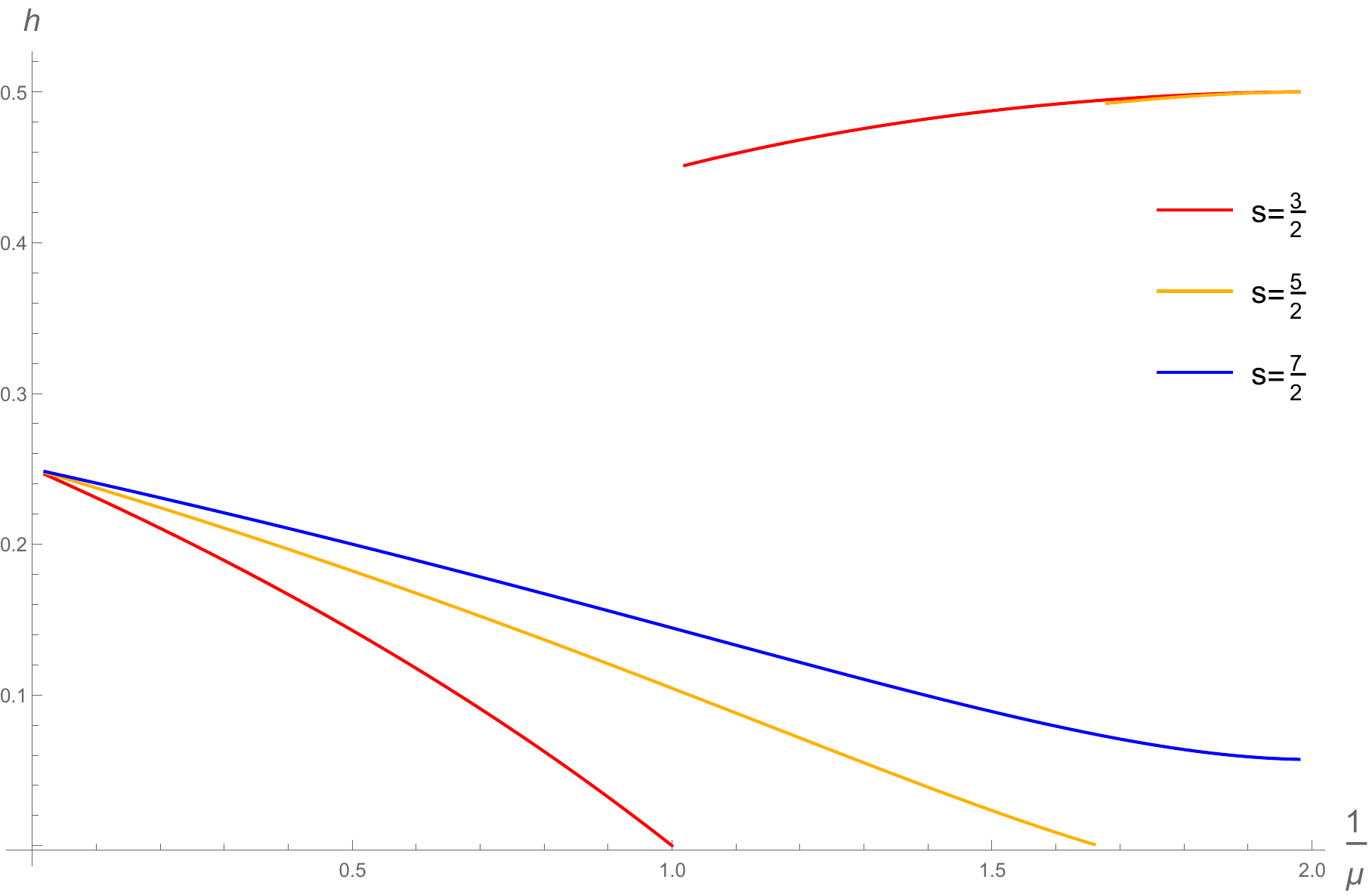}
		\caption{The operators with the lowest dimension in the $\bar\f\l$ sector at $q=2$. There is no higher-spin conserved operators in this sector. }\label{fig:fermifl}
		
	\end{subfigure}%
	~ 
	\begin{subfigure}[t]{0.5\textwidth}
		\centering
		\includegraphics[width=0.9\linewidth]{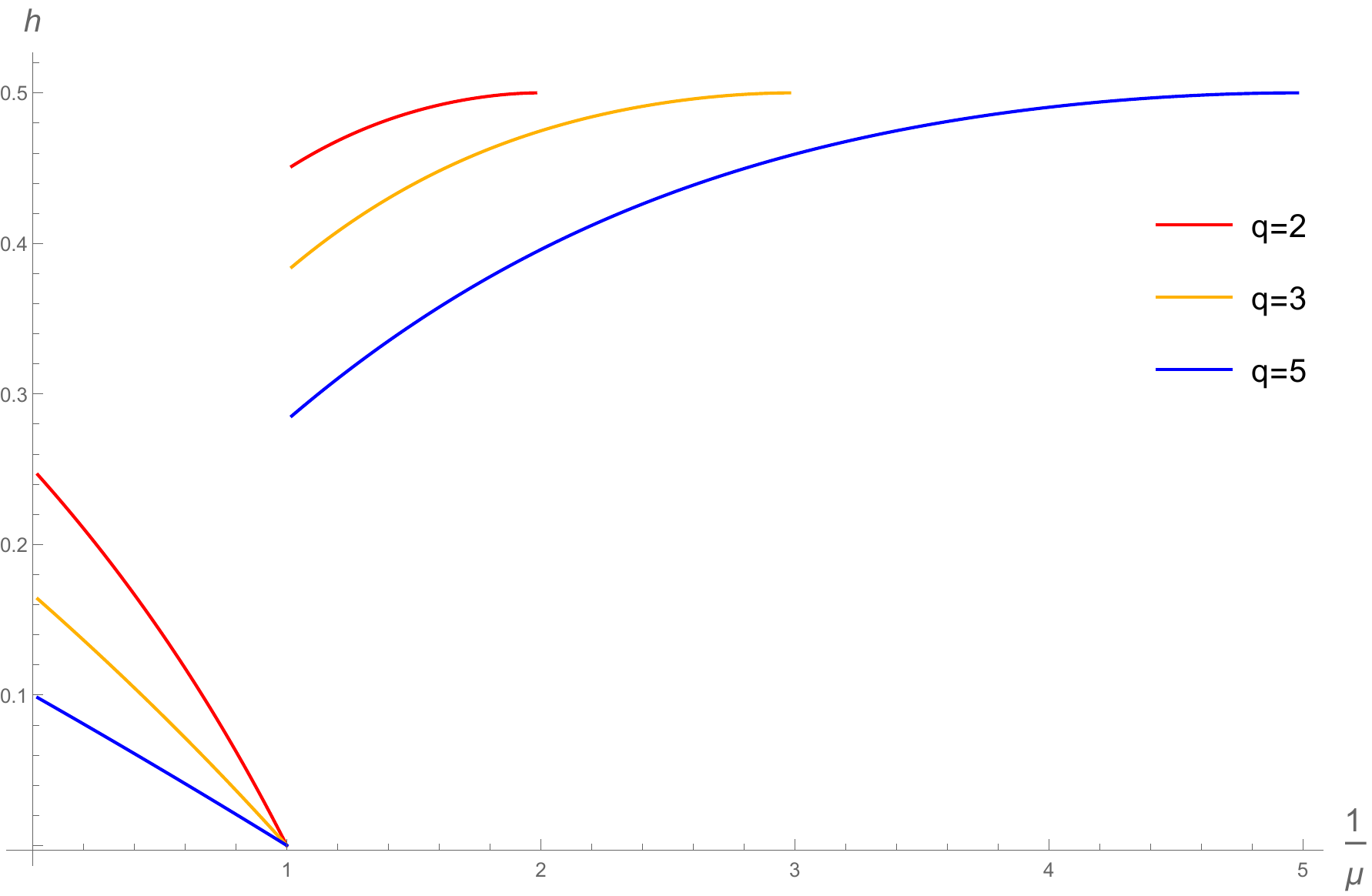}
		\caption{The $\bar\f\l$ channel at general $q$. The curves represent the dimensions of the spin-3 operators in models with different $q$. }\label{fig:fermifl2}
	\end{subfigure}
	\caption{The lowest dimensions of the fermionic operators in the $\<\f^i\bar{\l}^i{\l}^j\bar{\f}^j\>$ and   $\<\f^i\bar{\l}^i{G}^j\bar{\psi}^j\>$ four point functions. The plots illustrate how does the dimension change as a function of $\m$. The jumps of the curves are due to the fact that as $\m$ becomes smaller than 1, the branch of operators with the lowest dimension when $\m\geq 1$ ceases to exist so the operators with the second lowest dimensions become the ``lowest" one at $\m<1$. }
	\label{fig:fermiflamall}
\end{figure}

We can also consider the $\<\f^i\bar{\l}^i{\l}^j\bar{\f}^j\>$ and   $\<\f^i\bar{\l}^i{G}^j\bar{\psi}^j\>$ correlators. There are 6 contributing kernels 
\bal
&K^{\bar\f\l\f\bar\l}(z_1,z_2,z_3,z_4)=-2(q-1)  J^2   G^\f(z_{13}) G^\l(z_{24}) (G^\f(z_{34}) )^{q-2} G^\psi(z_{34})\\
&K^{\l\bar\f\bar\l\f}(z_1,z_2,z_3,z_4)=2(q-1)  J^2   G^\l(z_{13}) G^\f(z_{24})  (G^\f(z_{34}) )^{q-2} G^\psi(z_{34})\\
&K^{\bar\f\l\psi\bar{G}}(z_1,z_2,z_3,z_4)= 2   J^2   G^\f(z_{13}) G^\l(z_{24}) (G^\f(z_{34}) )^{q-1} \\
&K^{\l\bar\f\bar{G}\psi}(z_1,z_2,z_3,z_4)=2   J^2   G^\l(z_{13}) G^\f(z_{24})  (G^\f(z_{34}) )^{q-1} \\
&K^{\bar\psi G\f\bar{\l}}(z_1,z_2,z_3,z_4)= 2   J^2    G^\psi(z_{13}) G^G(z_{24}) (G^\f(z_{34}) )^{q-1} \\
&K^{G \bar\psi\bar{\l}\f}(z_1,z_2,z_3,z_4)=2   J^2    G^G(z_{13}) G^\psi(z_{24})  (G^\f(z_{34}) )^{q-1} \ .
\eal
With a similar set of eigenfunctions, the above kernels act as a multiplication of the following
\bal
k^{\bar\f\l\f\bar\l}&=\frac{(q-1)^2 q (\mu  q-1) \Gamma \left(\frac{1-q}{q^2 \mu -1}\right) \Gamma \left(\frac{(q-1) q \mu }{q^2 \mu -1}\right) \Gamma \left(\frac{\mu  q+q+h \left(2-2 q^2 \mu \right)-2}{2 q^2 \mu -2}\right) \Gamma \left(\frac{2 \tilde{h} \left(q^2 \mu -1\right)-(q-1) (q \mu -1)}{2 q^2 \mu -2}\right)}{\left(\mu  q^2-1\right)^2 \Gamma \left(\frac{q-1}{q^2 \mu -1}\right) \Gamma \left(\frac{q \mu -1}{q^2 \mu -1}\right) \Gamma \left(\frac{2 h \mu  q^2-4 \mu  q^2+\mu  q+q-2 h+2}{2-2 q^2 \mu }\right) \Gamma \left(\frac{(q-1) (q \mu -1)+2 \tilde{h} \left(q^2 \mu -1\right)}{2 q^2 \mu -2}\right)}\\
k^{\l\bar\f\bar\l\f}&=\frac{(q-1)^2 q (\mu  q-1) \Gamma \left(\frac{1-q}{q^2 \mu -1}\right) \Gamma \left(\frac{(q-1) q \mu }{q^2 \mu -1}\right) \Gamma \left(\frac{\mu  q+q+h \left(2-2 q^2 \mu \right)-2}{2 q^2 \mu -2}\right) \Gamma \left(\frac{2 \tilde{h} \left(q^2 \mu -1\right)-(q-1) (q \mu -1)}{2 q^2 \mu -2}\right)}{\left(\mu  q^2-1\right)^2 \Gamma \left(\frac{q-1}{q^2 \mu -1}\right) \Gamma \left(\frac{q \mu -1}{q^2 \mu -1}\right) \Gamma \left(\frac{2 h \mu  q^2-4 \mu  q^2+\mu  q+q-2 h+2}{2-2 q^2 \mu }\right) \Gamma \left(\frac{(q-1) (q \mu -1)+2 \tilde{h} \left(q^2 \mu -1\right)}{2 q^2 \mu -2}\right)}\\
k^{\bar\f\l\psi\bar{G}}&=-\frac{(q-1) q \Gamma \left(\frac{1-q}{q^2 \mu -1}\right) \Gamma \left(\frac{(q-1) q \mu }{q^2 \mu -1}\right) \Gamma \left(\frac{\mu  q+q+h \left(2-2 q^2 \mu \right)-2}{2 q^2 \mu -2}\right) \Gamma \left(\frac{2 \tilde{h} \left(q^2 \mu -1\right)-(q-1) (q \mu -1)}{2 q^2 \mu -2}\right)}{2 \left(\mu  q^2-1\right) \Gamma \left(\frac{q-1}{q^2 \mu -1}\right) \Gamma \left(\frac{q \mu -1}{q^2 \mu -1}\right) \Gamma \left(\frac{2 h \mu  q^2-4 \mu  q^2+\mu  q+q-2 h+2}{2-2 q^2 \mu }\right) \Gamma \left(\frac{(q-1) (q \mu -1)+2 \tilde{h} \left(q^2 \mu -1\right)}{2 q^2 \mu -2}\right)}\\
k^{\l\bar\f\bar{G}\psi}&=\frac{(q-1) q \Gamma \left(\frac{1-q}{q^2 \mu -1}\right) \Gamma \left(\frac{(q-1) q \mu }{q^2 \mu -1}\right) \Gamma \left(\frac{\mu  q+q+h \left(2-2 q^2 \mu \right)-2}{2 q^2 \mu -2}\right) \Gamma \left(\frac{2 \tilde{h} \left(q^2 \mu -1\right)-(q-1) (q \mu -1)}{2 q^2 \mu -2}\right)}{2 \left(\mu  q^2-1\right) \Gamma \left(\frac{q-1}{q^2 \mu -1}\right) \Gamma \left(\frac{q \mu -1}{q^2 \mu -1}\right) \Gamma \left(\frac{2 h \mu  q^2-4 \mu  q^2+\mu  q+q-2 h+2}{2-2 q^2 \mu }\right) \Gamma \left(\frac{(q-1) (q \mu -1)+2 \tilde{h} \left(q^2 \mu -1\right)}{2 q^2 \mu -2}\right)}\\
k^{\bar\psi G\f\bar{\l}}&=\frac{2 (q-1)^2 q (\mu  q-1) \Gamma \left(\frac{1-q}{q^2 \mu -1}\right) \Gamma \left(\frac{(q-1) q \mu }{q^2 \mu -1}\right) \Gamma \left(\frac{2 \mu  q^2+\mu  q+q+h \left(2-2 q^2 \mu \right)-4}{2 q^2 \mu -2}\right) \Gamma \left(\frac{2 \tilde{h} \left(q^2 \mu -1\right)-(q-1) (q \mu -1)}{2 q^2 \mu -2}\right)}{\left(\mu  q^2-1\right)^3 \Gamma \left(\frac{\mu  q^2+q-2}{q^2 \mu -1}\right) \Gamma \left(\frac{\mu  q^2+\mu  q-2}{q^2 \mu -1}\right) \Gamma \left(\frac{q ((2 q-1) \mu -1)+h \left(2-2 q^2 \mu \right)}{2 q^2 \mu -2}\right) \Gamma \left(\frac{(q-1) (q \mu -1)+2 \tilde{h} \left(q^2 \mu -1\right)}{2 q^2 \mu -2}\right)}\\
k^{G \bar\psi\bar{\l}\f}&=-\frac{2 (q-1)^2 q (\mu  q-1) \Gamma \left(\frac{1-q}{q^2 \mu -1}\right) \Gamma \left(\frac{(q-1) q \mu }{q^2 \mu -1}\right) \Gamma \left(\frac{2 \mu  q^2+\mu  q+q+h \left(2-2 q^2 \mu \right)-4}{2 q^2 \mu -2}\right) \Gamma \left(\frac{2 \tilde{h} \left(q^2 \mu -1\right)-(q-1) (q \mu -1)}{2 q^2 \mu -2}\right)}{\left(\mu  q^2-1\right)^3 \Gamma \left(\frac{\mu  q^2+q-2}{q^2 \mu -1}\right) \Gamma \left(\frac{\mu  q^2+\mu  q-2}{q^2 \mu -1}\right) \Gamma \left(\frac{q ((2 q-1) \mu -1)+h \left(2-2 q^2 \mu \right)}{2 q^2 \mu -2}\right) \Gamma \left(\frac{(q-1) (q \mu -1)+2 \tilde{h} \left(q^2 \mu -1\right)}{2 q^2 \mu -2}\right)}
\eal
The final eigenvalues are obtained from diagonalizing the following matrix
\bal
\begin{pmatrix}
	0 & 0 & k^{\bar\f\l\f\bar\l} & k^{\bar\f\l\psi\bar{G}} \\
	0 & 0 & k^{\bar\psi G\f\bar{\l}} & 0\\
	k^{\l\bar\f\bar\l\f} & k^{\l\bar\f\bar{G}\psi} & 0 & 0\\
	k^{G \bar\psi\bar{\l}\f} & 0 & 0 & 0
\end{pmatrix}
\eal
Once more we omit the analytic expression of the eigenvalues but only give the dimensions of the  operators that is of most interest to us. This is obtained in a similar manner as above and the result is present in figure~\ref{fig:fermiflamall}. From the plot, we indeed observe that there is no half-integer higher-spin operators in the limit \eqref{freechiral} at any $q$. These behavior is compatible with the results from the bosonic operator spectrum since we do not expect conserved anti-holomorphic fermionic operators due to the absence of supersymmetry in the left-moving sector.  
The only conserved operator appears at $\m=1$, $s=\frac{3}{2}$, which correspond to the left-moving supercharges emerging as the model develops an enhanced $\cn=(2,2)$ supersymmetry at $\m=1$, as we expected.


\begin{thebibliography}{99}

\bibitem{Sachdev:1992fk}
S.~Sachdev and J.~Ye\emph{},
\href{https://doi.org/10.1103/PhysRevLett.70.3339}{\emph{Phys. Rev. Lett.}
	{\bfseries 70} (1993) 3339},
[\href{https://arxiv.org/abs/cond-mat/9212030}{{\ttfamily
		cond-mat/9212030}}].

\bibitem{Parcollet:1997ysb}
O.~Parcollet, A.~Georges, G.~Kotliar and A.~Sengupta\emph{},
\href{https://doi.org/10.1103/PhysRevB.58.3794}{\emph{Phys. Rev.} {\bfseries
		B58} (1998) 3794}, 

\bibitem{PG}
O.~{Parcollet} and A.~{Georges}\emph{},
\href{https://doi.org/10.1103/PhysRevB.59.5341}{\emph{Phys. Rev.} {\bfseries
		B59} (1999) 5341}, [\href{https://arxiv.org/abs/cond-mat/9806119}{{\ttfamily
		cond-mat/9806119}}].

\bibitem{KitaevTalk1}
A.~{Kitaev}\emph{}, {{ ``Hidden correlations in the Hawking
		radiation and thermal noise"} }.

\bibitem{KitaevTalk2}
A.~{Kitaev}\emph{}, {{``A simple
		model of quantum holography"} }.

\bibitem{Maldacena:2016hyu}
J.~Maldacena and D.~Stanford\emph{},
\href{https://doi.org/10.1103/PhysRevD.94.106002}{\emph{Phys. Rev.}
	{\bfseries D94} (2016) 106002},
[\href{https://arxiv.org/abs/1604.07818}{{\ttfamily 1604.07818}}].

\bibitem{Kitaev:2017awl}
A.~Kitaev and S.~J. Suh\emph{},
\href{https://arxiv.org/abs/1711.08467}{{\ttfamily 1711.08467}}.

\bibitem{Polchinski:2016xgd}
J.~Polchinski and V.~Rosenhaus\emph{},
\href{https://doi.org/10.1007/JHEP04(2016)001}{\emph{JHEP} {\bfseries 04}
	(2016) 001}, [\href{https://arxiv.org/abs/1601.06768}{{\ttfamily
		1601.06768}}].

\bibitem{Jevicki2016}
A.~Jevicki and K.~Suzuki\emph{},
\href{https://doi.org/10.1007/JHEP11(2016)046}{\emph{JHEP} {\bfseries 11}
	(2016) 046}, [\href{https://arxiv.org/abs/1608.07567}{{\ttfamily
		1608.07567}}].

\bibitem{Gross:2017hcz}
D.~J. Gross and V.~Rosenhaus\emph{},
\href{https://doi.org/10.1007/JHEP05(2017)092}{\emph{JHEP} {\bfseries 05}
	(2017) 092}, [\href{https://arxiv.org/abs/1702.08016}{{\ttfamily
		1702.08016}}].

\bibitem{Gross:2017aos}
D.~J. Gross and V.~Rosenhaus\emph{},
\href{https://arxiv.org/abs/1710.08113}{{\ttfamily 1710.08113}}.

\bibitem{Bagrets:2016cdf}
D.~Bagrets, A.~Altland and A.~Kamenev\emph{},
\href{https://doi.org/10.1016/j.nuclphysb.2016.08.002}{\emph{Nucl. Phys.}
	{\bfseries B911} (2016) 191--205}.

\bibitem{Stanford:2017thb}
D.~Stanford and E.~Witten\emph{},
\href{https://doi.org/10.1007/JHEP10(2017)008}{\emph{JHEP} {\bfseries 10}
	(2017) 008}, [\href{https://arxiv.org/abs/1703.04612}{{\ttfamily
		1703.04612}}].

\bibitem{Mertens:2017mtv}
T.~G. Mertens, G.~J. Turiaci and H.~L. Verlinde\emph{},
\href{https://doi.org/10.1007/JHEP08(2017)136}{\emph{JHEP} {\bfseries 08}
	(2017) 136}, [\href{https://arxiv.org/abs/1705.08408}{{\ttfamily
		1705.08408}}].

\bibitem{Jevicki:2016ito}
A.~Jevicki and K.~Suzuki\emph{},
\href{https://doi.org/10.1007/JHEP11(2016)046}{\emph{JHEP} {\bfseries 11}
	(2016) 046}, [\href{https://arxiv.org/abs/1608.07567}{{\ttfamily
		1608.07567}}].

\bibitem{Strominger:1998yg}
A.~Strominger\emph{},
\href{https://doi.org/10.1088/1126-6708/1999/01/007}{\emph{JHEP} {\bfseries
		01} (1999) 007}, [\href{https://arxiv.org/abs/hep-th/9809027}{{\ttfamily
		hep-th/9809027}}].

\bibitem{Maldacena:1998uz}
J.~M. Maldacena, J.~Michelson and A.~Strominger\emph{},
\href{https://doi.org/10.1088/1126-6708/1999/02/011}{\emph{JHEP} {\bfseries
		02} (1999) 011}. 

\bibitem{Almheiri:2014cka}
A.~Almheiri and J.~Polchinski\emph{},
\href{https://doi.org/10.1007/JHEP11(2015)014}{\emph{JHEP} {\bfseries 11}
	(2015) 014}, [\href{https://arxiv.org/abs/1402.6334}{{\ttfamily 1402.6334}}].

\bibitem{Maldacena:2016upp}
J.~Maldacena, D.~Stanford and Z.~Yang\emph{},
\href{https://doi.org/10.1093/ptep/ptw124}{\emph{PTEP} {\bfseries 2016}
	(2016) 12C104}, [\href{https://arxiv.org/abs/1606.01857}{{\ttfamily
		1606.01857}}].

\bibitem{Engelsoy:2016xyb}
J.~Engelsoy, T.~G. Mertens and H.~Verlinde\emph{},
\href{https://doi.org/10.1007/JHEP07(2016)139}{\emph{JHEP} {\bfseries 07}
	(2016) 139}, [\href{https://arxiv.org/abs/1606.03438}{{\ttfamily
		1606.03438}}].

\bibitem{Cvetic:2016eiv}
M.~Cvetic and I.~Papadimitriou\emph{},
\href{https://doi.org/10.1007/JHEP12(2016)008,
	10.1007/JHEP01(2017)120}{\emph{JHEP} {\bfseries 12} (2016) 008},
[\href{https://arxiv.org/abs/1608.07018}{{\ttfamily 1608.07018}}].

\bibitem{Grumiller:2017qao}
D.~Grumiller, R.~McNees, J.~Salzer, C.~Valc��rcel and
D.~Vassilevich\emph{},
\href{https://doi.org/10.1007/JHEP10(2017)203}{\emph{JHEP} {\bfseries 10}
	(2017) 203}, [\href{https://arxiv.org/abs/1708.08471}{{\ttfamily
		1708.08471}}].

\bibitem{Maldacena:2018lmt}
J.~Maldacena and X.-L. Qi\emph{},
\href{https://arxiv.org/abs/1804.00491}{{\ttfamily 1804.00491}}.

\bibitem{Bagrets:2017pwq}
D.~Bagrets, A.~Altland and A.~Kamenev\emph{},
\href{https://doi.org/10.1016/j.nuclphysb.2017.06.012}{\emph{Nucl. Phys.}
	{\bfseries B921} (2017) 727--752}.

\bibitem{Shenker:2013pqa}
S.~H. Shenker and D.~Stanford\emph{},
\href{https://doi.org/10.1007/JHEP03(2014)067}{\emph{JHEP} {\bfseries 03}
	(2014) 067}, [\href{https://arxiv.org/abs/1306.0622}{{\ttfamily 1306.0622}}].

\bibitem{Shenker:2014cwa}
S.~H. Shenker and D.~Stanford\emph{},
\href{https://doi.org/10.1007/JHEP05(2015)132}{\emph{JHEP} {\bfseries 05}
	(2015) 132}, [\href{https://arxiv.org/abs/1412.6087}{{\ttfamily 1412.6087}}].

\bibitem{Maldacena:2015waa}
J.~Maldacena, S.~H. Shenker and D.~Stanford\emph{},
\href{https://doi.org/10.1007/JHEP08(2016)106}{\emph{JHEP} {\bfseries 08}
	(2016) 106}, [\href{https://arxiv.org/abs/1503.01409}{{\ttfamily
		1503.01409}}].

\bibitem{Jensen:2016pah}
K.~Jensen\emph{},
\href{https://doi.org/10.1103/PhysRevLett.117.111601}{\emph{Phys. Rev. Lett.}
	{\bfseries 117} (2016) 111601},
[\href{https://arxiv.org/abs/1605.06098}{{\ttfamily 1605.06098}}].

\bibitem{Sachdev:2010um}
S.~Sachdev\emph{},
\href{https://doi.org/10.1103/PhysRevLett.105.151602}{\emph{Phys. Rev. Lett.}
	{\bfseries 105} (2010) 151602},
[\href{https://arxiv.org/abs/1006.3794}{{\ttfamily 1006.3794}}].

\bibitem{Sachdev:2015efa}
S.~Sachdev\emph{}, \href{https://doi.org/10.1103/PhysRevX.5.041025}{\emph{Phys.
		Rev.} {\bfseries X5} (2015) 041025},
[\href{https://arxiv.org/abs/1506.05111}{{\ttfamily 1506.05111}}].

\bibitem{Fu:2016yrv}
W.~Fu and S.~Sachdev\emph{},
\href{https://doi.org/10.1103/PhysRevB.94.035135}{\emph{Phys. Rev.}
	{\bfseries B94} (2016) 035135},
[\href{https://arxiv.org/abs/1603.05246}{{\ttfamily 1603.05246}}].

\bibitem{Jevicki:2016bwu}
A.~Jevicki, K.~Suzuki and J.~Yoon\emph{},
\href{https://doi.org/10.1007/JHEP07(2016)007}{\emph{JHEP} {\bfseries 07}
	(2016) 007}, [\href{https://arxiv.org/abs/1603.06246}{{\ttfamily
		1603.06246}}].

\bibitem{Garcia-Alvarez:2016wem}
L.~Garcia-Alvarez, I.~L. Egusquiza, L.~Lamata, A.~del Campo,
J.~Sonner and E.~Solano\emph{},
\href{https://doi.org/10.1103/PhysRevLett.119.040501}{\emph{Phys. Rev. Lett.}
	{\bfseries 119} (2017) 040501},
[\href{https://arxiv.org/abs/1607.08560}{{\ttfamily 1607.08560}}].

\bibitem{Garcia-Garcia:2016mno}
A.~M. Garcia-Garcia and J.~J.~M. Verbaarschot\emph{},
\href{https://doi.org/10.1103/PhysRevD.94.126010}{\emph{Phys. Rev.}
	{\bfseries D94} (2016) 126010},

\bibitem{Cotler:2016fpe}
J.~S. Cotler, G.~Gur-Ari, M.~Hanada, J.~Polchinski, P.~Saad, S.~H. Shenker
et~al.\emph{}, \href{https://doi.org/10.1007/JHEP05(2017)118}{\emph{JHEP}
	{\bfseries 05} (2017) 118},
[\href{https://arxiv.org/abs/1611.04650}{{\ttfamily 1611.04650}}].

\bibitem{Garcia-Garcia:2017pzl}
A.~M. Garcia-Garcia and J.~J.~M. Verbaarschot\emph{},
\href{https://doi.org/10.1103/PhysRevD.96.066012}{\emph{Phys. Rev.}
	{\bfseries D96} (2017) 066012},

\bibitem{Kourkoulou:2017zaj}
I.~Kourkoulou and J.~Maldacena\emph{},
\href{https://arxiv.org/abs/1707.02325}{{\ttfamily 1707.02325}}.

\bibitem{Sonner:2017hxc}
J.~Sonner and M.~Vielma\emph{},
\href{https://doi.org/10.1007/JHEP11(2017)149}{\emph{JHEP} {\bfseries 11}
	(2017) 149}, [\href{https://arxiv.org/abs/1707.08013}{{\ttfamily
		1707.08013}}].

\bibitem{a:2018kvh}
A.~M. García-García, Y.~Jia and J.~J.~M. Verbaarschot\emph{},
\href{https://arxiv.org/abs/1801.02696}{{\ttfamily 1801.02696}}.

\bibitem{Peng:2017kro}
C.~Peng\emph{}, \href{https://doi.org/10.1007/JHEP05(2017)129}{\emph{JHEP}
	{\bfseries 05} (2017) 129},
[\href{https://arxiv.org/abs/1704.04223}{{\ttfamily 1704.04223}}].

\bibitem{Banerjee:2016ncu}
S.~Banerjee and E.~Altman\emph{},
\href{https://doi.org/10.1103/PhysRevB.95.134302}{\emph{Phys. Rev.}
	{\bfseries B95} (2017) 134302},
[\href{https://arxiv.org/abs/1610.04619}{{\ttfamily 1610.04619}}].

\bibitem{Bi:2017yvx}
Z.~Bi, C.-M. Jian, Y.-Z. You, K.~A. Pawlak and C.~Xu\emph{},
\href{https://doi.org/10.1103/PhysRevB.95.205105}{\emph{Phys. Rev.}
	{\bfseries B95} (2017) 205105},

\bibitem{Jian:2017jfl}
C.-M. Jian, Z.~Bi and C.~Xu\emph{},
\href{https://doi.org/10.1103/PhysRevB.96.115122}{\emph{Phys. Rev.}
	{\bfseries B96} (2017) 115122},
[\href{https://arxiv.org/abs/1703.07793}{{\ttfamily 1703.07793}}].

\bibitem{Song:2017pfw}
X.-Y. Song, C.-M. Jian and L.~Balents\emph{},
\href{https://doi.org/10.1103/PhysRevLett.119.216601}{\emph{Phys. Rev. Lett.}
	{\bfseries 119} (2017) 216601},

\bibitem{Luo:2017bno}
Z.~Luo, Y.-Z. You, J.~Li, C.-M. Jian, D.~Lu, C.~Xu et~al.\emph{},
\href{https://arxiv.org/abs/1712.06458}{{\ttfamily 1712.06458}}.

\bibitem{Nosaka:2018iat}
T.~Nosaka, D.~Rosa and J.~Yoon\emph{},
\href{https://arxiv.org/abs/1804.09934}{{\ttfamily 1804.09934}}.

\bibitem{Mondal:2018xwy}
S.~Mondal\emph{},  \href{https://arxiv.org/abs/1801.09669}{{\ttfamily
		1801.09669}}.

\bibitem{Gross:2017vhb}
D.~J. Gross and V.~Rosenhaus\emph{},
\href{https://doi.org/10.1007/JHEP07(2017)086}{\emph{JHEP} {\bfseries 07}
	(2017) 086}, [\href{https://arxiv.org/abs/1706.07015}{{\ttfamily
		1706.07015}}].

\bibitem{Taylor:2017dly}
M.~Taylor\emph{},  \href{https://arxiv.org/abs/1706.07812}{{\ttfamily
		1706.07812}}.

\bibitem{Das:2017hrt}
S.~R. Das, A.~Ghosh, A.~Jevicki and K.~Suzuki\emph{},
\href{https://arxiv.org/abs/1711.09839}{{\ttfamily 1711.09839}}.

\bibitem{Das:2017wae}
S.~R. Das, A.~Ghosh, A.~Jevicki and K.~Suzuki\emph{},
\href{https://arxiv.org/abs/1712.02725}{{\ttfamily 1712.02725}}.

\bibitem{Maldacena:2017axo}
J.~Maldacena, D.~Stanford and Z.~Yang\emph{},
\href{https://doi.org/10.1002/prop.201700034}{\emph{Fortsch. Phys.}
	{\bfseries 65} (2017) 1700034},

\bibitem{Murata:2017rbp}
K.~Murata\emph{}, \href{https://doi.org/10.1007/JHEP11(2017)049}{\emph{JHEP}
	{\bfseries 11} (2017) 049},
[\href{https://arxiv.org/abs/1708.09493}{{\ttfamily 1708.09493}}].

\bibitem{deBoer:2017xdk}
J.~de~Boer, E.~Llabr��s, J.~F. Pedraza and D.~Vegh\emph{},
\href{https://arxiv.org/abs/1709.01052}{{\ttfamily 1709.01052}}.

\bibitem{Cai:2017nwk}
R.-G. Cai, S.-M. Ruan, R.-Q. Yang and Y.-L. Zhang\emph{},
\href{https://arxiv.org/abs/1709.06297}{{\ttfamily 1709.06297}}.

\bibitem{Kitaev:2017hnr}
A.~Kitaev\emph{},  \href{https://arxiv.org/abs/1711.08169}{{\ttfamily
		1711.08169}}.

\bibitem{Qi:2018rqm}
Y.-H. Qi, Y.~Seo, S.-J. Sin and G.~Song\emph{},
\href{https://arxiv.org/abs/1804.06164}{{\ttfamily 1804.06164}}.

\bibitem{Gonzalez:2018enk}
H.~A. González, D.~Grumiller and J.~Salzer\emph{},
\href{https://arxiv.org/abs/1802.01562}{{\ttfamily 1802.01562}}.

\bibitem{Tarnopolsky:2018env}
G.~Tarnopolsky\emph{},  \href{https://arxiv.org/abs/1801.06871}{{\ttfamily
		1801.06871}}.

\bibitem{Witten:2016iux}
E.~Witten\emph{},  \href{https://arxiv.org/abs/1610.09758}{{\ttfamily
		1610.09758}}.

\bibitem{Gurau:2011xq}
R.~Gurau\emph{}, \href{https://doi.org/10.1007/s00023-011-0118-z}{\emph{Annales
		Henri Poincare} {\bfseries 13} (2012) 399--423},
[\href{https://arxiv.org/abs/1102.5759}{{\ttfamily 1102.5759}}].

\bibitem{Bonzom:2012hw}
V.~Bonzom, R.~Gurau and V.~Rivasseau\emph{},
\href{https://doi.org/10.1103/PhysRevD.85.084037}{\emph{Phys. Rev.}
	{\bfseries D85} (2012) 084037},
[\href{https://arxiv.org/abs/1202.3637}{{\ttfamily 1202.3637}}].

\bibitem{Carrozza:2015adg}
S.~Carrozza and A.~Tanasa\emph{},
\href{https://doi.org/10.1007/s11005-016-0879-x}{\emph{Lett. Math. Phys.}
	{\bfseries 106} (2016) 1531--1559},
[\href{https://arxiv.org/abs/1512.06718}{{\ttfamily 1512.06718}}].

\bibitem{Gurau:2016lzk}
R.~Gurau\emph{},
\href{https://doi.org/10.1016/j.nuclphysb.2017.01.015}{\emph{Nucl. Phys.}
	{\bfseries B916} (2017) 386--401},
[\href{https://arxiv.org/abs/1611.04032}{{\ttfamily 1611.04032}}].

\bibitem{Klebanov:2016xxf}
I.~R. Klebanov and G.~Tarnopolsky\emph{},
\href{https://doi.org/10.1103/PhysRevD.95.046004}{\emph{Phys. Rev.}
	{\bfseries D95} (2017) 046004},
[\href{https://arxiv.org/abs/1611.08915}{{\ttfamily 1611.08915}}].

\bibitem{Nishinaka:2016nxg}
T.~Nishinaka and S.~Terashima\emph{},
\href{https://doi.org/10.1016/j.nuclphysb.2017.11.012}{\emph{Nucl. Phys.}
	{\bfseries B926} (2018) 321--334},
[\href{https://arxiv.org/abs/1611.10290}{{\ttfamily 1611.10290}}].

\bibitem{Krishnan:2016bvg}
C.~Krishnan, S.~Sanyal and P.~N. Bala~Subramanian\emph{},
\href{https://doi.org/10.1007/JHEP03(2017)056}{\emph{JHEP} {\bfseries 03}
	(2017) 056}. 

\bibitem{Ferrari:2017ryl}
F.~Ferrari\emph{},  \href{https://arxiv.org/abs/1701.01171}{{\ttfamily
		1701.01171}}.

\bibitem{Gurau:2017xhf}
R.~Gurau\emph{}, \href{https://doi.org/10.1209/0295-5075/119/30003}{\emph{EPL}
	{\bfseries 119} (2017) 30003},
[\href{https://arxiv.org/abs/1702.04228}{{\ttfamily 1702.04228}}].

\bibitem{Bonzom:2017pqs}
V.~Bonzom, L.~Lionni and A.~Tanasa\emph{},
\href{https://doi.org/10.1063/1.4983562}{\emph{J. Math. Phys.} {\bfseries 58}
	(2017) 052301}, [\href{https://arxiv.org/abs/1702.06944}{{\ttfamily
		1702.06944}}].

\bibitem{Itoyama:2017emp}
H.~Itoyama, A.~Mironov and A.~Morozov\emph{},
\href{https://doi.org/10.1016/j.physletb.2017.05.043}{\emph{Phys. Lett.}
	{\bfseries B771} (2017) 180--188}.

\bibitem{Krishnan:2017ztz}
C.~Krishnan, K.~V.~P. Kumar and S.~Sanyal\emph{},
\href{https://doi.org/10.1007/JHEP06(2017)036}{\emph{JHEP} {\bfseries 06}
	(2017) 036}. 

\bibitem{Itoyama:2017xid}
H.~Itoyama, A.~Mironov and A.~Morozov\emph{},
\href{https://doi.org/10.1007/JHEP06(2017)115}{\emph{JHEP} {\bfseries 06}
	(2017) 115}, [\href{https://arxiv.org/abs/1704.08648}{{\ttfamily
		1704.08648}}].

\bibitem{Narayan:2017qtw}
P.~Narayan and J.~Yoon\emph{},
\href{https://doi.org/10.1007/JHEP08(2017)083}{\emph{JHEP} {\bfseries 08}
	(2017) 083}, [\href{https://arxiv.org/abs/1705.01554}{{\ttfamily
		1705.01554}}].

\bibitem{Chaudhuri:2017vrv}
S.~Chaudhuri, V.~I. Giraldo-Rivera, A.~Joseph, R.~Loganayagam and
J.~Yoon\emph{}, \href{https://arxiv.org/abs/1705.01930}{{\ttfamily
		1705.01930}}.

\bibitem{Gurau:2017qna}
R.~Gurau\emph{},  \href{https://arxiv.org/abs/1705.08581}{{\ttfamily
		1705.08581}}.

\bibitem{Dartois:2017xoe}
S.~Dartois, H.~Erbin and S.~Mondal\emph{},
\href{https://arxiv.org/abs/1706.00412}{{\ttfamily 1706.00412}}.

\bibitem{Klebanov:2017nlk}
I.~R. Klebanov and G.~Tarnopolsky\emph{},
\href{https://doi.org/10.1007/JHEP10(2017)037}{\emph{JHEP} {\bfseries 10}
	(2017) 037}, [\href{https://arxiv.org/abs/1706.00839}{{\ttfamily
		1706.00839}}].

\bibitem{Mironov:2017aqv}
A.~Mironov and A.~Morozov\emph{},
\href{https://doi.org/10.1016/j.physletb.2017.09.063}{\emph{Phys. Lett.}
	{\bfseries B774} (2017) 210--216},
[\href{https://arxiv.org/abs/1706.03667}{{\ttfamily 1706.03667}}].

\bibitem{Gurau:2017qya}
R.~Gurau\emph{},  \href{https://arxiv.org/abs/1706.05328}{{\ttfamily
		1706.05328}}.

\bibitem{Krishnan:2017txw}
C.~Krishnan and K.~V.~P. Kumar\emph{},
\href{https://doi.org/10.1007/JHEP10(2017)099}{\emph{JHEP} {\bfseries 10}
	(2017) 099}, [\href{https://arxiv.org/abs/1706.05364}{{\ttfamily
		1706.05364}}].

\bibitem{deMelloKoch:2017bvv}
R.~de~Mello~Koch, R.~Mello~Koch, D.~Gossman and L.~Tribelhorn\emph{},
\href{https://doi.org/10.1007/JHEP09(2017)011}{\emph{JHEP} {\bfseries 09}
	(2017) 011}, [\href{https://arxiv.org/abs/1707.01455}{{\ttfamily	1707.01455}}].

\bibitem{Giombi:2017dtl}
S.~Giombi, I.~R. Klebanov and G.~Tarnopolsky\emph{},
\href{https://doi.org/10.1103/PhysRevD.96.106014}{\emph{Phys. Rev.}
	{\bfseries D96} (2017) 106014}

\bibitem{Azeyanagi:2017drg}
T.~Azeyanagi, F.~Ferrari and F.~I. Schaposnik~Massolo\emph{},
\href{https://arxiv.org/abs/1707.03431}{{\ttfamily 1707.03431}}.

\bibitem{Bulycheva:2017ilt}
K.~Bulycheva, I.~R. Klebanov, A.~Milekhin and G.~Tarnopolsky\emph{},
\href{https://arxiv.org/abs/1707.09347}{{\ttfamily 1707.09347}}.

\bibitem{Choudhury:2017tax}
S.~Choudhury, A.~Dey, I.~Halder, L.~Janagal, S.~Minwalla and R.~Poojary\emph{},
\href{https://arxiv.org/abs/1707.09352}{{\ttfamily 1707.09352}}.

\bibitem{Krishnan:2017lra}
C.~Krishnan, K.~V.~P. Kumar and D.~Rosa\emph{},
\href{https://arxiv.org/abs/1709.06498}{{\ttfamily 1709.06498}}.

\bibitem{Azeyanagi:2017mre}
T.~Azeyanagi, F.~Ferrari, P.~Gregori, L.~Leduc and G.~Valette\emph{},
\href{https://arxiv.org/abs/1710.07263}{{\ttfamily 1710.07263}}.

\bibitem{Itoyama:2017wjb}
H.~Itoyama, A.~Mironov and A.~Morozov\emph{},
\href{https://arxiv.org/abs/1710.10027}{{\ttfamily 1710.10027}}.

\bibitem{Benedetti:2017fmp}
D.~Benedetti, S.~Carrozza, R.~Gurau and A.~Sfondrini\emph{},
\href{https://arxiv.org/abs/1710.10253}{{\ttfamily 1710.10253}}.

\bibitem{Halmagyi:2017leq}
N.~Halmagyi and S.~Mondal\emph{},
\href{https://arxiv.org/abs/1711.04385}{{\ttfamily 1711.04385}}.

\bibitem{BenGeloun:2017jbi}
J.~Ben~Geloun and V.~Rivasseau\emph{},
\href{https://arxiv.org/abs/1711.05967}{{\ttfamily 1711.05967}}.

\bibitem{Benedetti:2017qxl}
D.~Benedetti, S.~Carrozza, R.~Gurau and M.~Kolanowski\emph{},
\href{https://arxiv.org/abs/1712.00249}{{\ttfamily 1712.00249}}.

\bibitem{Benedetti:2018goh}
D.~Benedetti and R.~Gurau\emph{},
\href{https://arxiv.org/abs/1802.05500}{{\ttfamily 1802.05500}}.

\bibitem{Krishnan:2018jsp}
C.~Krishnan and K.~V. Pavan~Kumar\emph{},
\href{https://arxiv.org/abs/1804.10103}{{\ttfamily 1804.10103}}.

\bibitem{Delporte:2018iyf}
N.~Delporte and V.~Rivasseau\emph{},  2018,
\href{https://arxiv.org/abs/1804.11101}{{\ttfamily 1804.11101}},

\bibitem{Maldacena:2018vsr}
J.~Maldacena and A.~Milekhin\emph{},
\href{https://doi.org/10.1007/JHEP04(2018)084}{\emph{JHEP} {\bfseries 04}
	(2018) 084}, [\href{https://arxiv.org/abs/1802.00428}{{\ttfamily
		1802.00428}}].

\bibitem{Klebanov:2018nfp}
I.~R. Klebanov, A.~Milekhin, F.~Popov and G.~Tarnopolsky\emph{},
\href{https://arxiv.org/abs/1802.10263}{{\ttfamily 1802.10263}}.

\bibitem{Gross:2016kjj}
D.~J. Gross and V.~Rosenhaus\emph{},
\href{https://doi.org/10.1007/JHEP02(2017)093}{\emph{JHEP} {\bfseries 02}
	(2017) 093}, [\href{https://arxiv.org/abs/1610.01569}{{\ttfamily
		1610.01569}}].

\bibitem{Fu:2016vas}
W.~Fu, D.~Gaiotto, J.~Maldacena and S.~Sachdev\emph{},
\href{https://doi.org/10.1103/PhysRevD.95.069904,
	10.1103/PhysRevD.95.026009}{\emph{Phys. Rev.} {\bfseries D95} (2017) 026009}.

\bibitem{Peng:2016mxj}
C.~Peng, M.~Spradlin and A.~Volovich\emph{},
\href{https://doi.org/10.1007/JHEP05(2017)062}{\emph{JHEP} {\bfseries 05}
	(2017) 062}, [\href{https://arxiv.org/abs/1612.03851}{{\ttfamily
		1612.03851}}].

\bibitem{Murugan:2017eto}
J.~Murugan, D.~Stanford and E.~Witten\emph{},
\href{https://doi.org/10.1007/JHEP08(2017)146}{\emph{JHEP} {\bfseries 08}
	(2017) 146}, [\href{https://arxiv.org/abs/1706.05362}{{\ttfamily
		1706.05362}}].

\bibitem{Peng:2017spg}
C.~Peng, M.~Spradlin and A.~Volovich\emph{},
\href{https://doi.org/10.1007/JHEP10(2017)202}{\emph{JHEP} {\bfseries 10}
	(2017) 202}, [\href{https://arxiv.org/abs/1706.06078}{{\ttfamily
		1706.06078}}].

\bibitem{Sannomiya:2016mnj}
N.~Sannomiya, H.~Katsura and Y.~Nakayama\emph{},
\href{https://doi.org/10.1103/PhysRevD.95.065001}{\emph{Phys. Rev.}
	{\bfseries D95} (2017) 065001}.

\bibitem{Li:2017hdt}
T.~Li, J.~Liu, Y.~Xin and Y.~Zhou\emph{},
\href{https://doi.org/10.1007/JHEP06(2017)111}{\emph{JHEP} {\bfseries 06}
	(2017) 111}, [\href{https://arxiv.org/abs/1702.01738}{{\ttfamily
		1702.01738}}].

\bibitem{Forste:2017kwy}
S.~Forste and I.~Golla\emph{},
\href{https://doi.org/10.1016/j.physletb.2017.05.039}{\emph{Phys. Lett.}
	{\bfseries B771} (2017) 157--161},
[\href{https://arxiv.org/abs/1703.10969}{{\ttfamily 1703.10969}}].

\bibitem{Kanazawa:2017dpd}
T.~Kanazawa and T.~Wettig\emph{},
\href{https://doi.org/10.1007/JHEP09(2017)050}{\emph{JHEP} {\bfseries 09}
	(2017) 050}, [\href{https://arxiv.org/abs/1706.03044}{{\ttfamily
		1706.03044}}].

\bibitem{Hunter-Jones:2017raw}
N.~Hunter-Jones, J.~Liu and Y.~Zhou\emph{},
\href{https://arxiv.org/abs/1710.03012}{{\ttfamily 1710.03012}}.

\bibitem{Hunter-Jones:2017crg}
N.~Hunter-Jones and J.~Liu\emph{},
\href{https://arxiv.org/abs/1710.08184}{{\ttfamily 1710.08184}}.

\bibitem{Narayan:2017hvh}
P.~Narayan and J.~Yoon\emph{},
\href{https://arxiv.org/abs/1712.02647}{{\ttfamily 1712.02647}}.

\bibitem{Forste:2017apw}
S.~Forste, J.~Kames-King and M.~Wiesner\emph{},
\href{https://arxiv.org/abs/1712.07398}{{\ttfamily 1712.07398}}.

\bibitem{Garcia-Garcia:2018ruf}
A.~M. Garcia-Garcia, Y.~Jia and J.~J.~M. Verbaarschot\emph{},
\href{https://doi.org/10.1103/PhysRevD.97.106003}{\emph{Phys. Rev.}
	{\bfseries D97} (2018) 106003},
[\href{https://arxiv.org/abs/1801.01071}{{\ttfamily 1801.01071}}].

\bibitem{Bulycheva2018}
K.~Bulycheva\emph{},  \href{https://arxiv.org/abs/1801.09006v2}{{\ttfamily
		1801.09006v2}}.

\bibitem{Gu:2016oyy}
Y.~Gu, X.-L. Qi and D.~Stanford\emph{},
\href{https://doi.org/10.1007/JHEP05(2017)125}{\emph{JHEP} {\bfseries 05}
	(2017) 125}, [\href{https://arxiv.org/abs/1609.07832}{{\ttfamily
		1609.07832}}].

\bibitem{Berkooz:2016cvq}
M.~Berkooz, P.~Narayan, M.~Rozali and J.~Simon\emph{},
\href{https://doi.org/10.1007/JHEP01(2017)138}{\emph{JHEP} {\bfseries 01}
	(2017) 138}, [\href{https://arxiv.org/abs/1610.02422}{{\ttfamily
		1610.02422}}].

\bibitem{Davison:2016ngz}
R.~A. Davison, W.~Fu, A.~Georges, Y.~Gu, K.~Jensen and S.~Sachdev\emph{},
\href{https://doi.org/10.1103/PhysRevB.95.155131}{\emph{Phys. Rev.}
	{\bfseries B95} (2017) 155131},
[\href{https://arxiv.org/abs/1612.00849}{{\ttfamily 1612.00849}}].

\bibitem{Turiaci:2017zwd}
G.~Turiaci and H.~Verlinde\emph{},
\href{https://doi.org/10.1007/JHEP10(2017)167}{\emph{JHEP} {\bfseries 10}
	(2017) 167}, [\href{https://arxiv.org/abs/1701.00528}{{\ttfamily
		1701.00528}}].

\bibitem{Berkooz:2017efq}
M.~Berkooz, P.~Narayan, M.~Rozali and J.~Simon\emph{},
\href{https://doi.org/10.1007/JHEP09(2017)057}{\emph{JHEP} {\bfseries 09}
	(2017) 057}, [\href{https://arxiv.org/abs/1702.05105}{{\ttfamily
		1702.05105}}].

\bibitem{Gu:2017ohj}
Y.~Gu, A.~Lucas and X.-L. Qi\emph{},
\href{https://doi.org/10.21468/SciPostPhys.2.3.018}{\emph{SciPost Phys.}
	{\bfseries 2} (2017) 018},
[\href{https://arxiv.org/abs/1702.08462}{{\ttfamily 1702.08462}}].

\bibitem{Jian:2017unn}
S.-K. Jian and H.~Yao\emph{},
\href{https://doi.org/10.1103/PhysRevLett.119.206602}{\emph{Phys. Rev. Lett.}
	{\bfseries 119} (2017) 206602},
[\href{https://arxiv.org/abs/1703.02051}{{\ttfamily 1703.02051}}].

\bibitem{Chen:2017dav}
X.~Chen, R.~Fan, Y.~Chen, H.~Zhai and P.~Zhang\emph{},
\href{https://doi.org/10.1103/PhysRevLett.119.207603}{\emph{Phys. Rev. Lett.}
	{\bfseries 119} (2017) 207603}.

\bibitem{Chen:2017dbb}
Y.~Chen, H.~Zhai and P.~Zhang\emph{},
\href{https://doi.org/10.1007/JHEP07(2017)150}{\emph{JHEP} {\bfseries 07}
	(2017) 150}, [\href{https://arxiv.org/abs/1705.09818}{{\ttfamily
		1705.09818}}].

\bibitem{Zhang:2017jvh}
P.~Zhang\emph{}, \href{https://doi.org/10.1103/PhysRevB.96.205138}{\emph{Phys.
		Rev.} {\bfseries B96} (2017) 205138},
[\href{https://arxiv.org/abs/1707.09589}{{\ttfamily 1707.09589}}].

\bibitem{Jian:2017tzg}
S.-K. Jian, Z.-Y. Xian and H.~Yao\emph{},
\href{https://arxiv.org/abs/1709.02810}{{\ttfamily 1709.02810}}.

\bibitem{Simmons-Duffin:2017nub}
D.~Simmons-Duffin, D.~Stanford and E.~Witten\emph{},
\href{https://arxiv.org/abs/1711.03816}{{\ttfamily 1711.03816}}.

\bibitem{Cai:2017vyk}
W.~Cai, X.-H. Ge and G.-H. Yang\emph{},
\href{https://arxiv.org/abs/1711.07903}{{\ttfamily 1711.07903}}.

\bibitem{Ge:2017fix}
X.-H. Ge, S.-J. Sin, Y.~Tian, S.-F. Wu and S.-Y. Wu\emph{},
\href{https://arxiv.org/abs/1712.00705}{{\ttfamily 1712.00705}}.

\bibitem{Bulycheva:2017uqj}
K.~Bulycheva\emph{}, \href{https://doi.org/10.1007/JHEP12(2017)069}{\emph{JHEP}
	{\bfseries 12} (2017) 069},
[\href{https://arxiv.org/abs/1706.07411}{{\ttfamily 1706.07411}}].

\bibitem{Maldacena2016a}
J.~Maldacena, S.~H. Shenker and D.~Stanford\emph{},
\href{https://doi.org/10.1007/JHEP08(2016)106}{\emph{JHEP} {\bfseries 08}
	(2016) 106}, [\href{https://arxiv.org/abs/1503.01409}{{\ttfamily
		1503.01409}}].

\bibitem{Witten:1993yc}
E.~Witten\emph{},
\href{https://doi.org/10.1016/0550-3213(93)90033-L}{\emph{Nucl. Phys.}
	{\bfseries B403} (1993) 159--222},
[\href{https://arxiv.org/abs/hep-th/9301042}{{\ttfamily hep-th/9301042}}].

\bibitem{Murugan2017}
J.~Murugan, D.~Stanford and E.~Witten\emph{},
\href{https://doi.org/10.1007/JHEP08(2017)146}{\emph{JHEP} {\bfseries 08}
	(2017) 146}, [\href{https://arxiv.org/abs/1706.05362}{{\ttfamily
		1706.05362}}].

\bibitem{Peng2017}
C.~Peng\emph{}, \href{https://doi.org/10.1007/JHEP05(2017)129}{\emph{JHEP}
	{\bfseries 05} (2017) 129},
[\href{https://arxiv.org/abs/1704.04223}{{\ttfamily 1704.04223}}].

\bibitem{Peng2017a}
C.~Peng, M.~Spradlin and A.~Volovich\emph{},
\href{https://doi.org/10.1007/JHEP10(2017)202}{\emph{JHEP} {\bfseries 10}
	(2017) 202}, [\href{https://arxiv.org/abs/1706.06078}{{\ttfamily
		1706.06078}}].

\bibitem{Gaberdiel:2017hcn}
M.~R. Gaberdiel, W.~Li, C.~Peng and H.~Zhang\emph{},
\href{https://arxiv.org/abs/1711.07449}{{\ttfamily 1711.07449}}.

\bibitem{Gaberdiel:2017toap}
M.~R. Gaberdiel, W.~Li and C.~Peng\emph{}, .

\bibitem{Henneaux:2012ny}
M.~Henneaux, G.~Lucena~Gómez, J.~Park and S.-J. Rey\emph{},
\href{https://doi.org/10.1007/JHEP06(2012)037}{\emph{JHEP} {\bfseries 06}
	(2012) 037}.

\bibitem{Hanaki:2012yf}
K.~Hanaki and C.~Peng\emph{},
\href{https://doi.org/10.1007/JHEP08(2013)030}{\emph{JHEP} {\bfseries 08}
	(2013) 030}, [\href{https://arxiv.org/abs/1203.5768}{{\ttfamily 1203.5768}}].

\bibitem{Peng:2012ae}
C.~Peng\emph{}, \href{https://doi.org/10.1007/JHEP03(2013)054}{\emph{JHEP}
	{\bfseries 03} (2013) 054},
[\href{https://arxiv.org/abs/1211.6748}{{\ttfamily 1211.6748}}].

\bibitem{Gaberdiel:2014yla}
M.~R. Gaberdiel and C.~Peng\emph{},
\href{https://doi.org/10.1007/JHEP05(2014)152}{\emph{JHEP} {\bfseries 05}
	(2014) 152}, [\href{https://arxiv.org/abs/1403.2396}{{\ttfamily 1403.2396}}].

\bibitem{Gaberdiel:2017dbk}
M.~R. Gaberdiel, R.~Gopakumar, W.~Li and C.~Peng\emph{},
\href{https://doi.org/10.1007/JHEP04(2017)152}{\emph{JHEP} {\bfseries 04}
	(2017) 152}.

\bibitem{Gaberdiel:2015uca}
M.~R. Gaberdiel, C.~Peng and I.~G. Zadeh\emph{},
\href{https://doi.org/10.1007/JHEP10(2015)101}{\emph{JHEP} {\bfseries 10}
	(2015) 101}, [\href{https://arxiv.org/abs/1506.02045}{{\ttfamily
		1506.02045}}].

\bibitem{Gubser:2002tv}
S.~S. Gubser, I.~R. Klebanov and A.~M. Polyakov\emph{},
\href{https://doi.org/10.1016/S0550-3213(02)00373-5}{\emph{Nucl. Phys.}
	{\bfseries B636} (2002) 99--114}.

\bibitem{Shenker2015}
S.~H. Shenker and D.~Stanford\emph{},
\href{https://doi.org/10.1007/JHEP05(2015)132}{\emph{JHEP} {\bfseries 05}
	(2015) 132}, [\href{https://arxiv.org/abs/1412.6087}{{\ttfamily 1412.6087}}].

\bibitem{Perlmutter:2016pkf}
E.~Perlmutter\emph{},
\href{https://doi.org/10.1007/JHEP10(2016)069}{\emph{JHEP} {\bfseries 10}
	(2016) 069}, [\href{https://arxiv.org/abs/1602.08272}{{\ttfamily
		1602.08272}}].

\bibitem{Seiberg:1999xz}
N.~Seiberg and E.~Witten\emph{},
\href{https://doi.org/10.1088/1126-6708/1999/04/017}{\emph{JHEP} {\bfseries
		04} (1999) 017}, [\href{https://arxiv.org/abs/hep-th/9903224}{{\ttfamily
		hep-th/9903224}}].

\bibitem{Maldacena:2000hw}
J.~M. Maldacena and H.~Ooguri\emph{},
\href{https://doi.org/10.1063/1.1377273}{\emph{J. Math. Phys.} {\bfseries 42}
	(2001) 2929--2960}.

\bibitem{Aharony:2006th}
O.~Aharony, Z.~Komargodski and S.~S. Razamat\emph{},
\href{https://doi.org/10.1088/1126-6708/2006/05/016}{\emph{JHEP} {\bfseries
		05} (2006) 016}.

\bibitem{Gaberdiel:2017oqg}
M.~R. Gaberdiel, R.~Gopakumar and C.~Hull\emph{},
\href{https://doi.org/10.1007/JHEP07(2017)090}{\emph{JHEP} {\bfseries 07}
	(2017) 090}, [\href{https://arxiv.org/abs/1704.08665}{{\ttfamily
		1704.08665}}].

\bibitem{Ferreira:2017pgt}
K.~Ferreira, M.~R. Gaberdiel and J.~I. Jottar\emph{},
\href{https://doi.org/10.1007/JHEP07(2017)131}{\emph{JHEP} {\bfseries 07}
	(2017) 131}, [\href{https://arxiv.org/abs/1704.08667}{{\ttfamily
		1704.08667}}].

\bibitem{Giribet:2018ada}
G.~Giribet, C.~Hull, M.~Kleban, M.~Porrati and E.~Rabinovici\emph{},
\href{https://arxiv.org/abs/1803.04420}{{\ttfamily 1803.04420}}.

\bibitem{Gaberdiel:2018rqv}
M.~R. Gaberdiel and R.~Gopakumar\emph{},
\href{https://arxiv.org/abs/1803.04423}{{\ttfamily 1803.04423}}.

\bibitem{Gaberdiel:2012uj}
M.~R. Gaberdiel and R.~Gopakumar\emph{},
\href{https://doi.org/10.1088/1751-8113/46/21/214002}{\emph{J. Phys.}
	{\bfseries A46} (2013) 214002},
[\href{https://arxiv.org/abs/1207.6697}{{\ttfamily 1207.6697}}].

\bibitem{Chang:2012kt}
C.-M. Chang, S.~Minwalla, T.~Sharma and X.~Yin\emph{},
\href{https://doi.org/10.1088/1751-8113/46/21/214009}{\emph{J. Phys.}
	{\bfseries A46} (2013) 214009}. 


\end{thebibliography}
\end{document}